\documentclass[a4paper,usenatbib,useAMS]{mnras}
\pdfoutput=1
\usepackage[T1]{fontenc}
\usepackage{ae,aecompl}
\usepackage{graphicx}	
\usepackage{amsmath}	
\usepackage{amssymb}	
\usepackage{bm}		
\usepackage{pdflscape} 	
\usepackage{lscape}

\usepackage[justification=centering]{caption}
\usepackage{longtable}
\usepackage{footnote}
\usepackage{natbib}
\usepackage{subfigure}
\usepackage{times}
\usepackage{txfonts}
\usepackage{color}
\usepackage{makecell}
\usepackage{caption}

\usepackage{etoolbox}
\usepackage{subfigure}
\usepackage{longtable}

\title{How Are Gamma-Ray Burst Radio Afterglows Populated?}

\author[Zhang et al.]{
K. Zhang$^{1, 3}$,
Z. B. Zhang$^{1}$\thanks{E-mail: astrophy0817@163.com},
Y. F. Huang$^{2}$\thanks{E-mail: hyf@nju.edu.cn},
L. M. Song$^{3}$,
S. J. Zheng$^{3}$,
and X. J. Li$^{1}$
\\
$^{1}$College of Physics and Engineering, Qufu Normal University, Qufu 273165, P. R. China\\
$^{2}$School of Astronomy and Space Science, Nanjing University, Nanjing 210023, P. R. China\\
$^{3}$Key Laboratory of Particle Astrophysics, Institute of High Energy Physics, Chinese Academy of Sciences, Beijing 100049, P. R. China\\
}
\date{Accepted XXX. Received YYY; in original form ZZZ}

\pubyear{2019}

\begin{document}
\label{firstpage}
\pagerange{\pageref{firstpage}--\pageref{lastpage}}
\maketitle

\begin{abstract}
We systematically analyze three GRB samples named as radio-loud, radio-quiet and radio-none afterglows, respectively. It is shown that dichotomy of the radio-loud afterglows is not necessary. Interestingly, we find that the intrinsic durations ($T_{int}$), isotropic energies of prompt gamma-rays ($E_{\gamma, iso}$) and redshifts ($z$) of their host galaxies are log-normally distributed for both the radio-loud and radio-quiet samples except those GRBs without any radio detections. Based on the distinct distributions of $T_{int}$, $E_{\gamma, iso}$, the circum-burst medium density ($n$) and the isotropic equivalent energy of radio afterglows ($L_{\nu,p}$), we confirm that the GRB radio afterglows are really better to be divided into the dim and the bright types. However, it is noticeable that the distributions of flux densities ($F_{host}$) from host galaxies of both classes of radio afterglows are intrinsically quite similar. Meanwhile, we point out that the radio-none sample is also obviously different from the above two samples with radio afterglows observed, according to the cumulative frequency distributions of the $T_{int}$ and the $E_{\gamma, iso}$, together with correlations between $T_{int}$ and $z$. In addition, a positive correlation between $E_{\gamma, iso}$ and $L_{\nu,p}$ is found in the radio-loud samples especially for the supernova-associated GRBs. Besides, we also find this positive correlation in the radio-quiet sample. A negative correlation between $T_{int}$ and $z$ is confirmed to hold for the radio-quiet sample too. The dividing line between short and long GRBs in the rest frame is at $T_{int}\simeq$1 s. Consequently, we propose that the radio-loud, the radio-quiet and the radio-none GRBs could be originated from different progenitors.
\end{abstract}


\section{Introduction}

Gamma-Ray Bursts(GRBs) are instantaneous brightening event of gamma rays in the distant universe. After it was reported in 1973 \citep{Klebesadel+0}, a lot of properties of progenitors have been investigated by many previously theoretical and observational researches, see review papers \citep[e.g.,][]{Piran+0,Zhang+1} for details. Study of GRB afterglows is crucial to understand the central engine and the environment of distinct progenitors. The general interpretation is that a sudden energy release will produce a high temperature fireball expanding at a relativistic speed. The internal dissipation of the fireball leads to the gamma-rays, and the blast wave against the external medium produce the afterglow (\citealt{Meszaros+0}, \citealt{Rees+0}, \citealt{Rees+1}). The hydrodynamic evolution of the jetted outflows from the ultra-relativistic phase to the non-relativistic phase has been studied by a few authors (e.g., \citealt{Huang+99,Huang+03}). But there are many questions remaining for GRBs, such as how the inner engine runs, the reason of flares in afterglow and so on (\citealt{Woosley+0}, \citealt{Paczynski+0}, \citealt{Duncan+0}, \citealt{Becerra+0}, \citealt{Hascoet+0}, \citealt{Mu+0}). As illustrated in \citet{Chandra+0}, the detecting rates of X-ray and optical afterglows are higher than that of radio afterglows. Due to the relatively longer timescale of radio afterglows, one can have more opportunities to observe the radio afterglows in detail at a later period. In particular, the rebrightening phenomena of some radio afterglows caused by multiple activities of the inner engine of GRBs \citep{Li+0}, energy injection\citep{Geng+0}, supernova (SN) components or the forward and reverse shock can be detected and utilized to constrain the above theoretical rebrightening models. At the same time, the statistical classifications of radio afterglows become more and more important and feasible with the data accumulation of the radio afterglows.

\citet{Chandra+0} sorted 304 GRBs radio afterglows, and found the detection rate to be about 31\% that is obviously lower than those of X-ray and optical afterglows even after the \textit{Swift} satellite was launched to detect more X-ray and optical afterglows than before. Also, they sorted radio afterglows at 8.5 GHz for detection and 3$\sigma$ upper limit between 5 and 10 days and found that there was only little difference between them. The tiny difference was thought to be resulted from the telescope sensitivity (\citealt{Chandra+0}). However, \citet{Hancock+0} pointed out that the instrumental sensitivity was not the intrinsic reason for the difference mentioned above and they found that 60\% $\sim$ 70\% of the radio selected GRB samples are truly radio bright, while the convinced  fraction of the radio faint GRBs is about one third. \citet{Chandra+0} found that there was an apparent correlation between the detectability and the energy of GRBs which may cause the diverse detection rates for the radio bright and faint GRB samples. To reduce the influence of many unknown reasons on classifications in terms of the radio brightness, \citet{Lloyd+0} and \citet{Lloyd+1} selected the GRBs with larger isotropic energy($E_{\gamma,iso}>10^{52}erg$) in prompt gamma-rays, and divided them into two sub-samples, that is radio-loud and radio-quiet types. They proposed that the two subsamples might be generated from different progenitors; that is the radio-loud GRBs might be produced from the He-merger while the radio-quiet GRBs may be interpreted by the core-collapse of massive stars.

Owing to the relatively less brightness of GRBs in radio bands, whether the radio afterglows can be classified into any subclasses is still controversial. With the increase of radio afterglow numbers, statistical study becomes more and more reliable and important. Motivated by the above incongruous results, we do a similar analysis but for different samples of GRB radio afterglows in very detail. In addition, we will examine the effects of surrounding mediums and GRB host galaxies on the GRB classifications in radio bands. In order to deduce their potential progenitors, several supernova-associated GRBs with radio afterglow measurements are also included. Simultaneously, we also pay attention to GW170817/GRB170817A detected by Laser Interferometer Gravitational-Wave Observatory (LIGO) and Fermi/Integral satellites (Goldstein et al. 2017; Savchenko et al. 2017) as the first short GRB associated with Kilonova originated from a binary neutron star merger system \citep{Abbott+0}.

\section{DATA PREPARATION}
\label{sec:degeneracy}

First of all, we define our sampling criteria in the following: (1) GRBs with radio flux density larger than 3$\sigma$ error bars constitute the radio-loud sample; (2) those radio afterglows with flux density lower than 3$\sigma$ levels belong to the radio-quiet (including upper limits)  sample; (3) other GRBs without any radio flux detections comprise our radio-none sample. The fact that no radio afterglow is reported (not even an upper limit) for a given burst could be because the telescope was down, or the PI ran out of their budget, or the burst did not fulfill the team's observational criteria which may be a bright optical/X-ray afterglows, or proximity, or something like that. However, each burst involved in our radio-none sample was indeed observed by some radio telescopes or array, but no meaningful flux densities were reported according to \cite{Chandra+0}. Most probably, radio afterglows of the radio-none sample could exist but are extremely too weaker to be detected by the current instruments due to sensitivity limits. We choose the GRBs with measured redshift ($z$) to calculate the intrinsic duration $T_{int}$ and isotropic equivalent energy $E_{\gamma,iso}$ (the intrinsic duration defined as $ T_{int}= T_{90}/(1+z) $, where $ T_{90} $ is defined as the time that the burst takes from 5 to 95 percent counts of the total gamma-rays, \citealt{Kouveliotou+0}).

\citet{Chandra+0} reported a large sample of GRB radio afterglows, of which the majority were detected by the Very Large Array (VLA) or Expanded Very Large Array (EVLA), and a small fraction of these radio afterglows were successfully observed by the Australia Telescope Compact Array (ATCA), Westerbork Synthesis Radio Telescope (WSRT), Giant Metrewave Radio Telescope (GMRT) and the Very Long Baseline Array (VLBA). Out of the 304 GRBs in \cite{Chandra+0}, we have selected 84 detections and 63 upper limits from the VLA-based afterglows, of which 79 radio-loud, 48 radio-quiet and 25 radio-none bursts with known redshift are involved (hereafter called the VLA-based sample). To compare with the recent high-frequency radio afterglows detected by the Arcminute Microkelvin Imager (AMI) telescope, we have taken 45 detections and 74 upper limits out of 139 bursts at 15.7 GHz from \cite{Anderson+0}, from which 21 radio-loud and 34 radio-quiet AMI afterglows with measured redshift are picked out to study the rest-frame features (hereafter called the AMI sample). It is likely that the lower $E_{\gamma,iso}$ bursts in the SN-associated GRB sample are relatively brighter in radio bands in contrast with other bursts. To explore the interesting issue, we have paid particular attention to the SN-associated GRBs and chosen 23 SN/GRBs as a unique subgroup including 21 radio-loud and 2 radio-quiet GRBs. It is noticeable that more than 90 percent of SN/GRB afterglows are radio-loud and the redshifts of all the SN/GRBs in our sample are well known. Moreover, \cite{Lloyd+0} and \cite{Lloyd+1} only chose those energetic bursts with $E_{\gamma,iso}>10^{52}$ ergs, which will inevitably bias the results of radio quiet afterglows since the $E_{\gamma,iso}$ and radio peak luminosity are positively correlated for different kinds of bursts as described in Sec. 3.6.

All the above samples of radio afterglows are compiled in Tables \ref{Table1:radio-loud} and \ref{Table2:radio-quiet}, in which the key parameters of radio-loud and radio-quiet GRBs are similarly presented. Column 1 gives the name of GRBs; Columns 2 and 3 are respectively the duration ($T_{90}$) and the redshift ($z$); In Column 4, we list the $k$-corrected isotropic energies ($E_{\gamma,iso}$) in $\gamma$-ray band; Column 5 gives the medium densities $n$; Column 6 provides the \textbf spectral peak luminosity ($L_{\nu,p}$) of radio afterglows at a frequency of 8.5 GHz or 15.7 GHz; In Columns 7 and 8, we present the peak radio flux density together $1\sigma$ RMS at 8.5 GHz or 15.7 GHz; Column 9 list the radio telescopes which were used to carry out observations; References are given in Column 10. In Table \ref{Table3:radio-none}, we only provide the values of $T_{90}$, $z$, $E_{\gamma,iso}$ and $n$ along with the employed radio telescope in order for the radio-none bursts. If there is no any parameters measured, we just leave them blank. To investigate the properties of host galaxies for different kinds of radio samples, we directly utilize the data of radio flux densities for host galaxies in \citet{Li+0} and \citet{Zhang+0}.

\section{RESULTS}
\subsection{Flux density of radio afterglows}
\begin{figure*}
	\centering
	\subfigure{
		\includegraphics[width=0.45\textwidth]{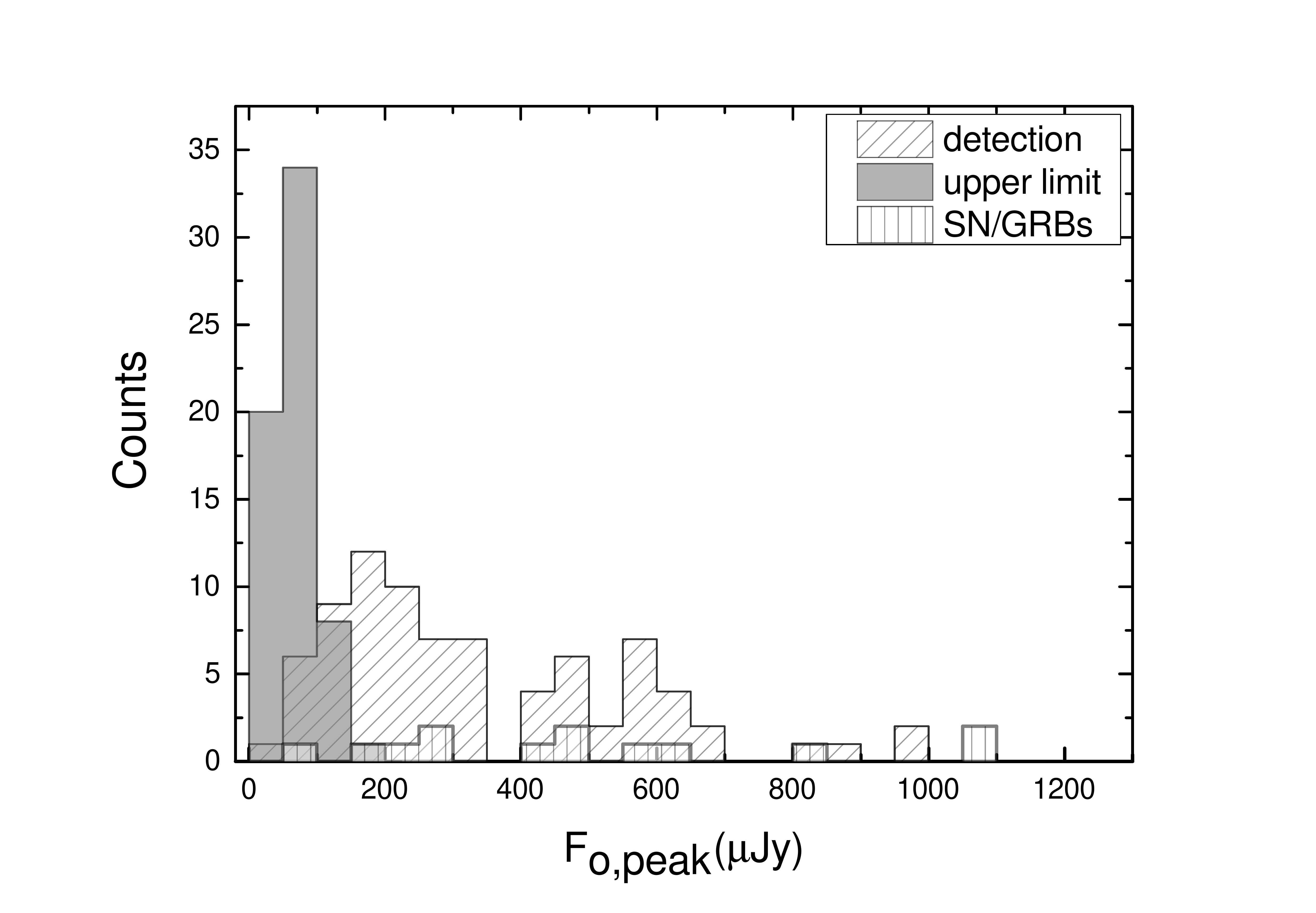}
	}
	\subfigure{
		\includegraphics[width=0.45\textwidth]{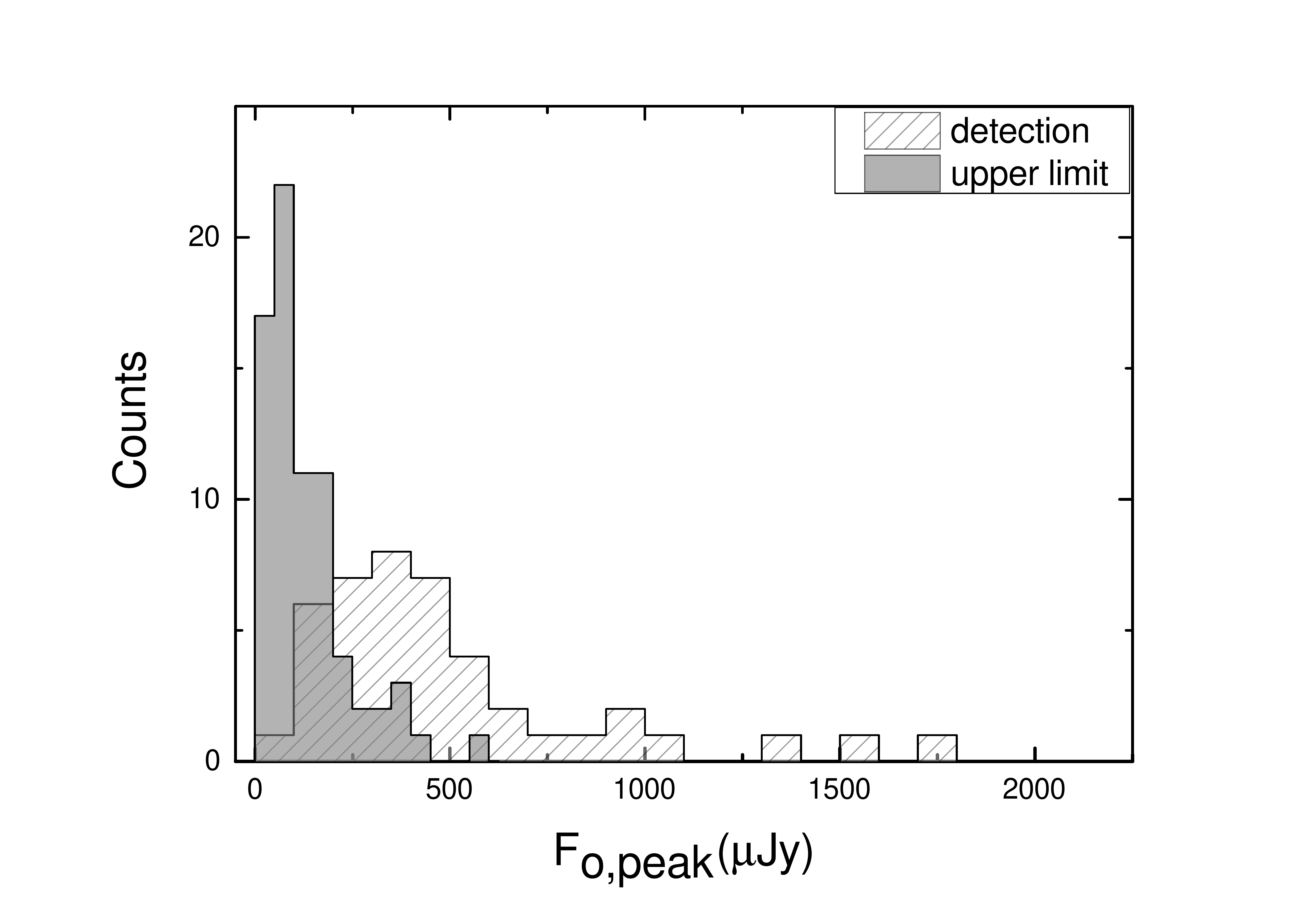}
	}
		\subfigure{
		\includegraphics[width=0.45\textwidth]{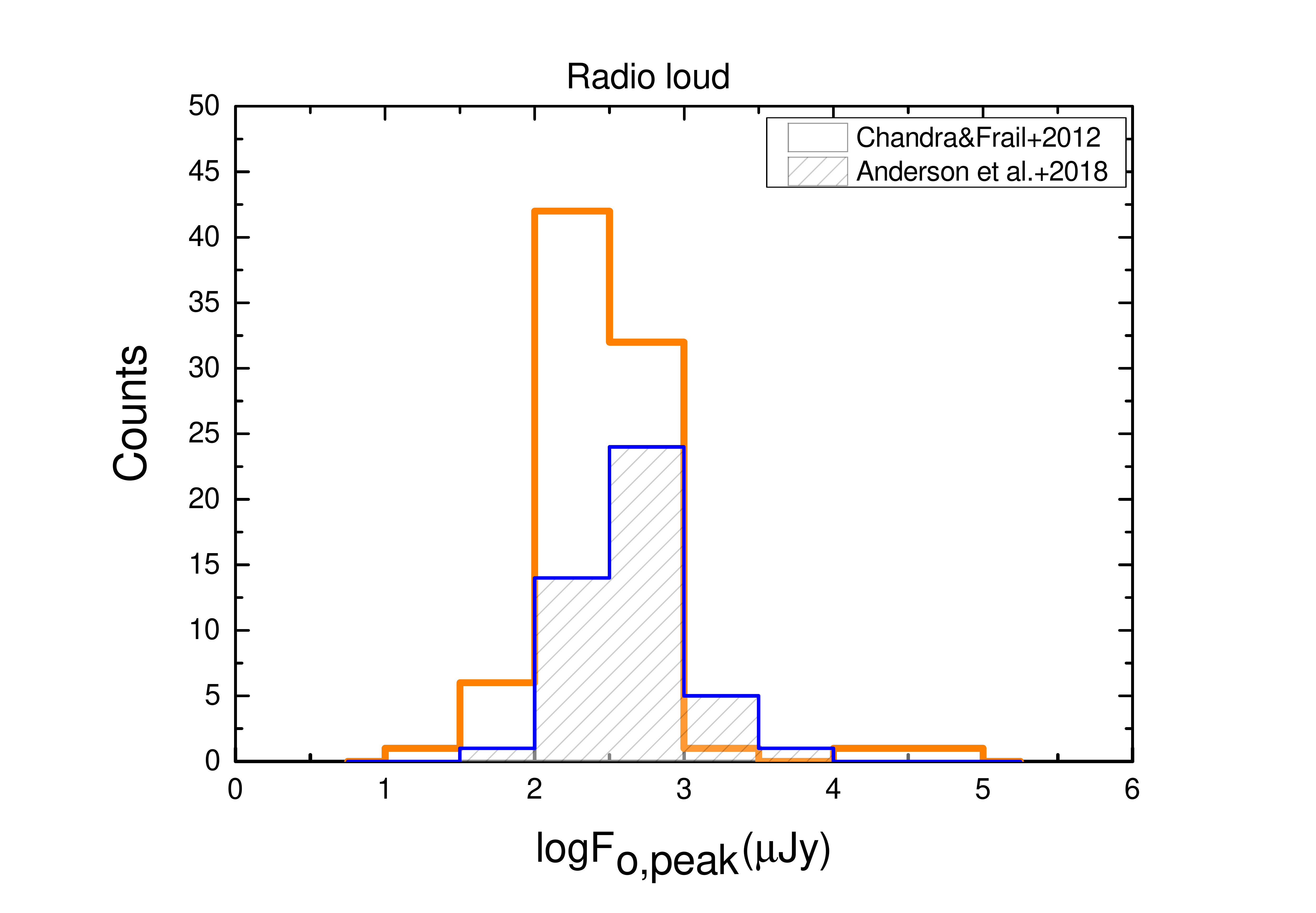}
	}
	\subfigure{
		\includegraphics[width=0.45\textwidth]{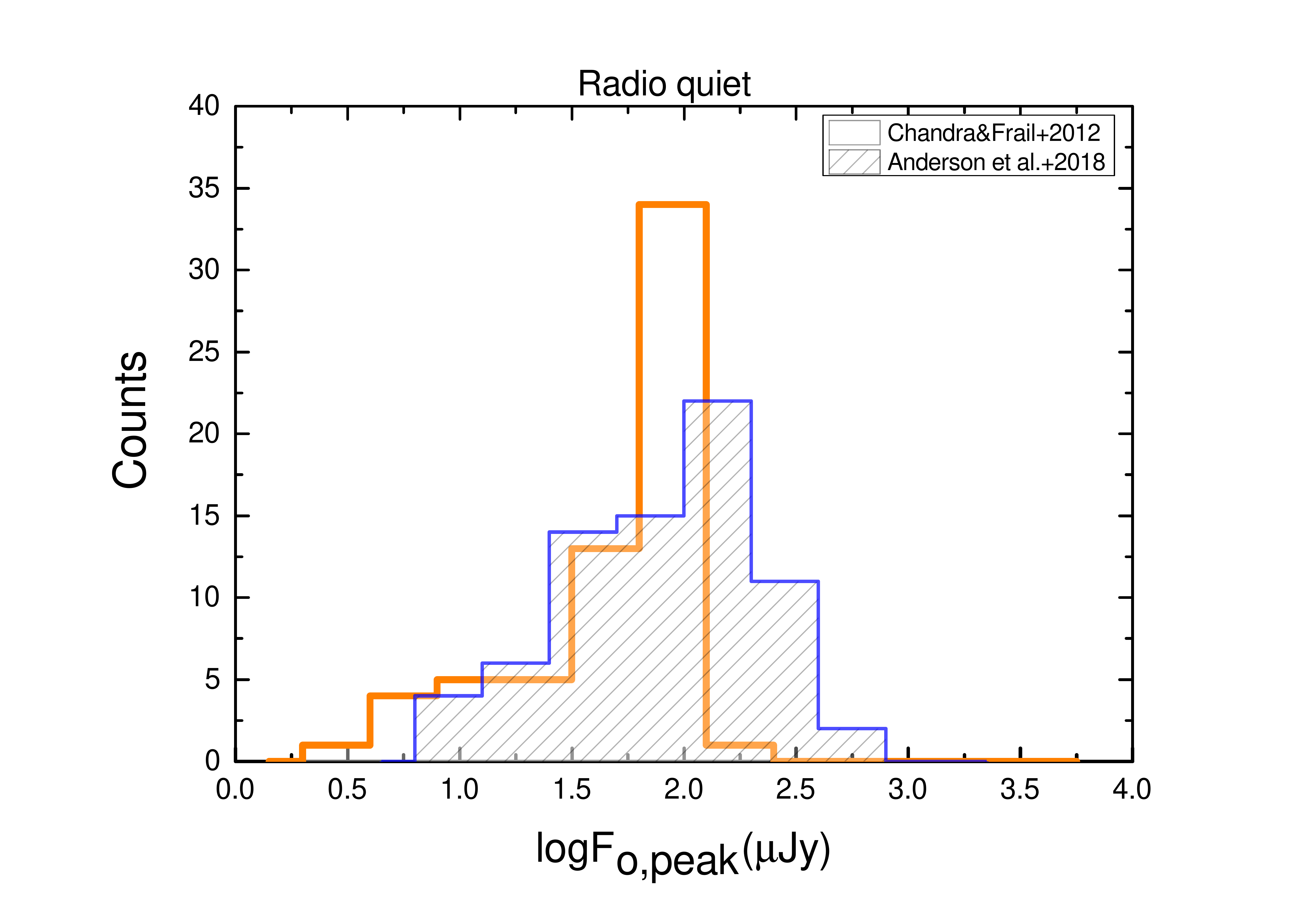}
}
	\caption{The distributions of peak flux densities for detections and upper limits of radio-loud (filled) and radio-quiet (hatched) afterglows between 0-10 days taken from \citet{Chandra+0} at 8.5 GHz and \citet{Anderson+0} at 15.7 GHz are shown on the top-left and top-right panels, respectively. The vertical shadow on the top-left panel represents the distributional histogram of SN/GRBs. Two bottom panels display the peak flux comparisons of radio afterglows at different frequencies for radio-loud (left) and radio-quiet (right) samples in the logarithmic scale.}
	\label{fig1---Flux Dis}
\end{figure*}

We first plot the distributions of radio afterglows for detections and 3$\sigma $ upper limits between 0 and 10 days at 8.5 GHz in top-left panel of  Figure \ref{fig1---Flux Dis}, where it is found that our distributions are similar to those in \citet{Chandra+0} and \citet{Hancock+0}, in which the upper limits are confirmed again to peak at 50-100 $\mu$Jy in and the detections peaked around 200 $\mu$Jy with a long extending tail. We also find that there is an obvious truncation at $\sim$ 400 $\mu$Jy in the VLA-based detection sample, which motivates us to examine whether the distribution of the flux densities less than 400 $\mu$Jy is associated with that of the upper-limit sample. For the purpose, we try to define the detection sample whose flux density larger than 400 $\mu$Jy as radio-loud I sample, and other detections with radio flux density less than 400 $\mu$Jy to be radio-loud II sample, temporally. It is interestingly found from the bottom panels of Figure \ref{fig1---Flux Dis} that the flux density distributions of radio-loud and radio-quiet AMI afterglows are also bimodally distributed and resemble those of the VLA-based sample. However, the AMI peak flux densities of both detections and upper limits are on average two times larger than those VLA-based ones, correspondingly.

\begin{figure*}
	\centering
	\subfigure{
		\includegraphics[width=0.45\textwidth]{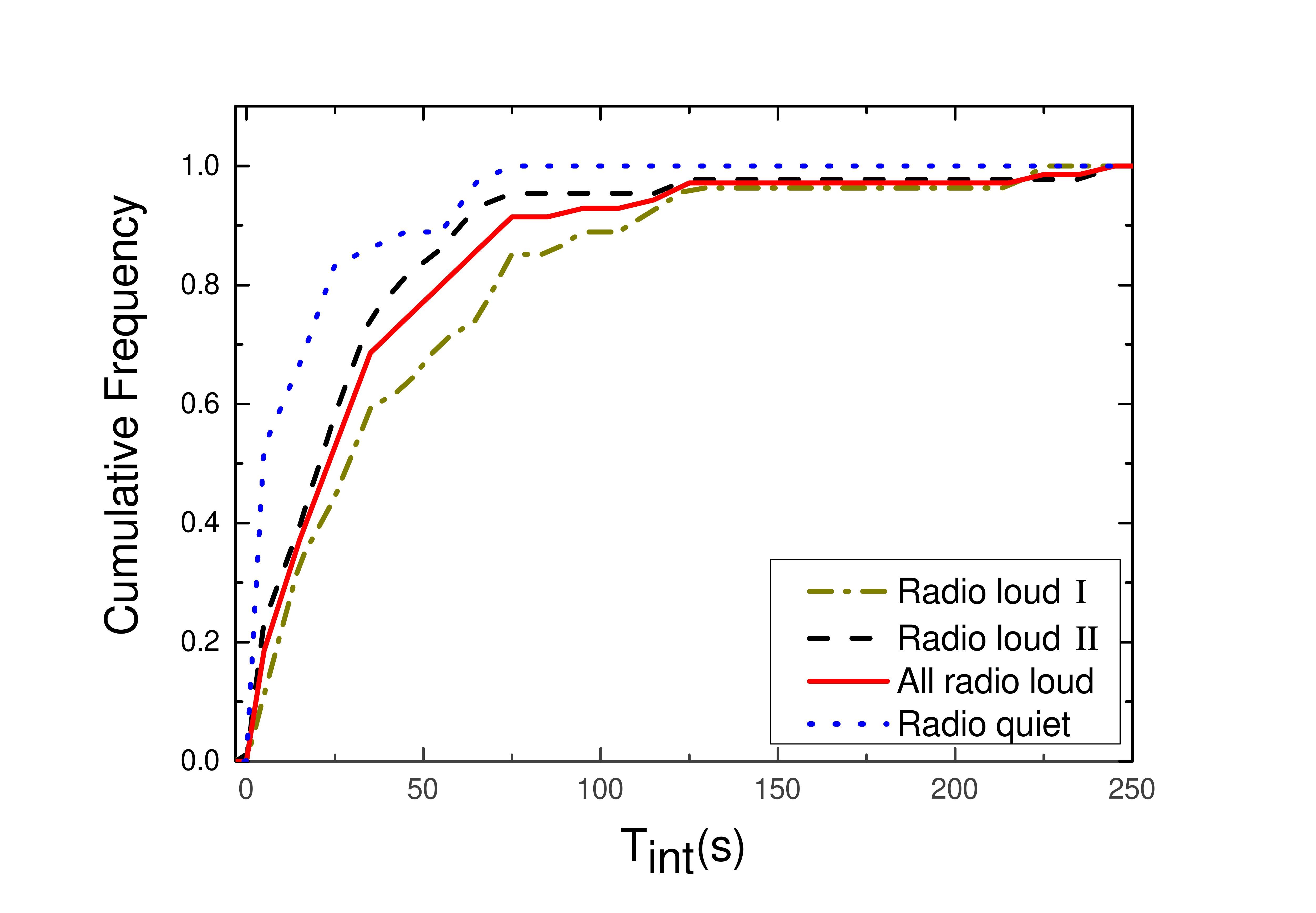}
	}
	\subfigure{
		\includegraphics[width=0.45\textwidth]{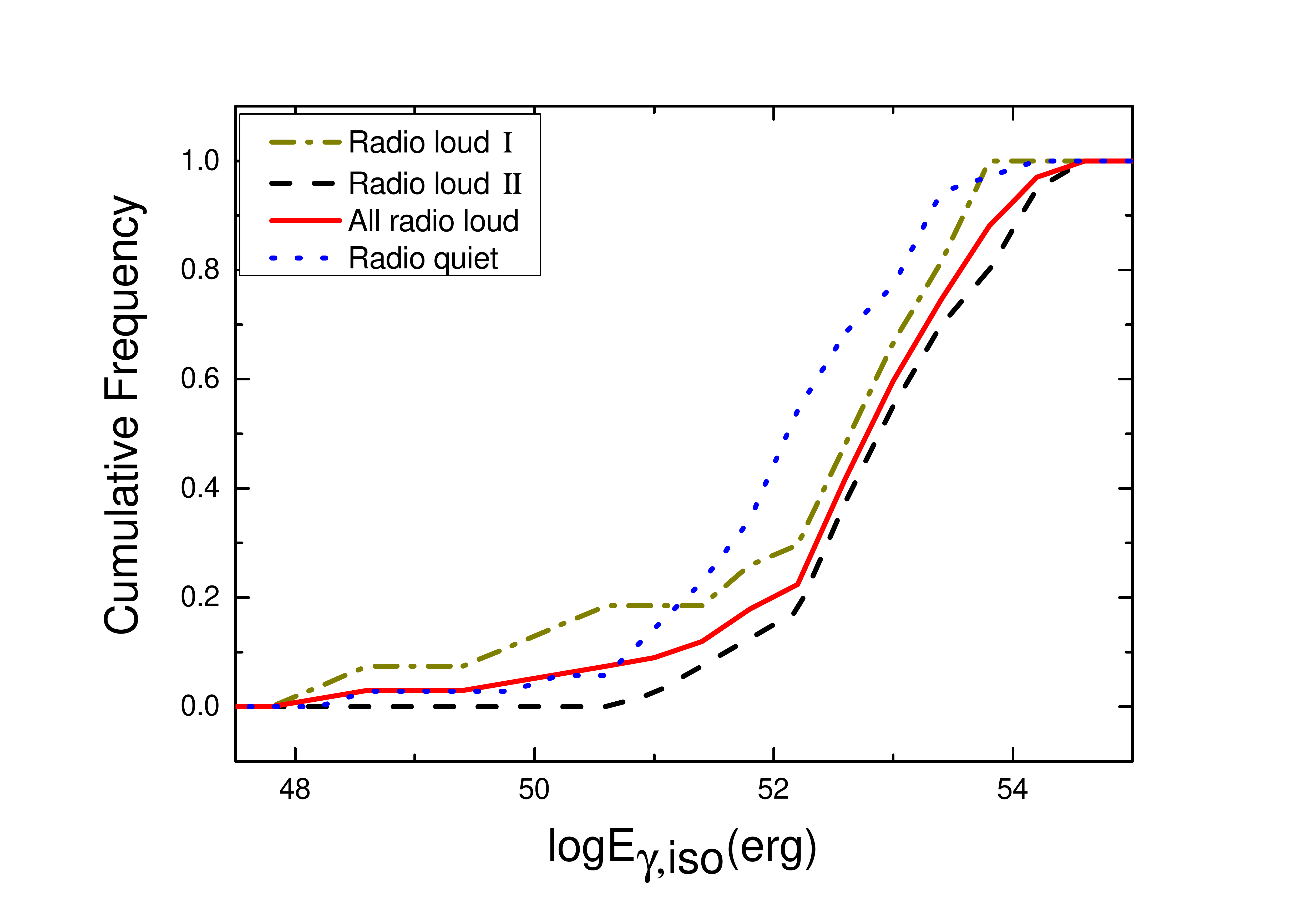}
	}
	\caption{The cumulative fractions of $T_{int}$ and $E_{\gamma,iso}$ are shown for different radio-selected GRBs in left and right panels, respectively.}
	\label{fig2--Tint/Eiso dis1}
\end{figure*}

To check if it is necessary to reclassify radio-loud GRBs into two subsamples, we display the cumulative fractions of the intrinsic duration $T_{int}$ and the $E_{\gamma,iso}$ for radio-loud I, radio-loud II and radio-quiet GRBs (upper limits) in Figure \ref{fig2--Tint/Eiso dis1}. As shown in Table \ref{Table5:k-s test}, the Kolmogorov-Smirnov (K-S) tests return the statistic $D=0.48$ (0.31) and $P=7\times10^{-4}$ (0.031) between the $T_{int}$ distributions of the radio-loud I (II) and the radio-quiet samples showing the radio-quiet bursts are different from either radio-loud I or II. Similarly, the statistic and p-value of the $E_{\gamma,iso}$ distributions are $D=0.35$ (0.38) and $P=0.037$ (0.004) for comparisons between the radio-quiet and the radio-loud I (II) samples. Surprisingly, the K-S test to the radio-loud I and the radio-loud II samples returns $D=0.27$ with $P=0.24$ for the $T_{int}$ distribution and $D=0.14$ with $P=0.92$ for the $E_{\gamma,iso}$ distribution, indicating that the two radio-loud sub-samples should be taken from the same parent distribution. In other words, dividing radio-loud bursts into two classes is not necessary. Consequently, we shall only investigate the radio-loud, the radio-quiet and the radio-none samples in the subsequent sections, and explore in statistics whether they are basically different kinds of bursts on basis of their observational properties.

\subsection{Distributions of $z$, $T_{int}$ and $E_{\gamma,iso}$ revisited}
Using the total sample of 206 GRBs including 151 VLA-based and 55 AMI bursts, we plot the histograms of $z$, $E_{\gamma,iso}$ and $T_{int}$ for radio-loud, radio-quiet and radio-none samples in Figure \ref{Figure12-histogram}, where one can find that the distributions of $z$, $T_{int}$ and $E_{\gamma,iso}$ of radio-loud and radio-quiet samples are well fitted by a gaussian function, but the radio-none sample seems to be eccentric (see Appendix for a detail). The fitting results are summarized in Table \ref{Table4:statistical para}, from which we notice that the mean values of $z$, $E_{\gamma,iso}$ and $T_{int}$ of radio-none GRBs are systematically smaller than those of the other two samples. In particular, the isotropic energies of radio-none bursts are on average one order of magnitude lower than the $E_{\gamma,iso}$ values of either radio-loud or radio-quiet bursts.

\begin{figure*}
	\centering
	\subfigure{
		\includegraphics[width=0.45\textwidth]{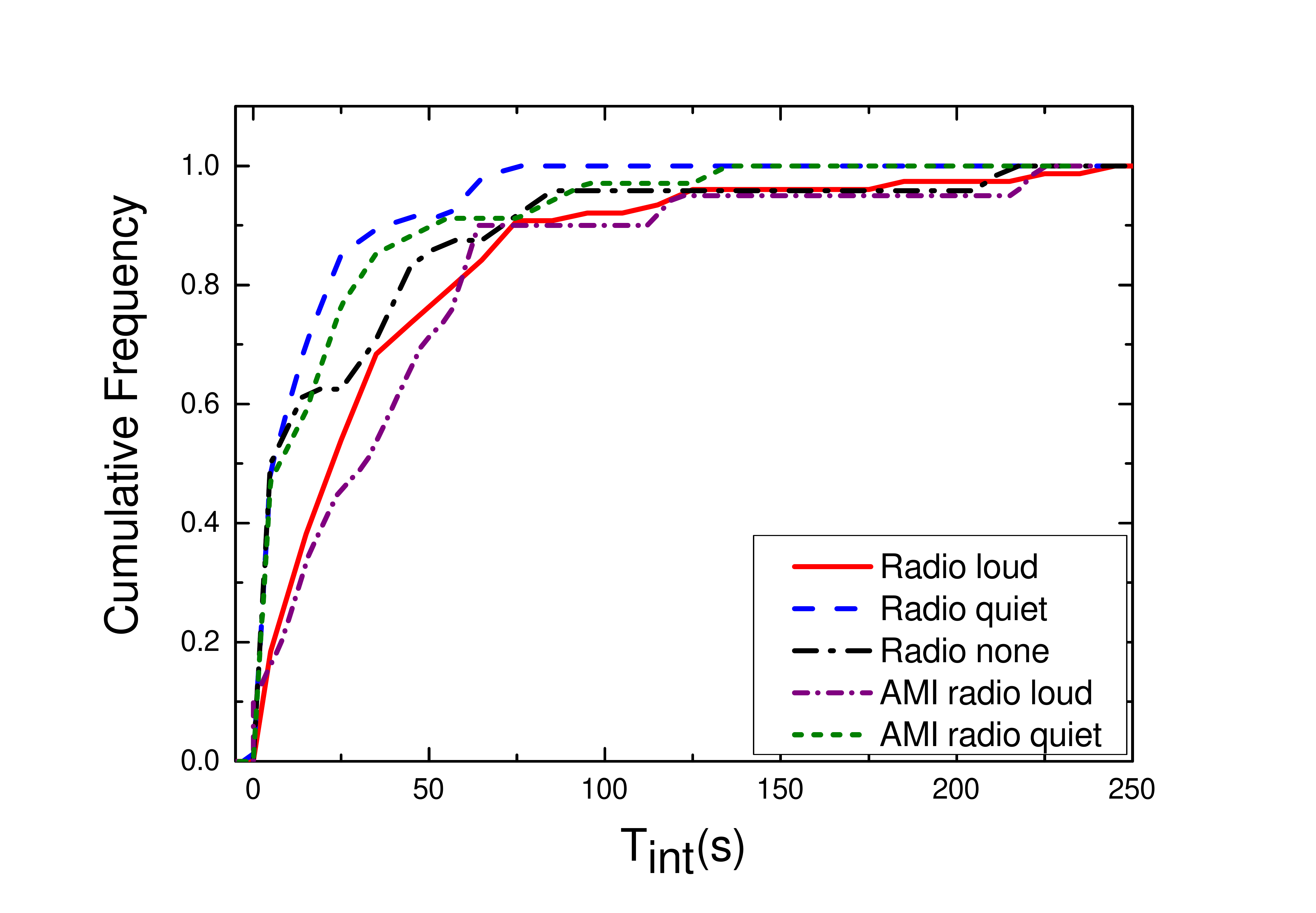}
	}
	\subfigure{
		\includegraphics[width=0.45\textwidth]{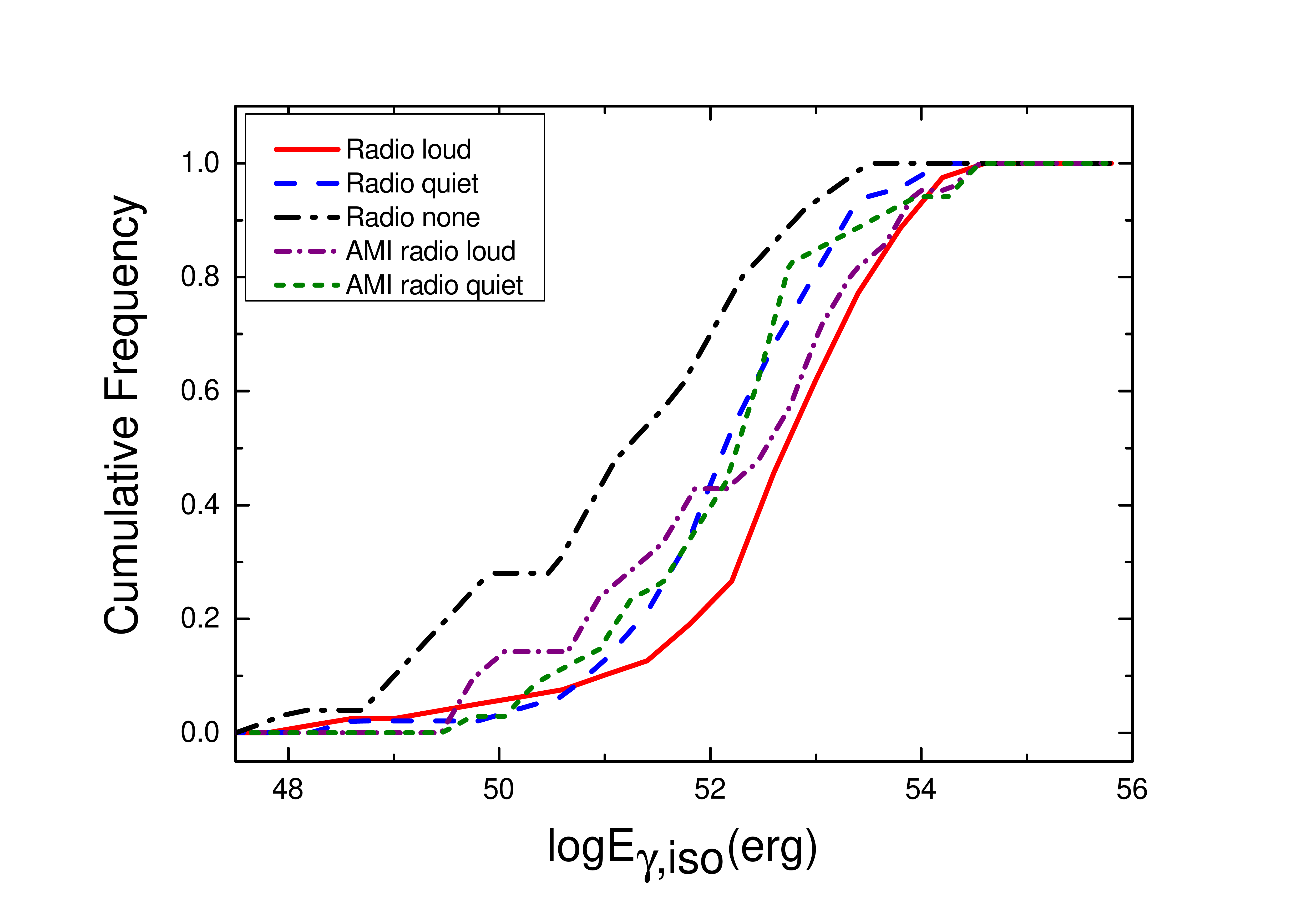}
	}
	\caption{Cumulative fractions of $T_{int}$ (left panel) and $E_{\gamma,iso}$ (right panel) are plotted for the VLA-based radio-loud (red solid line), radio-quiet (blue dashed line), radio-none (black dot-dashed line) bursts and for the AMI Radio-loud (purple short dash-dotted line) and AMI Radio-quiet (green short dashed line) GRBs.}
	\label{Fig3--Tint/Eiso dis2}
\end{figure*}

Following \citet{Hancock+0} and \citet{Lloyd+1}, we also analyze the cumulative fractions of the $T_{int}$ and the $E_{\gamma,iso}$ but for different radio-loud, radio-quiet and radio-none VLA-based samples in Figure \ref{Fig3--Tint/Eiso dis2} and Table \ref{Table5:k-s test}, where we see that the radio-quiet samples are evidently different from the radio-loud ones in terms of the $T_{int}$ distribution, on average the radio-loud GRBs have relatively longer $T_{int}$ as found before \citep{Hancock+0,Lloyd+1}. However, the K-S test demonstrates that the $T_{int}$ distributions of radio-quiet and radio-none GRBs are indistinguishable. Regarding the $E_{\gamma,iso}$ distributions, we also perform the K-S tests to any two of the above three VLA-based samples and find from Table \ref{Table5:k-s test} that they are drawn from different parent distributions. In addition, the median discrepancy of $E_{\gamma,iso}$ between the radio-loud and the radio-none bursts is about two orders of magnitude. 
With the increase of frequency, it is interestingly found that two AMI samples of radio-loud and radio-quiet GRBs at 15.7 GHz are consistent with being drawn from the same parent $E_{\gamma,iso}$ distribution.
\subsection{Radio fluxes of host galaxies}
\begin{figure*}
	\centering
	\includegraphics[width=0.5\textwidth]{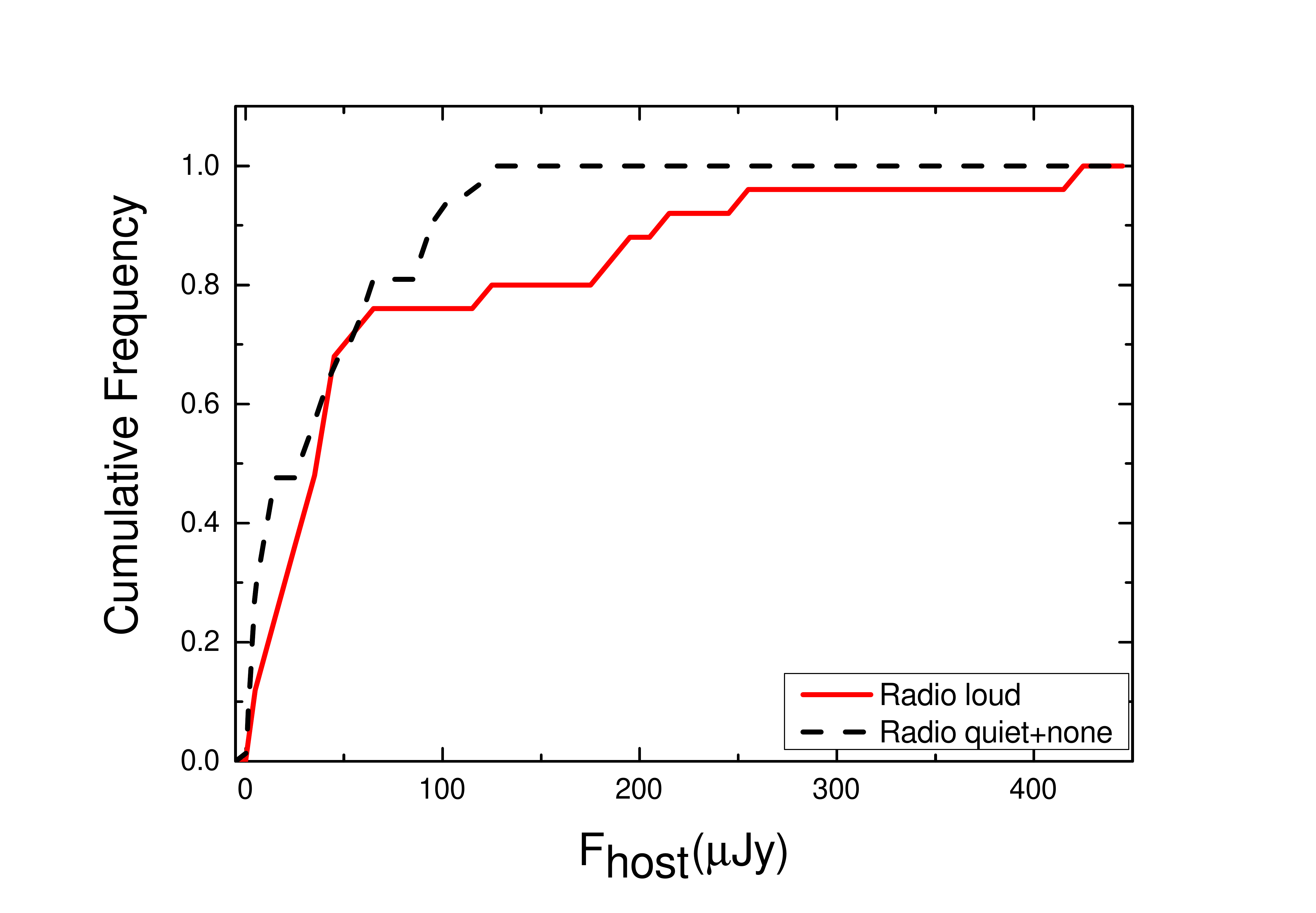}
	\caption{Cumulative fractions of $F_{host}$ for radio-loud (solid red line), radio-faint (black dashed line) samples.}
	\label{Fig4--Fhost dis}
\end{figure*}

We notice that some GRBs with radio flux densities of host galaxies in \citet{Zhang+0} were not included in our initial radio-loud, radio-quiet or none samples. To increase the reliability in statistics, we assume them to be radio-none or radio-quiet because they don't have radio afterglows detected. In order to analyze the radio flux density of host galaxies for the three samples, we combine the data of radio-none and radio-quiet into a simple radio-faint sample. Then we plot the cumulative fractions for the radio-loud, radio-faint samples in Figure \ref{Fig4--Fhost dis} where we find when the radio (flux density of the host  galaxies, $F_{host}$) is less than 50 $\mu$Jy the radio-loud, the radio-faint samples share the same distribution, but when it is more than 50$\mu$Jy the Cumulative fractions of these two samples are significantly different. A K-S test shows that the probability of those two samples from the same distribution is 0.19, so that in terms of host galaxies the two samples might be taken from the same distribution.

\begin{figure*}
	\centering
	\subfigure{
		\includegraphics[width=0.4\textwidth]{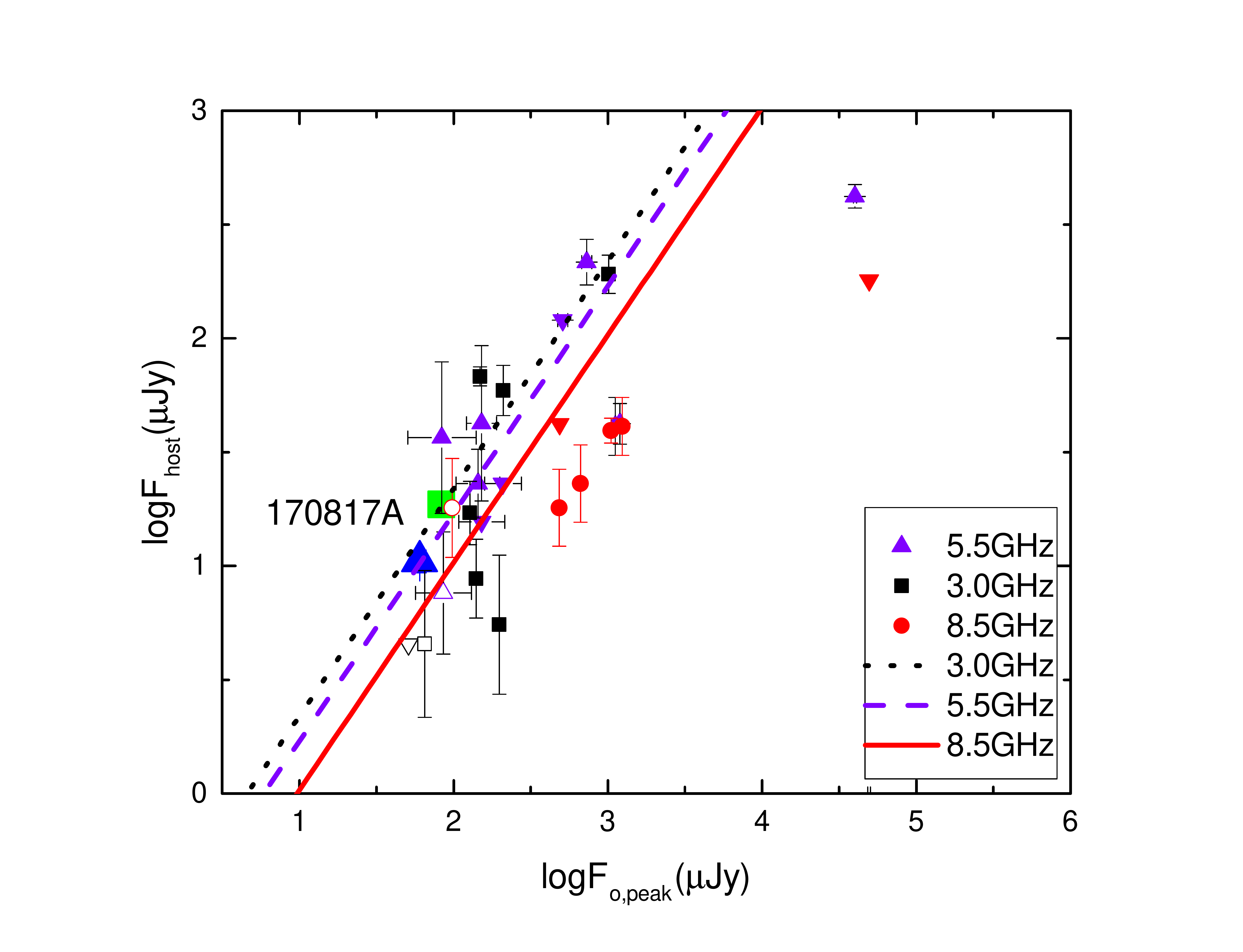}
	}
	\subfigure{
		\includegraphics[width=0.4\textwidth]{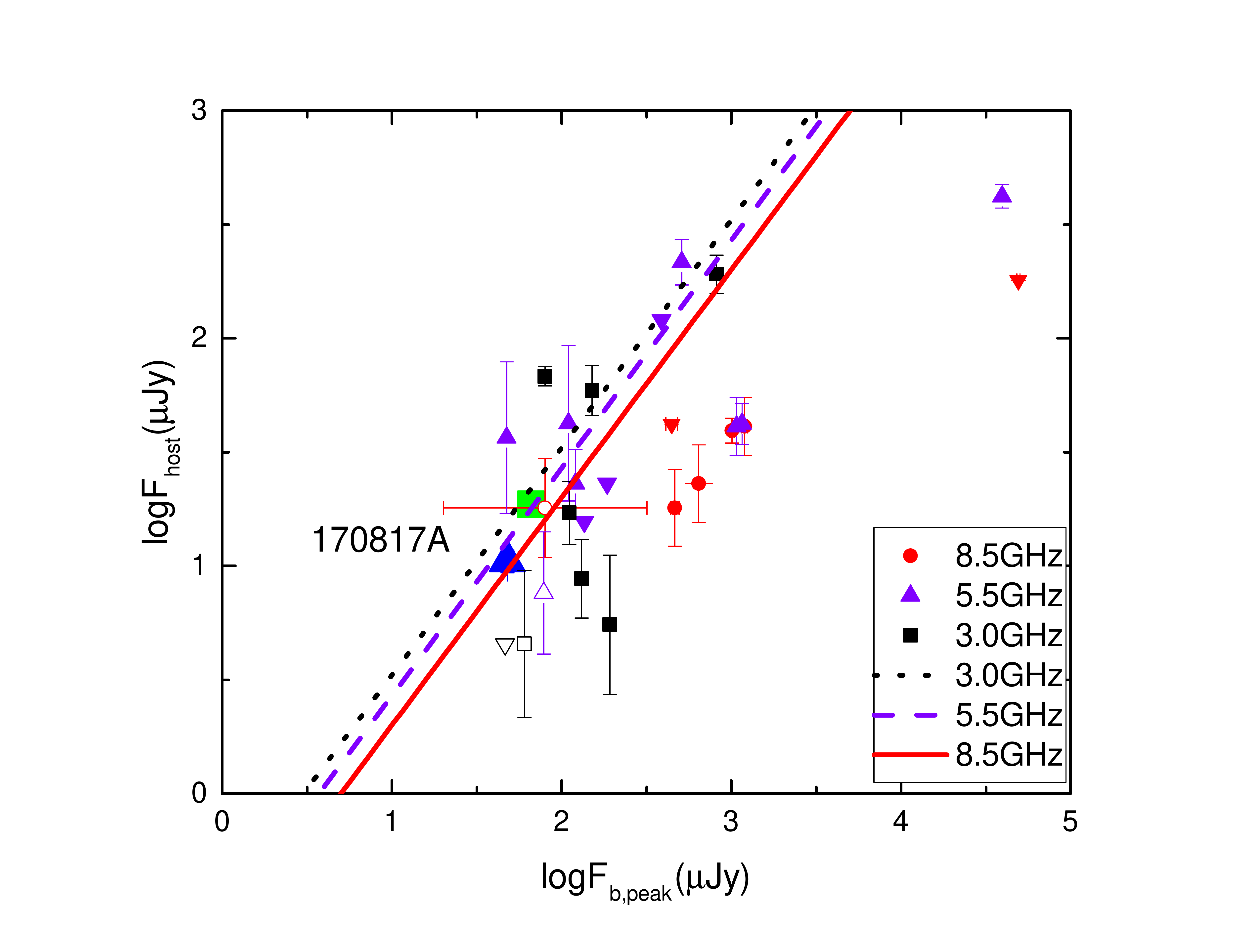}
	}
	\caption{The relations of $F_{host}$ versus $F_{o,peak}$ and $F_{host}$ versus $F_{b,peak}$ are displayed on left and right panels respectively for different radio frequencies. Three correspondingly empirical power-low relations at frequencies of $\nu$=8.5GHz (solid line), 5.5GHz (dashed line) and 3.0GHz (dotted line) are compared. Those upper limits are marked with different downward arrows. The radio-loud and the radio-quiet bursts are denoted with filled and empty symbols, individually. The off-axis GRB 170817A is symbolized with large triangle for $\nu=5$ GHz and large square for $\nu=3$ GHz.}
	\label{Fig5--Fhost vs. Flux}
\end{figure*}

\citet{Li+0} found the host flux density $F_{host}$ is positively correlated with the observed peak flux density ($F_{o,peak}$) or the pure flux density($F_{b,peak}$) of GRBs at a given radio frequency $\nu$ as follows
\begin{equation}
\label{eq1}
F_{host} = (a+b\nu)F_{o,peak},
\end{equation}
and
\begin{equation}
\label{eq2}
F_{host} = \frac{a+b\nu}{1-(a+b\nu)}F_{b,peak},
\end{equation}
where $a\simeq0.3$, $b\simeq-0.02$, and $F_{o,peak}=F_{b,peak}+F_{host}$. The Eq. (1) can be used to estimate the host flux density once the peak values of radio afterglows are measured. Figure \ref{Fig5--Fhost vs. Flux} displays the relationships of $F_{host}$ with $F_{o,peak}$ or $F_{b,peak}$ for the radio-loud and radio-quiet samples. One can find that the radio flux densities of the radio-quiet GRBs and their host galaxies are relatively lower than those of the radio-loud ones. It is noticeable that the very famous nearby short GRB (sGRB) 170817A seen off-axis with an estimated viewing angle of $20^{\circ}\sim40^{\circ}$ \citep{Alexander+01} is the first electromagnetic counterpart of gravitational-wave event. It has peak flux densities of $\sim$84.5 $\mu$Jy and $\sim$58.6 $\mu$Jy observed correspondingly at $\nu=$3 GHz and 5.5 GHz around 130 days since the merge of double neutron stars \citep{Li+18}. Using the above Eq. (\ref{eq1}) and (\ref{eq2}), one can easily predict the host flux densities to be about 20.3 $\mu$Jy at $\nu=$3 GHz and 11.1 $\mu$Jy at $\nu=$5.5 GHz. Interestingly, GW 170817/sGRB170817A as a radio-loud burst has relatively weaker radio afterglows and lower host fluxes in contrast with other normal radio-loud GRBs. However, it is located near the radio-quiet bursts as shown in Figure \ref{Fig5--Fhost vs. Flux}, which indicates that galactic types or circum-burst environment of different radio-selected GRBs could be diverse although their dominant radiation mechanisms might be the same.
\subsection{The surrounding medium density}

\begin{figure*}
	\centering
	\includegraphics[width=0.5\textwidth]{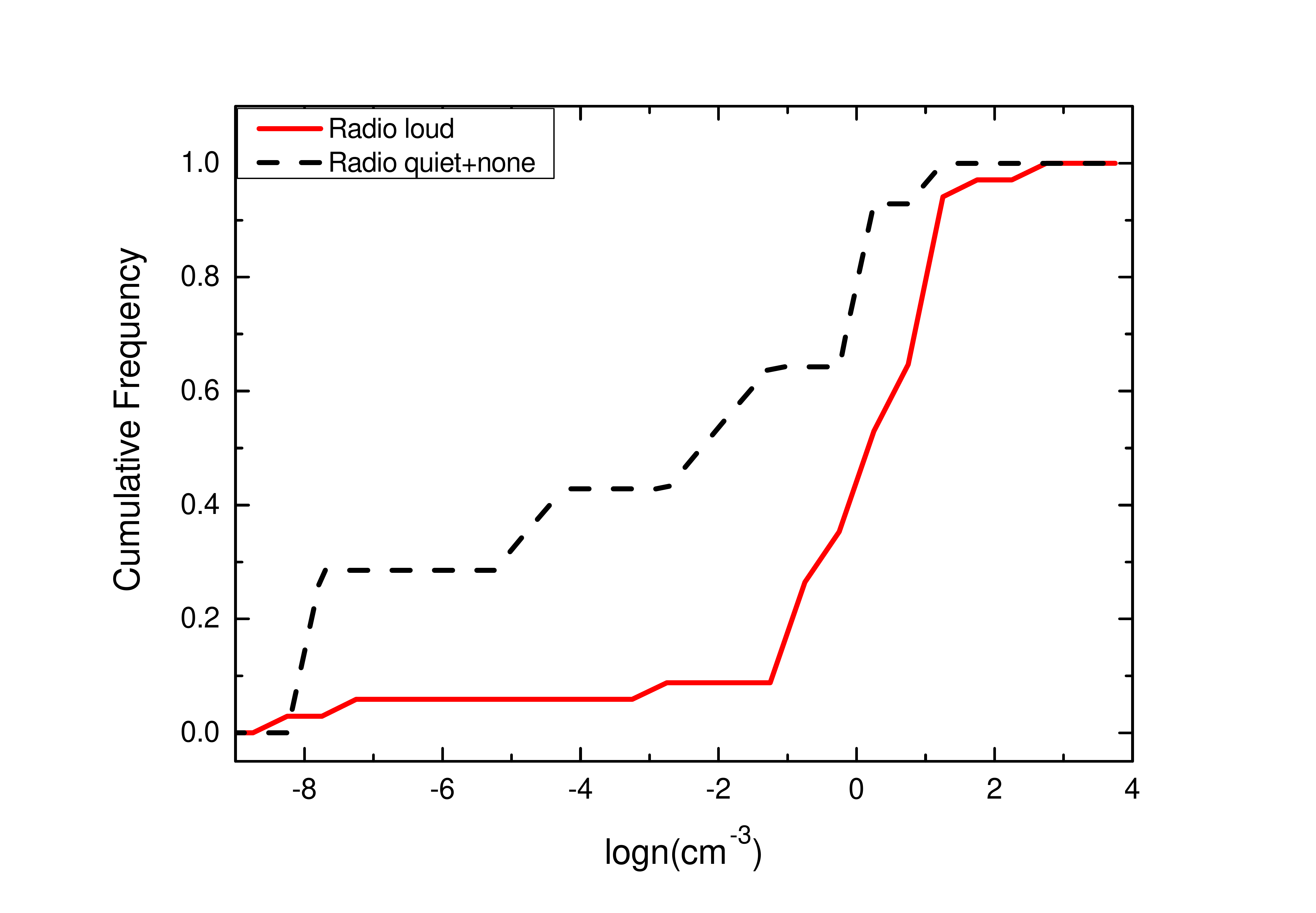}
	\caption{Cumulative fractions of log$n$ for radio-loud (red solid line) and radio faint (black dashed line) samples.}
	\label{Fig6-medium dis}
\end{figure*}

As pointed out by \citet{Chandra+0}, the centimeter radio afterglow emission is the brightest for circum-burst densities from 1 to 10 cm$^{-3}$. Beyond the narrow density range, the flux density will become weak due to either a low intrinsic emission strength (for lower densities) or the increased synchrotron self-absorption (for higher densities). From the literatures, it is well known that the circum-burst medium densities ($n$) of GRBs usually span serval orders of magnitude and are hard to be determined \citep[e.g.][]{Wijers+99,Chandra+0,Fong+0,Zhang+0}. In our samples, the circum-burst densities are distributed in a fairly wide scope spanning $\sim$10 orders of magnitude seen from Table \ref{Table1:radio-loud} to \ref{Table3:radio-none}. Because the number of radio-none GRBs with estimated densities is extremely limited, we thus combine the radio-quiet and the radio-none samples into a newly-formed radio faint sample in order to increase the statistical confidence level. Then we plot the cumulative fractions for the two samples in Figure \ref{Fig6-medium dis} and apply a K-S test to get $D$= 0.55 with a probability of 0.002, which demonstrates that the radio-loud and radio faint samples are significantly incongruous with each other. In contrast, the medium densities of the radio-loud host galaxies are relatively larger than those of the radio faint ones. Furthermore, the fraction of low densities of $n\leq$ 0.1 cm$^{-3}$ for the radio faint sample is around six times more than that for the radio-loud sample. On the contrary, about 90 percent of radio-loud afterglows are surrounded by relatively denser mediums of $n\simeq10^{-1}-10^{2}$ cm$^{-3}$.

\subsection{Spectral luminosity of radio afterglows}
\begin{figure*}
	\centering
	\subfigure{
		\includegraphics[width=0.45\textwidth]{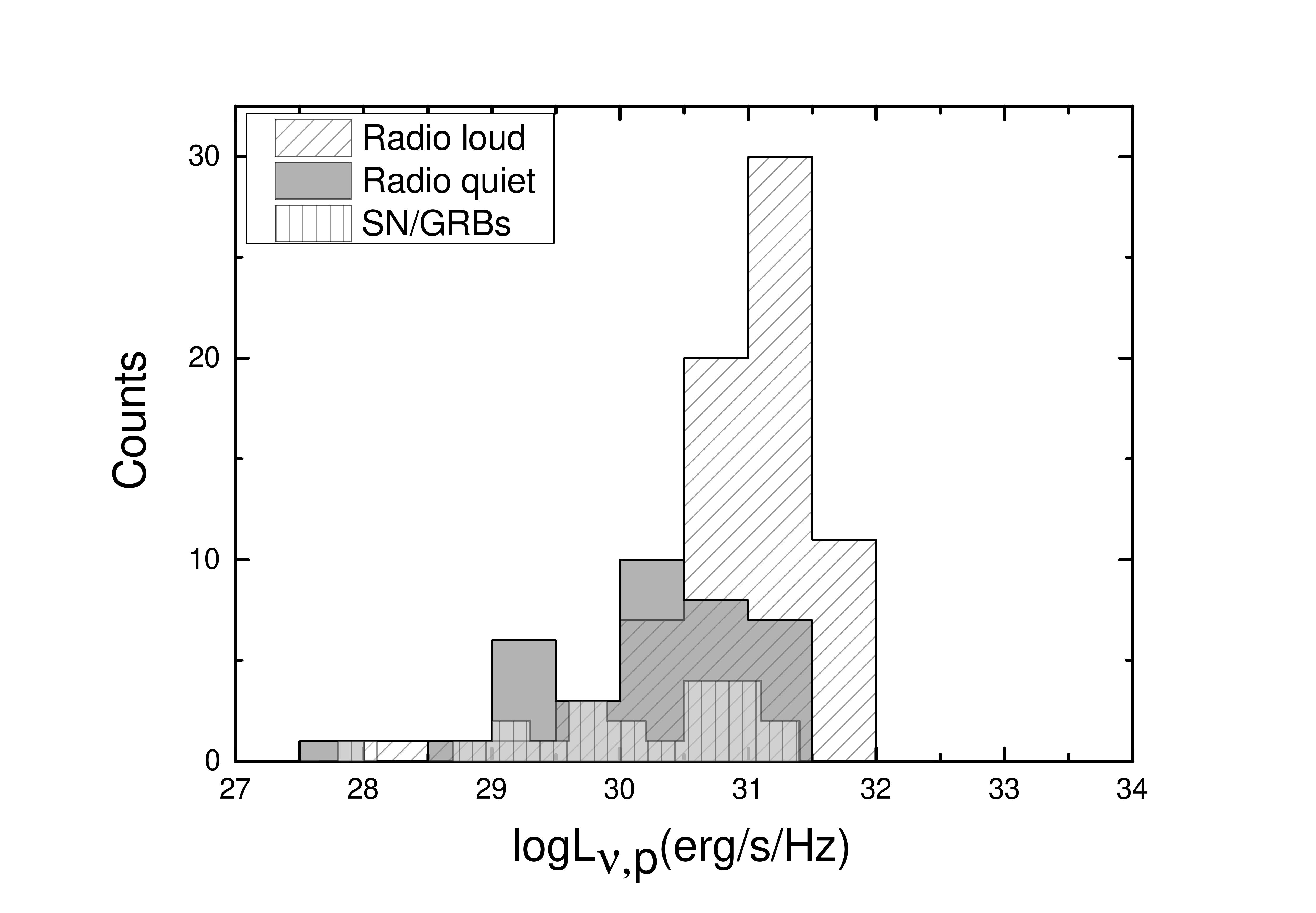}
	}
	\subfigure{
		\includegraphics[width=0.45\textwidth]{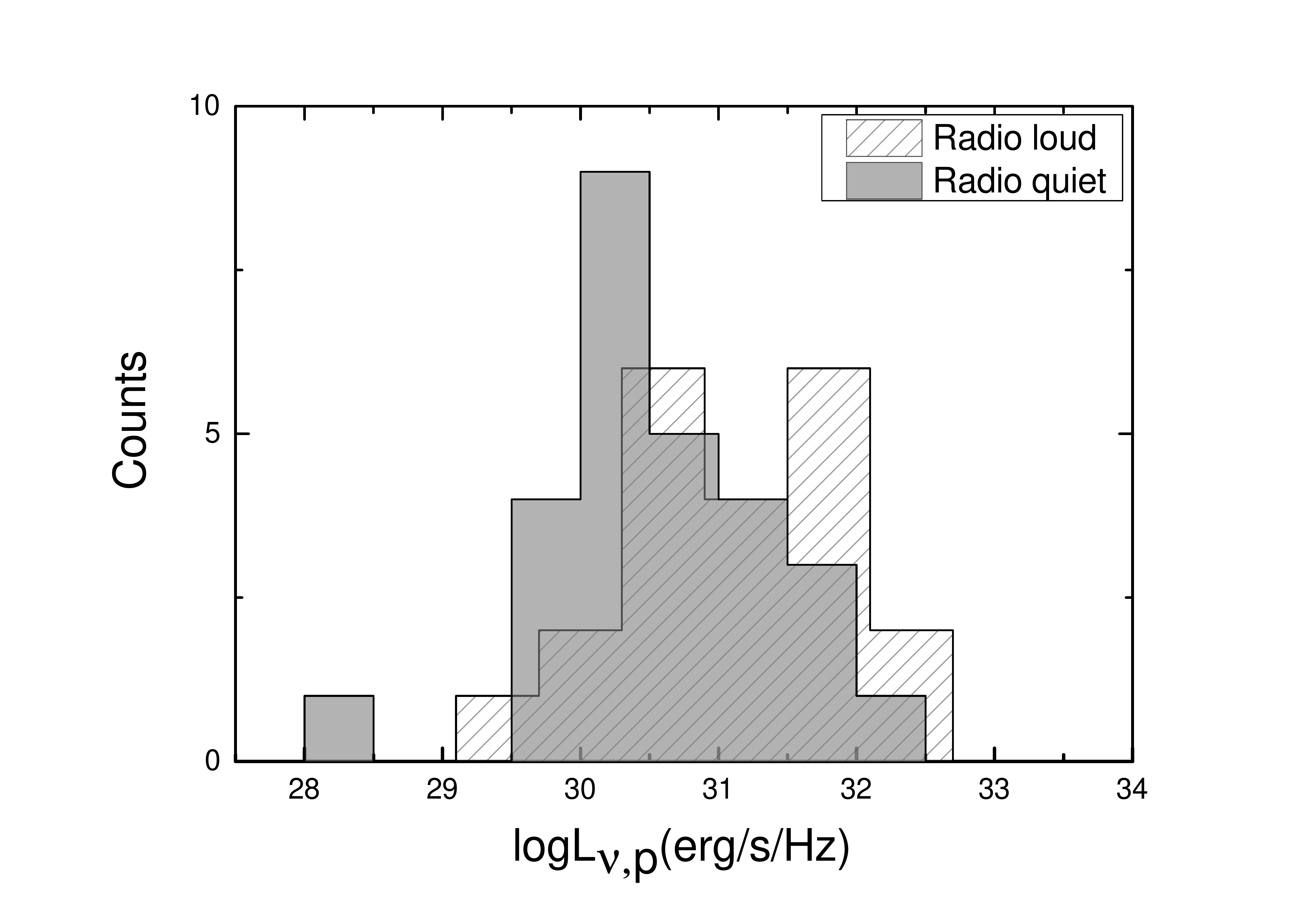}
	}
	\caption{The left panel shows the distribution of spectral peak luminosity of the VLA-based radio afterglows. The right panel displays the distribution of spectral peak luminosity of the AMI GRBs. The filled and hatched histograms respectively represent the upper limit and detection samples, and the vertical-line hatched histogram corresponds to the SN/GRBs.}
	\label{Fig7-radioluminoisitydis}
\end{figure*}

\begin{figure*}
	\centering
	\includegraphics[width=0.5\textwidth]{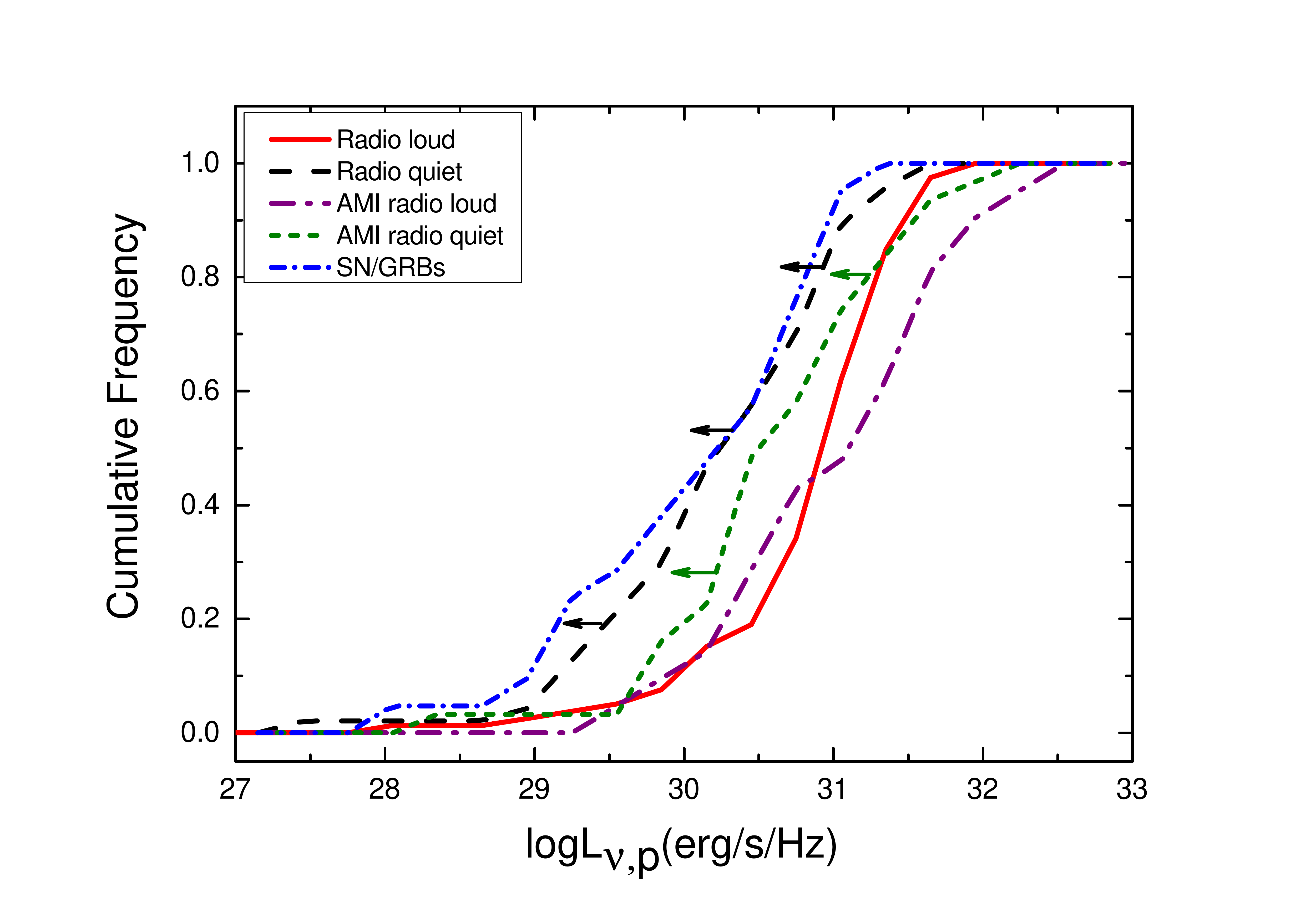}
	\caption{Cumulative fractions of $L_{\nu,p}$ for the VLA-based radio-loud (red solid line)and radio-quiet (black dashed line) samples, the AMI radio-loud (purple dash-dotted line), the AMI radio-quiet (green dotted line) and the SN-associated GRBs (blue short dash-dotted line). Note that the arrows denote that the $L_{\nu,p}$ distributions of radio-quiet GRBs are just the upper limits.}
	\label{Fig8-luminosity dis}
\end{figure*}

We utilize all the GRBs with measured isotropic $\gamma$-ray energy instead of $E_{\gamma,iso}$ $\textgreater$ $10^{52}$ \citep{Lloyd+1} only to ensure our samples to be as complete as possible. Simultaneously, we calculate the spectral peak luminosity at radio band ($L_{\nu,p}$) for the radio-loud and the radio-quiet (or upper limit) samples as \citep{Zhang+0}
\begin{equation}
\label{Eq3-spectralLumninosity}
L_{\nu,\text{p}} = 4 \pi D_{L}^{2} f_{m,radio} (1+z)^{-1}k,
\end{equation}	
where $f_{m,radio}$ denotes the peak flux density $F_{o,peak}$ of the radio-loud afterglows or the upper limits of radio-quiet afterglows, $k$ is a $K$-correction factor determined by
\begin{equation}
k = (1+z)^{\alpha-\beta},
\end{equation}
where $\alpha$ $\sim$ 0 and $\beta$ $\sim$ 1/3 are assumed to be the normal temporal and spectral indexes, respectively.
$D_{L}$ denoting the luminosity distance of a burst is given by
\begin{equation}
D_{L} = cH_{0}^{-1} (1+z) \int_{0}^{z} dz^{'}[(1+z^{'})^{3} \Omega_{M} +\Omega_{\Lambda}]^{-1/2},
\end{equation}
in which $c$=$3.0 \times 10^{8}$ m/s is the speed of light, $H_{0}$ is the Hubble constant taken as 70 km/s/Mpc, other cosmological parameters $\Omega_{M}$=0.27 and $\Omega_{\Lambda}$=0.73 have been assumed for a flat universe \citep{Schaefer+0}. Consequently, the $L_{\nu,p}$ values can be obtained from Eq. (\ref{Eq3-spectralLumninosity}) for the VLA-based GRBs at 8.5 GHz since most afterglows were detected at this frequency. For the AMI bursts reported in \citet{Anderson+0}, their $L_{\nu,p}$ values are calculated at a frequency of 15.7 GHz. Owing to lack of measurement of the radio afterglows with the upper limits, the $L_{\nu,p}$ values of radio-quiet afterglows can be only estimated as the upper limits too. Figure \ref{Fig7-radioluminoisitydis} displays the $L_{\nu,p}$ distributions of radio-loud, radio-quiet and SN-associated GRBs respectively. On average, the peak luminosity of radio-loud bursts is relatively larger than the other two, while the mean values of radio-quiet and SN-associated GRBs are comparable. The cumulative fractions of all the above samples are shown in Figure \ref{Fig8-luminosity dis}. A K-S test to them shows that the luminosity distributions of radio-loud and radio-quiet GRBs are largely different for the VLA-based samples since $D=0.4$ (>$D_{\alpha=0.05}=0.25$) with $P\simeq6.6\times10^{-5}$ and are however consistent with each other for the AMI samples. It needs to be emphasized that the distributional consistency of $L_{\nu,p}$ for different kinds of radio-selected GRBs is similar to that of the $E_{\gamma,iso}$ distributions in Figure \ref{Fig3--Tint/Eiso dis2}. Moreover, the actual deviation between them would become more significant since the accumulative line of the radio-quiet sample consisted of the upper limits should move leftward in a certain sense. The median $L_{\nu,p}$ of radio-quiet sample is about one order of magnitude smaller than that of radio-loud sample. Interestingly, this is similar to the one order of magnitude difference between radio fluxes of host galaxies and GRB afterglows \citep{Zhang+0}. Hence, we conclude that the majority of radio-quiet emissions should be contributed by their surrounding host galaxies.

\subsection{The $L_{\nu,p}$-$E_{\gamma,iso}$ relationship}
\begin{figure*}
	\centering
	\subfigure{
		\includegraphics[width=0.45\textwidth]{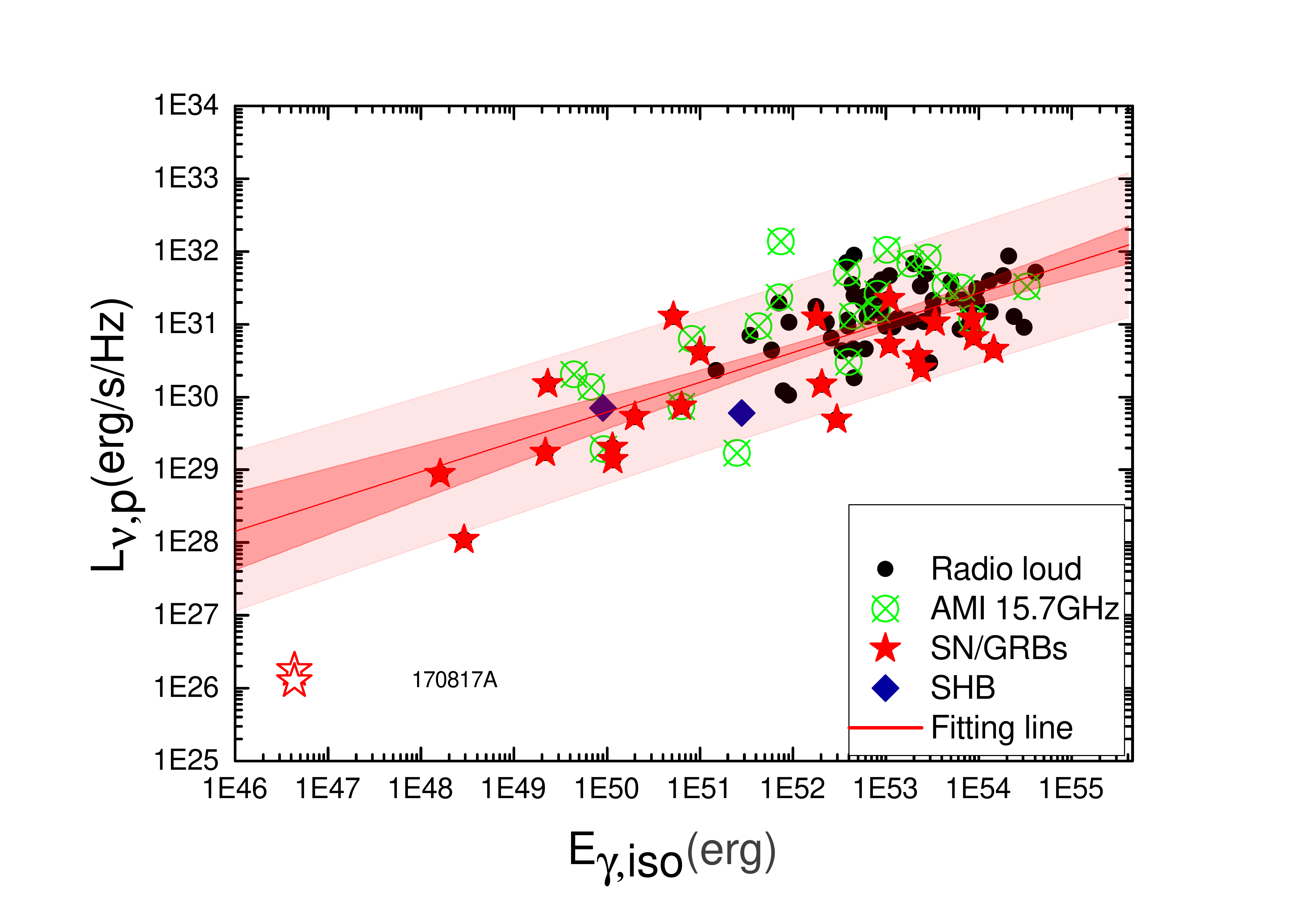}
	}
	\subfigure{
		\includegraphics[width=0.45\textwidth]{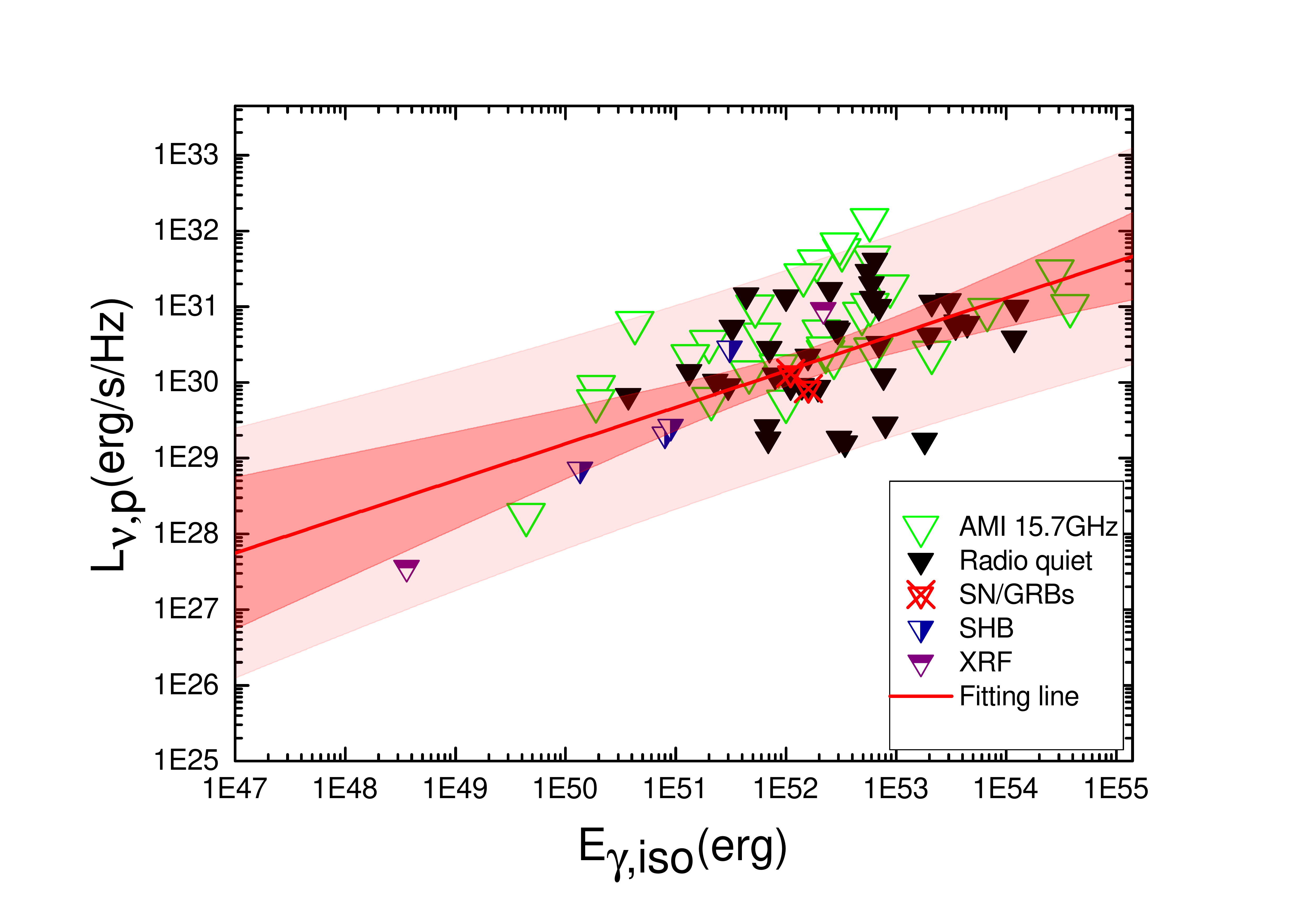}
	}
	\caption{The correlations of $L_{\nu,p}$ versus $E_{\gamma,iso}$ for radio-loud sample (the left panel) and radio-quiet sample (the right panel). All illustrations are marked on the insert. Note that the inverted triangles stand for the upper limits. Except the AMI radio-loud (cross circles) and radio-quiet (empty triangles) GRBs, all other symbols represent the VLA-based bursts. GRB 170817A detected at $\nu=$3 and 5.5 GHz has been marked with two empty stars. The solid lines are the best power-law fits to all bursts but GRB 170817A. The light shaded regions are 2$\sigma$ confidence ranges and the heavy shaded areas show the 2$\sigma$ prediction ranges. }
	\label{Fig9-LpvsEiso}
\end{figure*}

As shown in Figures \ref{fig2--Tint/Eiso dis1}, \ref{Fig3--Tint/Eiso dis2}, \ref{Fig7-radioluminoisitydis} and \ref{Fig8-luminosity dis}, the averaged energies of $E_{\gamma,iso}$ and $L_{\nu,p}$ of radio-loud bursts are larger than the corresponding values of radio-quiet ones. In the section, we will testify the possible correlation between the $E_{\gamma,iso}$ and the $L_{\nu,p}$ of radio-loud (N=100) and radio-quiet (N=76) GRB samples. For this purpose, the radio peak flux densities at 8.5 GHz and 15.7 GHz have been utilized. Figure \ref{Fig9-LpvsEiso} displays the relations of $E_{\gamma,iso}$ with $L_{\nu,p}$ for all the radio-loud/quiet VLA-based bursts including 95 long GRBs (lGRBs), 23 SN/GRBs, 2 X-Ray Flashes (XRFs) and 6 short GRBs (sGRBs), and 50 AMI GRBs. Interestingly, we find on the left panel that $E_{\gamma,iso}$ is positively correlated with $L_{\nu,p}$ with a Pearson correlation coefficient of $R=$0.76 ($SL=2.2\times10^{-16}$) or Spearman rank correlation coefficient of 0.55 ($SL=1.08\times10^{-7}$). The correlation function can be roughly written as \textbf{$L_{\nu,p}\propto E_{\gamma,iso}^{0.41\pm0.04}$} for the whole radio-loud sample with \textbf{\textbf{$\chi^{2}_{\nu}=0.23$}}. On the right panel, a positive correlation,\textbf{ $L_{\nu,p}\propto E_{\gamma,iso}^{0.48\pm0.09}$ }with a \textbf{\textbf{$\chi^{2}_{\nu}=0.42$}}, weakly exists for the radio-quiet bursts, of which the Pearson and the Spearman correlation coefficients are respectively $R=$0.62 ($SL=5.48\times10^{-6}$) and $R=$0.47 ($SL=1.2\times10^{-3}$) that are very close to those of the radio-loud bursts. This demonstrates that the radio peak luminosities and the prompt $\gamma$-ray energies are highly associated. It is notable that our finding here is different from \citet{Chandra+0}, where they claimed no obvious correlation between $E_{\gamma,iso}$ and $L_{\nu,p}$ in their Figure 20 possibly owing to the limit of sample size. Recently, \cite{Tang+19} found that the X-ray peak luminosity is positively correlated with the $E_{\gamma,iso}$ as $L_X\propto E_{\gamma,iso}^{0.97}$. It is valuable to mention that the radio peak luminosities of 21 SN/GRBs in our sample and 6 SN/GRBs in \citet{Chandra+0} exhibit a consistent dependence of $E_{\gamma,iso}$. This may imply these SN/GRBs should undergo with the same processes of energy dissipations. Data points of the sGRBs and the XRFs are too limited to show if they behave a positive interdependency as the lGRBs did.

\subsection{The correlation between $T_{int}$ and $1+z$}
\citet{Lloyd+1} found that there was a negative correlation between $T_{int}$ and $1+z$ for the radio-loud rather than radio-quiet GRB sample. They concluded that if this negative correlation indeed exists, other than affected by the selection effect, it could reflect that the systems at higher redshift have less angular momentum or less materials accreted to the GRB disks. Recently, \citet{Zhang+0} investigated the correlations between the intrinsic peak times of radio afterglows at 8.5 GHz and the redshift factor $(1+z)$ and found that they are fully uncorrelated, which seems to conflict with the negative correlation of $T_{int}$ versus $1+z$. Meanwhile, the $T_{int}$ distribution of Swift/BAT bursts was still bimodal in that all the durations move towards to the short end once the $T_{90}$ over 1+z was considered \citep{Zhang+08}. It is well known that the sGRBs are usually observed at nearby universe unlike the lGRBs. Strictly speaking, the negative dependence of the $T_{int}$ on the redshift is hard to understand unless a fraction of sGRBs have extremely small redshift while parts of lGRBs have very high redshift.

\begin{figure*}
	\centering
	\subfigure{
		\includegraphics[width=0.4\textwidth]{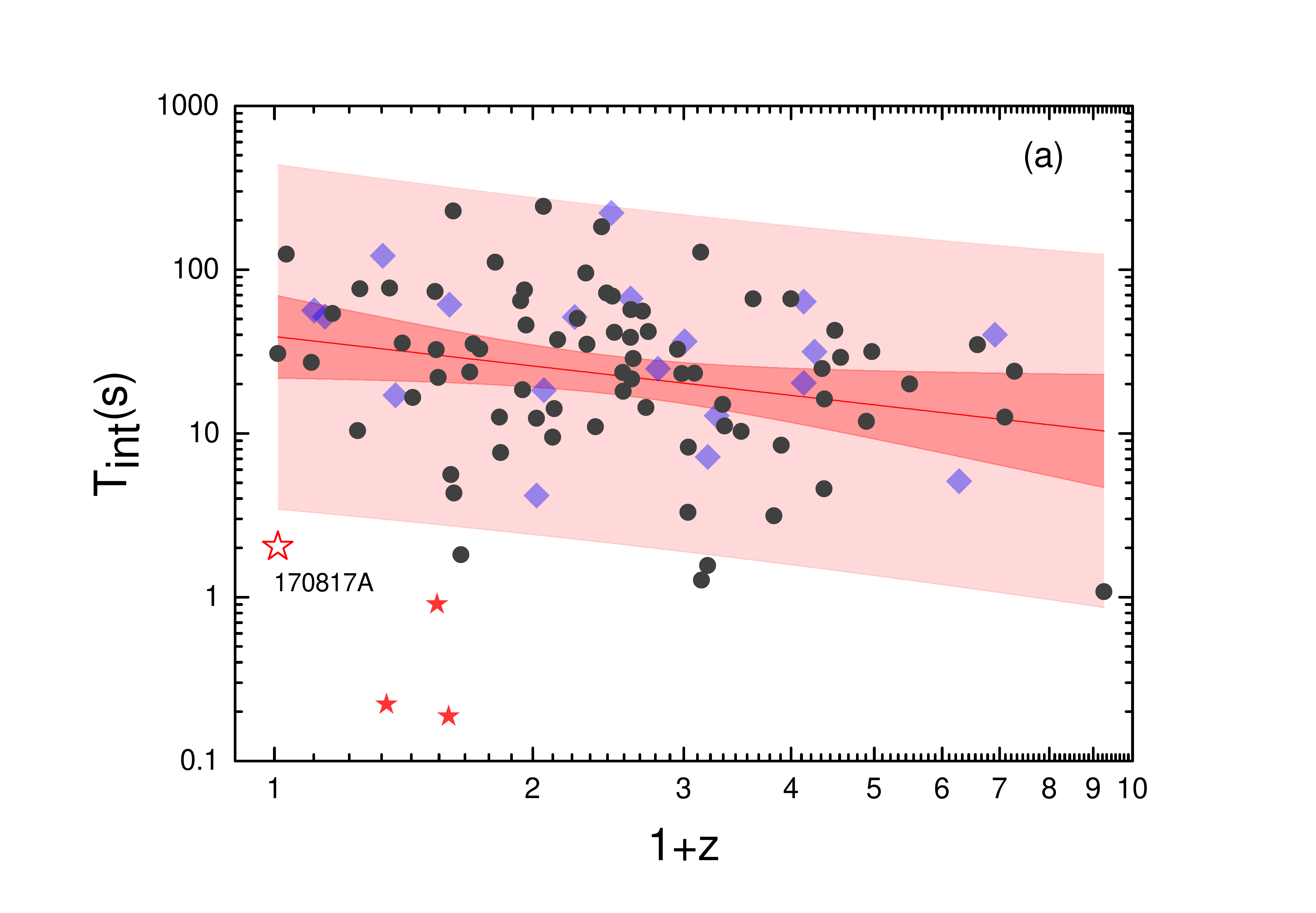}
	}
	\subfigure{
		\includegraphics[width=0.4\textwidth]{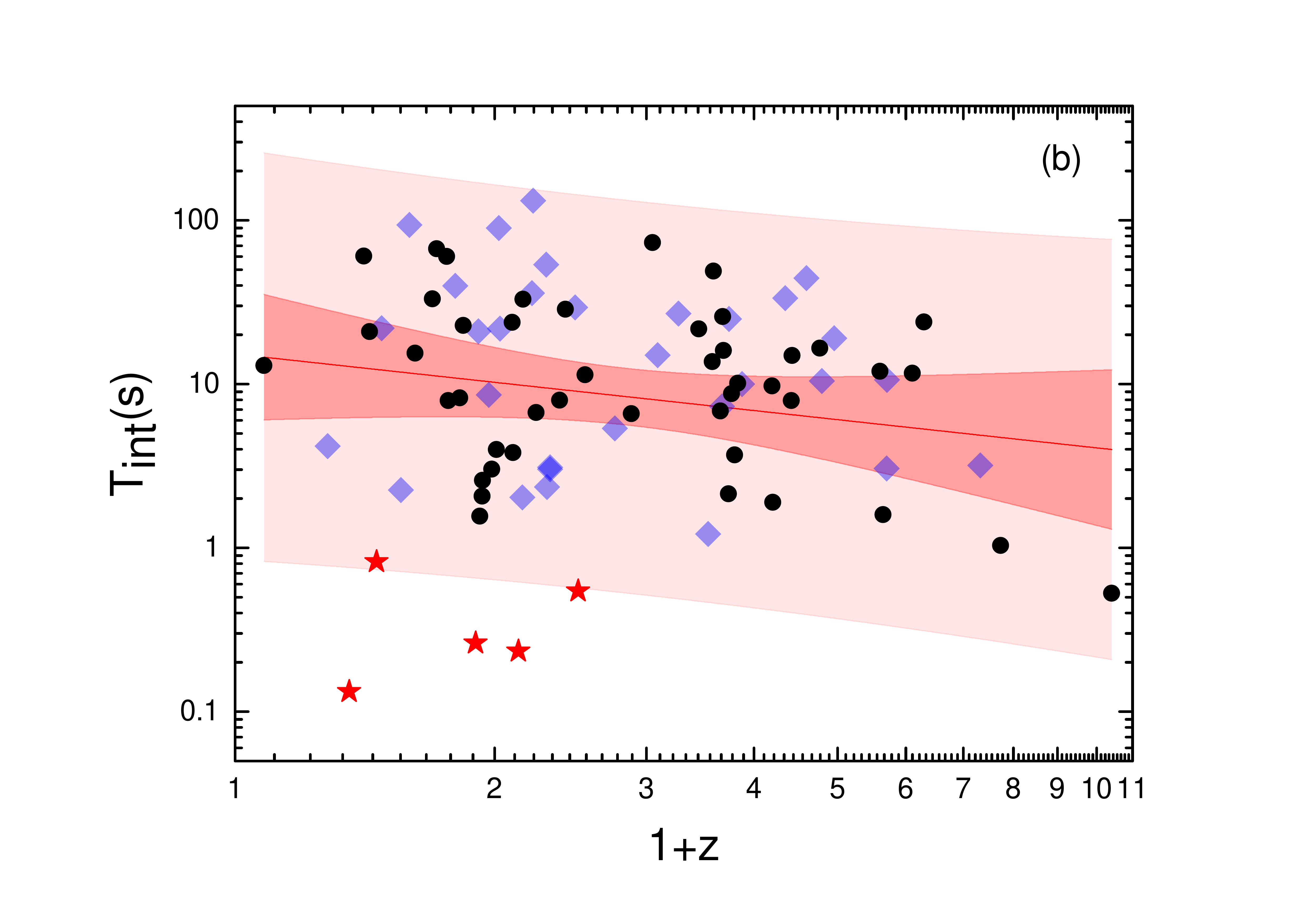}
	}
	\subfigure{
		\includegraphics[width=0.4\textwidth]{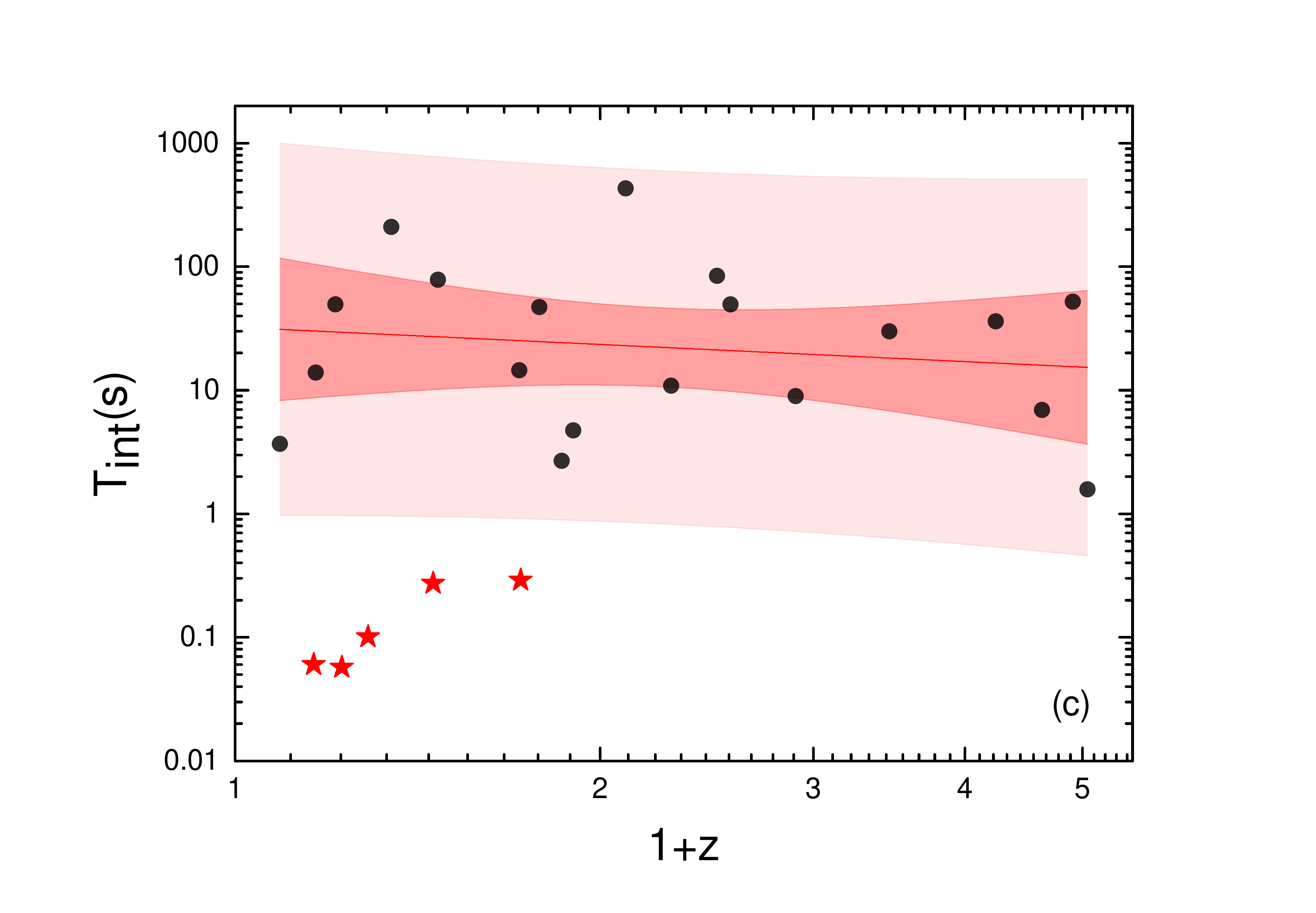}
	}
	\caption{Panel (a), (b) and (c) respectively show the correlations between $T_{int}$ and $1+z$ for the radio-loud, radio-quiet and radio-none samples. The filled circles and diamonds denote lGRBs and the filled stars represent sGRBs. The red solid lines are the best fits to the observed data with a confidence level of 95.4\% (heavy shadow region) and a prediction of 2$\sigma$ range (light shadow region). The AMI data at 15.7 GHz have been symbolized with blue diamonds. The radio-loud sGRB 170817A characterized with empty star is displayed in Panel (a). }
	\label{Fig10-Tint vs z}		
\end{figure*}

As mentioned in Section 2, our current samples as an expansion of \citet{Lloyd+1} are relatively complete. Therefore, it is timely and essential to check if the correlations between $T_{int}$ and $1+z$ coexist in both radio-loud, radio-quiet and radio-none samples as plotted in Figure \ref{Fig10-Tint vs z}. In statistics, the Pearson correlations of $T_{int}$ vs. $1+z$ for the radio-loud and the radio-quiet samples give the R-indexes as -0.29 ($SL$=0.012), -0.35 ($SL$=0.018) and -0.15 ($SL$=0.53) for the radio-loud, the radio-quiet and radio-none lGRB samples, respectively. This demonstrates that the radio-loud GRBs do have a weaker negative correlation of $T_{int}$ with redshift, of which this result is in good agreement with \citet{Lloyd+1}. Additionally, our radio-quiet sample also hold the similar anti-correlation with a 95.4\% confidence level like the radio-loud GRBs. It is surprisingly found that there is very weak correlation between $T_{int}$ and $1+z$ for the radio-none sample. We notice that the sGRBs in any case of our samples are outliers of the $T_{int}$-$(1+z)$ correlation of the lGRBs and the sGRBs with smaller $T_{int}$ and $1+z$ are systematically located at the bottom-left side of plane. Particularly, the radio-loud sGRB 170817A is situated in the region of normal sGRBs. Hence, the $T_{int}$ distributions of two kinds of GRBs are well in agreement with \cite{Zhang+08}.

\section{Conclusion and Discussion}
Based on the above systematic investigations of a relatively ``complete'' sample of radio-selected GRBs, we briefly summarize our main results as follows:
\begin{itemize}
\item According to the distributions of $z$, $T_{int}$, $E_{\gamma,iso}$ and $L_{\nu,p}$, we find that the radio-loud, the radio-quiet and the radio-none samples observationally differ with each other, particulary for the two energies $E_{\gamma,iso}$ and $L_{\nu,p}$. The radio-loud sample is not required to be redivided into two subgroups.
\item It is also supported that the radio-loud and the radio-faint (radio-quiet plus radio-none) GRBs have largely different distributions of the radio isotropic energies and the surrounding medium densities, and could be thus originated from diverse central engines.
\item Although the radio flux density distributions of host galaxies for the radio-loud and the radio-faint samples are not significantly different, the flux densities of the radio-quiet GRBs and their host galaxies are relatively lower than those of the radio-loud ones, which indicates the host types of the radio-loud and the radio-faint GRBs might be diverse in essence.
\item The mean values of $z$, $F_{o,peak}$, $T_{int}$, $E_{\gamma,iso}$, $n$ and $L_{\nu,p}$ for the radio-faint GRBs are comparatively smaller than those of the radio-loud sample correspondingly. Especially, it can be seen from Figures \ref{Fig3--Tint/Eiso dis2}, \ref{Fig10-Tint vs z} and \ref{Figure12-histogram} that the radio-none GRBs with the lowest means of $z$ and $T_{int}$ are unique and different from other radio-selected bursts.
\item Interestingly, we find $E_{\gamma,iso}$ and $L_{\nu,p}$ are correlated with the power law relations of $L_{\nu,p}\propto E_{\gamma,iso}^{0.41}$ for the radio-loud sample and $L_{\nu,p}\propto E_{\gamma,iso}^{0.48}$ for the radio-quiet sample, which were not distinguished by \citet{Chandra+0} for the correlation between $E_{\gamma,iso}$ and the peak radio spectral luminosity.
\item We follow \citet{Lloyd+1} to study the dependencies of $T_{int}$ with $z$ for different radio-selected samples. Excitingly, we not only gain the anti-correlation between   $T_{int}$ and $z$ for the radio-loud sample as \citet{Lloyd+1} proposed, but also find that this dependency holds for the radio-quiet instead of the radio-none sample.
\item Despite of the AMI radio afterglows detected at higher frequency, all the above conclusions based on the VLA-based GRB samples are well supported.
\end{itemize}

Most of our radio-selected GRBs are lGRBs that are thought be produced from core collapse of massive stars to form a black hole (\citealt{Woosley+0}, \citealt{MacFadyen+0}). In the collapsar model, the intrinsic time $T_{int}$ relies on the accretion rate that is related with the momentum of the progenitor system, namely larger momentum corresponds to longer $T_{int}$, and the masses forming the accretion disk \citep{Janiuk+0}. The collapsing progress exits in either a single stellar system or a binary system with three scenarios \citep{Fryer+1}, i.e. Scenario I: a single star evolves off main sequence and its winds blow off the hydrogen envelope to form a helium core, and then this helium core collapses to produce the GRBs; Scenario II: a binary system with primary evolving off main sequence evolves into a common envelope phase, and then after the H envelope was ejected the primary becomes a helium core collapsing and accreting the secondary to produce GRBs; Scenario III: this is also a binary system with primary evolving off main sequence into a common system, and then the secondary evolving off main sequence too, subsequently the system enter into a double-helium-star binary system. Finally, the two helium stars merge into one helium star and then the helium core collapse to cause the GRBs \citep{Fryer+1}. Because $T_{int}$ is tightly determined by the momentum of collapsing systems, together with more masses accreted on the disk, Scenario I would readily lead to the longer $T_{int}$ even though its angular momentum is expected to be less than the other two Scenarios (\citealt{Fryer+0}, \citealt{Zhang+2})).

\begin{figure*}
	\centering
	\includegraphics[width=0.6\textwidth]{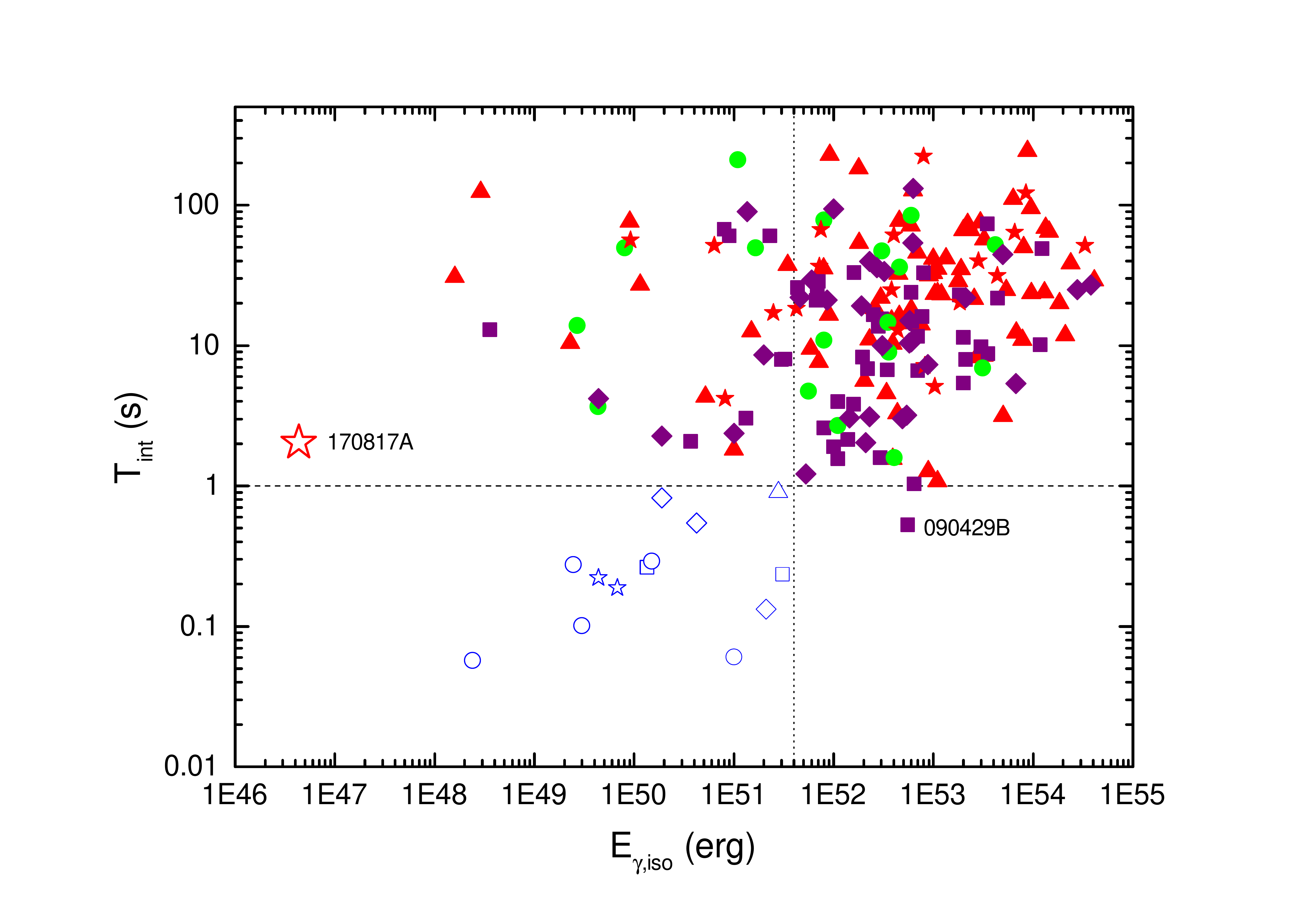}
	\caption{Relations of $T_{int}$ with $E_{\gamma,iso}$ for the radio-loud (triangles), radio-quiet (squares) and radio-none (circles) lGRB (filled symbols) and sGRB (empty symbols) in the VLA-based samples. Vertical and horizontal lines stand for $E_{\gamma,iso}=4\times10^{51}$ erg and $T_{int}=1$ s, correspondingly. The radio-loud and radio-quiet bursts in the AMI samples are identified with small stars and diamonds, respectively. sGRB 170817A is marked with a large star. }
	\label{Fig11-TintvsEiso}		
\end{figure*}

Note that the soft lGRBs associated with core-collapse supernovae \citep{Galama+98,Woosley+99,Fryer+1,Stanek+03,Hjorth+03,Campana+06,Xu+13} are generally believed to result from the deaths of massive stars. However, the hard sGRBs are usually thought to occur owing to the coalescence of two compact stars, such as double neutron stars, or a neutron star plus black hole system \citep{Lee+07,Berger+14}. Therefore, the sGRBs with lower redshifts and isotropic $\gamma$-ray energies would be expected to have relatively shorter $T_{int}$ in comparison with the lGRBs. We investigate the association of $E_{\gamma,iso}$ with $T_{int}$ in Figure \ref{Fig11-TintvsEiso}, from which we can see that there are no any correlations for either the sGRBs or the lGRBs. However, they can be separated by a horizontal line of $T_{int}=1$ s and a vertical line of $E_{\gamma,iso}=4\times10^{51}$ erg. All sGRBs but GRB 170817A possessing smaller $E_{\gamma,iso}$ and $T_{int}$ are located at the bottom-left corner. In comparison, the lGRBs with longer $T_{int}$ relatively generate larger $E_{\gamma,iso}$ spanning from $\sim10^{48}$ erg to $\sim10^{55}$ erg. Even though some lGRBs and sGRBs have comparable $E_{\gamma,iso}$, their $T_{int}$ values are completely distinct. It is valuable to focus on GRB 090429B, lying at the bottom-right corner, that is the farthest burst detected so far with $z\approx9.4$ and $T_{90}=5.5$ s \citep{Cucchiara+11}, whose progenitor is expected to be different from other lower redshift, especially short GRBs. Furthermore, we caution that sGRB 170817A differs from both the normal sGRBs and the low energy lGRBs as depicted in Figure \ref{Fig11-TintvsEiso}. Very recently, \cite{Tang+19} found that the $E_{\gamma,iso}$ and the $T_{90}$ are positively correlated, and they explained that this might happen when the observed intensities of $\gamma$-rays were constrained within a certain range. We nevertheless find that the positive correlation trend disappears for the lGRBs in the co-moving frame. In principal, one may pursue to convert the observed $E_{\gamma,iso}$ into the co-moving quantity by use of $E_{\gamma,iso}^{'}\simeq E_{\gamma,iso}/\Gamma$, where $\Gamma$ is the bulk Lorentz factor \citep{Ghirlanda+12}. Unfortunately, the Lorentz factor is still very hard to be determined precisely and uniquely although many authors have made great efforts (e.g., \citealt{Sari+99,Pe'er+07,Liang+10,Zou+10,Zou+15,Ghirlanda+18}), which will be confirmed by further observations of the next-generation telescopes.

\section{acknowledge}

This work is supported by the Research Foundation of China (grant Nos. ZR2018MA030, XKJJC201901 and 201909118), the National Natural Science Foundation of China (grant No. U1938201, 11873030, 11673023, U1838201, U1838202 and U1838104), the Strategic Priority Research Program of the Chinese Academy of Sciences ("Multiwaveband Gravitational Wave Universe", grant No.XDB23040000; Grant No. XDA15360300) and the National Key R\&D Program of China (2016YFA0400800). We thank Poonam Chandra for kindly offering the data of GRB radio afterglows observed by VLA. We also acknowledge E. W. Liang, L. B. Li and H. Y. Chang for helpful discussions.


\clearpage
\onecolumn

\begin{longtable}{p{1cm}<{\centering}p{1cm}<{\centering}p{1cm}<{\centering}p{1cm}<{\centering}p{2cm}<{\centering}p{2.5cm}<{\centering}p{1cm}<{\centering}p{1.5cm}<{\centering}p{1.5cm}<{\centering}p{1.5cm}<{\centering}}

    \caption{\textbf{Physical parameters of radio-loud GRBs}}
	\label{Table1:radio-loud} \\
	\hline
	\hline
	\multicolumn{1}{c}{GRB}&\multicolumn{1}{c}{$T_{90}$}&\multicolumn{1}{c}{$z$}&\multicolumn{1}{c}{$E_{\gamma,iso}$}&\multicolumn{1}{c}{$n$}&\multicolumn{1}{c}{$L_{\nu,p}$}&\multicolumn{1}{c}{$f_{p,radio}$}&\multicolumn{1}{c}{$RMS$}&\multicolumn{1}{c}{Radio Telescope}&\multicolumn{1}{c}{reference}\\
	\multicolumn{1}{c}{}&\multicolumn{1}{c}{$(s)$}&\multicolumn{1}{c}{}&\multicolumn{1}{c}{$(erg)$}&\multicolumn{1}{c}{($cm^{-3}$)}&\multicolumn{1}{c}{$(erg/s/Hz)$}&\multicolumn{1}{c}{$(uJy)$}&\multicolumn{1}{c}{$(uJy)$}&\multicolumn{1}{c}{}&\multicolumn{1}{c}{}\\
	\hline
	\endhead
970508&14&0.835&7.10E+51&1&9.40E+30&1270&33&VLA&1,4\\
970828&147&0.958&2.96E+53&$\cdots$&2.93E+30&147&33&VLA&1\\
980329&58&2-3.9&2.10E+54&20$_{-10}^{+10}$&4.76E+31&465&16&VLA&1,5\\
980425$^{\star}$&31&0.009&1.60E+48&$\cdots$&8.56E+28&49400&1000&ATCA&1\\
980519&30&$\cdots$&$\cdots$&0.14$_{-0.03}^{+0.32}$&1.95E+31&1050&20&VLA&1,5\\
980703&90&0.966&6.90E+52&28$_{-10}^{+10}$&1.95E+31&1050&55&VLA&1,6\\
981226&20&1.11&5.90E+51&$\cdots$&4.40E+30&169&28&VLA&1\\
990123&100&1.6&2.39E+54&$\cdots$&1.28E+31&260&32&VLA&1\\
990506&220&1.307&9.49E+53&$\cdots$&2.02E+31&581&45&VLA&1\\
990510&75&1.619&1.78E+53&0.29$_{-0.15}^{+0.11}$&1.14E+31&127&30&ATCA&1,7\\
$991208^{\star}$&60&0.706&1.10E+53&18$_{-6}^{+18}$&2.23E+31&1990&33&VLA&1,7\\
991216&25&1.02&6.75E+53&4.7$_{-1.8}^{+6.8}$&2.14E+31&960&67&VLA&1,7\\
000131&110&4.5&1.84E+54&$\cdots$&4.64E+31&207&46&ATCA&1\\
000210&10&0.85&2.00E+53&$\cdots$&1.48E+30&93&21&VLA&1\\
000301C&10&2.034&4.37E+52&27$_{-5}^{+5}$&2.29E+31&483&41&VLA&1\\
000418&30&1.119&7.51E+52&27$_{-14}^{+250}$&2.26E+31&1240&33&VLA&1,7\\
$000911^{\star}$&500&1.059&8.80E+53&$\cdots$&6.65E+30&278&36&VLA&1\\
000926&25&2.039&2.70E+53&27$_{-3}^{+3}$&4.84E+31&666&60&VLA&1,7\\
001007&375&$\cdots$&$\cdots$&$\cdots$&$\cdots$&222&33&VLA&1\\
001018&31&$\cdots$&$\cdots$&$\cdots$&$\cdots$&405&50&VLA&1\\
010222&170&1.477&1.33E+54&1.7&1.48E+31&344&39&VLA&1,7\\
010921&24&0.45&9.00E+51&$\cdots$&1.06E+30&229&22&VLA&1\\
011030&$\cdots$&$\textless$3&$\cdots$&$\cdots$&2.26E+31&219&20&VLA&1\\
011121&105&0.362&4.55E+52&$\cdots$&1.83E+30&610&39&ATCA&1\\
011211&400&2.14&6.30E+52&$\cdots$&1.18E+31&163&17&VLA&1\\
020305&247&$\cdots$&$\cdots$&$\cdots$&$\cdots$&76&15&VLA&1\\
$020405^{\star}$&40&0.69&1.10E+53&8&5.22E+30&487&34&VLA&1,5\\
020813&113&1.254&8.00E+53&$\cdots$&1.04E+31&323&39&VLA&1\\
020819B&50&0.41&7.90E+51&$\cdots$&1.22E+30&315&18&VLA&1\\
$020903^{\star}$&13&0.25&2.30E+49&$\cdots$&1.51E+30&1058&19&VLA&1\\
021004&50&2.33&3.80E+52&30$_{-27}^{+270}$&5.35E+31&691&33&VLA&1,8\\
021206&20&$\cdots$&$\cdots$&$\cdots$&$\cdots$&1377&47&VLA&1\\
030115&36&2.5&3.91E+52&$\cdots$&9.34E+30&94&22&VLA&1\\
030226&69&1.986&1.20E+53&$\cdots$&9.14E+30&131&27&VLA&1\\
030323&20&3.372&3.39E+52&$\cdots$&4.28E+30&530&170&VLA&1\\
$030329^{\star}$&63&0.169&1.80E+52&1.8&1.01E+31&19150&80&VLA&1,9\\
030723&31&$\cdots$&$\cdots$&$\cdots$&$\cdots$&219&22&VLA&1\\
$031203^{\star}$&30&0.105&1.15E+50&0.6&1.34E+29&811&40&VLA&1,10\\
040812&19&$\cdots$&$\cdots$&$\cdots$&$\cdots$&450&80&VLA&1\\
041219A&6&$\cdots$&$\cdots$&$\cdots$&$\cdots$&518&150&VLA&1\\
050315&96&1.95&5.70E+52&$\cdots$&2.03E+31&300&62&VLA&1\\
050401&33&2.898&3.20E+53&10&1.51E+31&122&33&VLA&1,3\\
050416A$^{\star}$&3&0.65&1.00E+51&3&4.12E+30&431&46&VLA&1,3\\
050509C&25&$\cdots$&$\cdots$&$\cdots$&$\cdots$&404&58&VLA&1\\
050525A$^{\star}$&9&0.606&2.04E+52&1.0$\times10^{-8}$&1.37E+30&178&46&VLA&1,3\\
050603&12&2.821&5.00E+53&$\cdots$&3.11E+31&316&45&VLA&1\\
050713B&125&$\cdots$&$\cdots$&$\cdots$&$\cdots$&426&45&VLA&1\\
050724&96&0.258&9.00E+49&0.1&7.08E+29&465&29&VLA&1,11\\
050730&157&3.968&9.00E+52&8&4.04E+31&212&35&VLA&1,3\\
050820A&240&2.615&2.00E+53&0.1&6.74E+31&634&62&VLA&1,12\\
050824&23&0.83&1.50E+51&1&2.32E+30&152&34&VLA&1,3\\
050904&174&6.29&1.30E+54&680&3.01E+31&116&18&VLA&1,13\\
050922C&5&2.199&3.90E+52&2&1.15E+31&140&42&VLA&1,3\\
051022&200&0.809&6.30E+53&$\cdots$&8.49E+30&585&49&VLA&1\\
051109A&37&2.346&2.30E+52&$\cdots$&1.06E+31&117&24&VLA&1\\
051111&46&1.55&6.00E+52&5.00$\times10^{-9}$&4.56E+30&98&28&VLA&1,3\\
051211B&80&$\cdots$&$\cdots$&$\cdots$&$\cdots$&68&19&VLA&1\\
051221A&1.4&0.547&2.80E+51&0.001&6.01E+29&88&26&VLA&1,14\\
060116&106&$\cdots$&$\cdots$&$\cdots$&$\cdots$&363&28&VLA&1\\
$060218^{\star}$&128&0.033&2.90E+48&5&1.09E+28&453&77&VLA&1,15\\
060418&103&1.49&1.00E+53&10&9.41E+30&216&48&VLA&1,3\\
061121&81&1.315&1.90E+53&3&1.07E+31&304&48&VLA&1,3\\
061222A&72&2.088&1.03E+53&$\cdots$&2.15E+31&285&68&VLA&1\\
070125&60&1.548&9.55E+53&42&2.61E+31&660&39&VLA&1,16\\
070612A&369&0.617&9.12E+51&$\cdots$&5.09E+30&589&54&VLA&1\\
071003&148&1.604&3.24E+53&$\cdots$&2.12E+31&431&51&VLA&1\\
071010B&36&0.947&2.60E+52&$\cdots$&6.43E+30&330&52&VLA&1\\
071020&4&2.146&8.91E+52&$\cdots$&1.47E+31&186&22&VLA&1\\
071021&229&$\textless$5.6&$\cdots$&$\cdots$&4.39E+31&149&44&VLA&1\\
071109&30&$\cdots$&$\cdots$&$\cdots$&$\cdots$&188&42&VLA&1\\
071122&80&1.14&3.47E+51&$\cdots$&6.96E+30&255&45&VLA&1\\
080229&64&$\cdots$&$\cdots$&$\cdots$&$\cdots$&635&44&VLA&1\\
$080319B^{\star}$&125&0.937&1.45E+54&10&4.43E+30&232&42&VLA&1,3\\
080603A&150&1.687&$\cdots$&$\cdots$&1.23E+31&230&29&VLA&1\\
080810&108&3.35&5.37E+53&$\cdots$&2.29E+31&151&50&VLA&1\\
081203B&23&$\cdots$&$\cdots$&$\cdots$&$\cdots$&162&44&VLA&1\\
081221&34&$\cdots$&$\cdots$&$\cdots$&$\cdots$&167&27&VLA&1\\
090313&71&3.375&4.57E+52&0.6&8.81E+31&576&44&VLA&1\\
090323&133&3.57&4.10E+54&0.1&3.72E+31&225&35&VLA&1,17\\
090328&57&0.736&1.00E+53&0.26&9.81E+30&809&39&VLA&1,17\\
090418&56&1.608&2.57E+53&$\cdots$&1.08E+31&219&44&VLA&1\\
090423&10&8.26&1.10E+53&0.9&4.63E+31&92.4&22.7&VLA&1,18\\
090424&50&0.544&4.47E+52&$\cdots$&4.54E+30&673&39&VLA&1\\
$090618^{\star}$&113&0.54&2.21E+53&$\cdots$&3.67E+30&551&51&VLA&1\\
090709A&89&\textless6.1&$\cdots$&$\cdots$&5.68E+31&174&53&VLA&1\\
090715B&265&3&2.36E+53&$\cdots$&3.33E+31&257&57&VLA&1\\
090902B&$\cdots$&1.883&3.09E+54&$\cdots$&8.33E+30&130&34&VLA&1\\
091020&39&1.71&4.56E+52&$\cdots$&2.47E+31&451&44&VLA&1\\
100413A&191&$\textless$3.5&$\cdots$&$\cdots$&2.56E+31&159&15&EVLA&1\\
100414A&26&1.368&7.79E+53&$\cdots$&1.56E+31&415&15&EVLA&1\\
100418A$^{\star}$&7&0.62&5.20E+50&$\cdots$&3.99E+30&458&22&EVLA&1\\
100805A&15&$\cdots$&$\cdots$&$\cdots$&$\cdots$&108&32&EVLA&1\\
100814A&175&1.44&5.97E+52&$\cdots$&1.90E+31&462&25&EVLA&1\\
100901A&439&1.408&1.78E+52&$\cdots$&1.74E+31&440&27&EVLA&1\\
100906A&114&1.727&1.34E+53&$\cdots$&1.20E+31&215&28&EVLA&1\\
101219B$^{\star}$&34&0.552&2.96E+52&$\cdots$&4.93E+29&71&15&EVLA&1\\
110428A&5.6&$\cdots$&$\cdots$&$\cdots$&$\cdots$&69&18&EVLA&1\\
120320A &25.74&$\cdots$&$\cdots$&$\cdots$&$\cdots$&380&80&AMI&19,20\\
120326A &69.6&1.798&3.82E+52&$\cdots$&5.12E+31&860&80&AMI&19,20\\
120514A  &164.4&$\cdots$&$\cdots$&$\cdots$&$\cdots$&460&130&AMI&19,20\\
121031A  &62.5&0.1126&$\cdots$&$\cdots$&1.91E+29&670&220&AMI&19,20\\
121128A  &23&2.2&8.20E+52&$\cdots$&2.62E+31&320&90&AMI&19,20\\
130216A  &6.5&$\cdots$&$\cdots$&$\cdots$&$\cdots$&990&100&AMI&19,20\\
130427A$^{\star}$ &162.83&0.338&8.50E+53&$\cdots$&1.19E+31&4540&80&AMI&19,20\\
130419A  &75.7&$\cdots$&$\cdots$&$\cdots$&$\cdots$&1700&120&AMI&19,20\\
130508A  &42&$\cdots$&$\cdots$&$\cdots$&$\cdots$&550&140&AMI&19,20\\
130603A  &$\cdots$&$\cdots$&$\cdots$&$\cdots$&$\cdots$&470&130&AMI&19,20\\
130604A  &37.7&1.06&$\cdots$&$\cdots$&9.34E+30&390&70&AMI&19,20\\
130606A  &276.58&5.91&2.83E+53&$\cdots$&8.17E+31&260&70&AMI&19,20\\
130608A  &44.4&$\cdots$&$\cdots$&$\cdots$&$\cdots$&240&80&AMI&19,20\\
130612A  &110&2.006&7.19E+51&$\cdots$&2.34E+31&330&90&AMI&19,20\\
130625A &38.1&$\cdots$&$\cdots$&$\cdots$&$\cdots$&590&110&AMI&19,20\\
130702A$^{\star}$ &59&0.145&6.36E+50&$\cdots$&7.42E+29&1560&130&AMI&19,20\\
130907A &115&1.238&3.30E+54&$\cdots$&3.29E+31&1040&100&AMI&19,20\\
131024B  &64&$\cdots$&$\cdots$&$\cdots$&$\cdots$&610&70&AMI&19,20\\
140108A  &97.8&0.6&4.00E+52&$\cdots$&3.03E+30&370&50&AMI&19,20\\
140209A &21.3&$\cdots$&$\cdots$&$\cdots$&$\cdots$&430&90&AMI&19,20\\
140215A  &84.2&$\cdots$&$\cdots$&$\cdots$&$\cdots$&240&50&AMI&19,20\\
140304A &32&5.28&1.03E+53&$\cdots$&1.04E+32&380&40&AMI&19,20\\
140305A &13.7&$\cdots$&$\cdots$&$\cdots$&$\cdots$&420&40&AMI&19,20\\
140318A &8.43&1.02&$\cdots$&$\cdots$&6.25E+30&280&40&AMI&19,20\\
140320B  &$\cdots$&$\cdots$&$\cdots$&$\cdots$&$\cdots$&470&30&AMI&19,20\\
140320C &$\cdots$&$\cdots$&$\cdots$&$\cdots$&$\cdots$&140&40&AMI&19,20\\
140423A  &134&3.26&4.38E+53&$\cdots$&3.35E+31&230&70&AMI&19,20\\
140430A  &173.6&1.6&$\cdots$&$\cdots$&1.37E+32&2800&110&AMI&19,20\\
140606A  &0.34&$\cdots$&$\cdots$&$\cdots$&$\cdots$&530&50&AMI&19,20\\
140606B  &23.6&0.384&2.50E+51&$\cdots$&1.69E+29&50&60&AMI&19,20\\
140607A &109.9&$\cdots$&$\cdots$&$\cdots$&$\cdots$&590&80&AMI&19,20\\
140629A &42&2.275&4.40E+52&$\cdots$&1.29E+31&150&50&AMI&19,20\\
140703A &84&3.14&1.84E+53&$\cdots$&6.78E+31&490&60&AMI&19,20\\
140709A &98.6&$\cdots$&$\cdots$&$\cdots$&$\cdots$&460&40&AMI&19,20\\
140713A &5.3&$\cdots$&$\cdots$&$\cdots$&$\cdots$&1370&40&AMI&19,20\\
140903A&0.3&0.351&4.40E+49&$\cdots$&2.04E+30&720&70&AMI&19,20\\
141015A  &11&$\cdots$&$\cdots$&$\cdots$&$\cdots$&280&60&AMI&19,20\\
141020A  &15.55&$\cdots$&$\cdots$&$\cdots$&$\cdots$&300&60&AMI&19,20\\
141109B  &54.2&$\cdots$&$\cdots$&$\cdots$&$\cdots$&910&250&AMI&19,20\\
141121A &549&1.47&8.00E+52&$\cdots$&1.57E+31&370&40&AMI&19,20\\
141212A  &0.3&0.596&6.80E+49&$\cdots$&1.37E+30&170&40&AMI&19,20\\
141212B  &10.5&$\cdots$&$\cdots$&$\cdots$&$\cdots$&110&30&AMI&19,20\\
150110B &10.6&$\cdots$&$\cdots$&$\cdots$&$\cdots$&530&40&AMI&19,20\\
150413A &263.6&3.139&6.53E+53&$\cdots$&3.18E+31&230&40&AMI&19,20\\
150213B  &181&$\cdots$&$\cdots$&$\cdots$&$\cdots$&140&40&AMI&19,20\\
161219B$^{\star}$&6.94&0.1475&1.16E+50&&1.37E+29&278.1&28.6&VLA&19,21\\
171205A$^{\star}$&189.4&0.0368&2.18E+49&&1.71E+29&5710&50&VLA&19,22\\
180720B$^{\star}$&49&0.654&3.40E+53&&1.06E+31&1096&62&AMI&19,23\\
190114C$^{\star}$&361.5&0.42&2.40E+53&&2.46E+30&607&17.3&VLA&19,23\\
190829A$^{\star}$&58.2&0.0785&2.00E+50&&5.36E+29&3889&197&AMI&19,24\\

	\hline	
\end{longtable}
	\footnotesize
	{\textbf{Note.} \textbf{In Column 1, $\star$ presents the SN/GRB. References are given in order for duration time ($T_{90}$), redshift($z$), isotropic equivalent energy($E_{\gamma,iso}$), peak flux density($f_{p,radio}$) and medium density($n$), repectively. [1]\citet{Chandra+0}; [2]\citet{Friedman+0}; [3]\citet{Ghisellini+0}};[4]\citet{Frail+0};[5]\citet{Bloom+03};[6]\citet{Frail+03};[7]\citet{Panaitescu+02};[8]\citet{Schaefer+03};[9]\citet{Berger+03};[10]\cite{Soderberg+04};[11]\cite{Berger+05};[12]\cite{Cenko+06};[13]\cite{Frail+06};[14]\cite{Soderberg+06a};[15]\cite{Soderberg+06b};[16]\cite{Chandra+08};[17]\cite{Cenko+11};[18]\cite{Chandra+10};[19]\textbf{\url{https://gcn.gsfc.nasa.gov/gcn3_archive.html}};[20]\cite{Anderson+0};[21]\citet{Laskar+01};[22]\citet{Urata+01};[23]\citet{Rhodes+01};[24]\cite{Laskar+02}.
}

\clearpage
\begin{longtable}{p{1cm}<{\centering}p{1cm}<{\centering}p{1cm}<{\centering}p{2cm}<{\centering}p{2cm}<{\centering}p{2cm}<{\centering}p{2cm}<{\centering}p{1cm}<{\centering}p{1cm}<{\centering}p{1cm}<{\centering}}
	
	\caption{Physical parameters of radio-quiet GRBs}
	\label{Table2:radio-quiet} \\
	\hline
	\hline
	\multicolumn{1}{c}{GRB}&\multicolumn{1}{c}{$T_{90}$}&\multicolumn{1}{c}{$z$}&\multicolumn{1}{c}{$E_{\gamma,iso}$}&\multicolumn{1}{c}{$n$}&\multicolumn{1}{c}{$L_{\nu,p}$}&\multicolumn{1}{c}{$f_{p,radio}$}&\multicolumn{1}{c}{RMS}&\multicolumn{1}{c}{Radio Telescope}&\multicolumn{1}{c}{reference}\\
	\multicolumn{1}{c}{}&\multicolumn{1}{c}{$(s)$}&\multicolumn{1}{c}{}&\multicolumn{1}{c}{$(erg)$}&\multicolumn{1}{c}{($cm^{-3}$)}&\multicolumn{1}{c}{$(erg/s/Hz)$}&\multicolumn{1}{c}{$(uJy)$}&\multicolumn{1}{c}{$(uJy)$}&\multicolumn{1}{c}{}&\multicolumn{1}{c}{}\\
	\hline
	\endhead
970228$^{\star}$&56&0.695&1.60E+52&$\cdots$&8.26E+29&76&50&VLA&1\\
971214&35&3.42&2.11E+53&$\cdots$&1.14E+31&73&50&VLA&1\\
980613&50&1.097&6.90E+51&$\cdots$&1.78E+29&7&28&VLA&1\\
990705&42&0.84&1.82E+53&$\cdots$&1.71E+29&11&36&ATCA&1\\
990712&30&0.433&6.72E+51&$\cdots$&2.58E+29&60&50&ATCA&1\\
000630&20&$\cdots$&$\cdots$&$\cdots$&$\cdots$&70&62&VLA&1\\
020124&41&3.2&3.00E+53&3&1.19E+31&84&30&VLA&1,7\\
020305&247&$\textless$2.8&$\cdots$&$\cdots$&$\cdots$&76&15&VLA&1\\
020410&1800&$\cdots$&$\cdots$&$\cdots$&$\cdots$&64&51&ATCA&1\\
021211$^{\star}$&8&1.01&1.10E+52&$\cdots$&1.32E+30&60&28&VLA&1\\
030131&124&$\cdots$&$\cdots$&$\cdots$&$\cdots$&8&35&VLA&1\\
030227&33&$\cdots$&$\cdots$&$\cdots$&$\cdots$&64&24&VLA&1\\
030418&110&$\cdots$&$\cdots$&$\cdots$&$\cdots$&69&27&VLA&1\\
030429&25&2.658&2.19E+52&$\cdots$&9.14E+30&84&54&VLA&1\\
040106&47&$\cdots$&$\cdots$&$\cdots$&$\cdots$&5&50&VLA&1\\
050215B&8&$\cdots$&$\cdots$&$\cdots$&$\cdots$&59&181&VLA&1\\
050306&158&$\cdots$&$\cdots$&$\cdots$&$\cdots$&56&28&VLA&1\\
050408&15&1.236&3.44E+52&0.01&1.58E+29&5&39&VLA&1,6\\
050607&26&$\cdots$&$\cdots$&$\cdots$&$\cdots$&59&23&VLA&1\\
050713A&120&$\cdots$&$\cdots$&3&$\cdots$&17&58&VLA&1,3\\
050801&19&1.38&3.24E+51&1.00$\times10^{-8}$&5.31E+30&139&50&VLA&1,5\\
050814&151&5.3&6.00E+52&$\cdots$&2.01E+31&73&36&VLA&1\\
050815&3&$\cdots$&$\cdots$&$\cdots$&$\cdots$&77&45&VLA&1\\
050915A&52&$\cdots$&$\cdots$&$\cdots$&$\cdots$&43&31&VLA&1\\
051016B&4&0.936&3.70E+50&$\cdots$&6.67E+29&35&13&VLA&1\\
051021A&27&$\cdots$&$\cdots$&$\cdots$&$\cdots$&36&25&VLA&1\\
051109B&14&0.08&3.60E+48&$\cdots$&3.58E+27&25&23&VLA&1\\
051227&115&0.714&8.00E+50&$\cdots$&2.06E+29&18&25&VLA&1\\
060105&54&$\cdots$&$\cdots$&3&$\cdots$&49&47&VLA&1,4\\
060108&14&$\textless$2.8&$\cdots$&$\cdots$&$\cdots$&12&25&VLA&1\\
060124&$\cdots$&$\cdots$&$\cdots$&$\cdots$&$\cdots$&$\textless$59&31&VLA&1\\
060522&71&5.11&7.00E+52&$\cdots$&1.00E+31&38&17&VLA&1\\
060604&95&2.68&4.37E+51&$\cdots$&1.43E+31&130&65&VLA&1\\
060605&79&3.773&2.50E+52&$\cdots$&1.67E+31&94&47&VLA&1\\
060707&66&3.43&6.10E+52&$\cdots$&1.28E+31&82&41&VLA&1\\
060719&67&$\textless$4.6&$\cdots$&$\cdots$&4.15E+31&180&60&ATCA&1\\
060801&0.5&1.131&3.09E+51&$\cdots$&2.83E+30&105&35&VLA&1\\
060825&8&$\cdots$&$\cdots$&$\cdots$&$\cdots$&94&47&VLA&1\\
060908&19&1.884&7.00E+52&10&3.27E+30&51&26&VLA&1,5\\
060912A&5&0.937&8.00E+51&$\cdots$&1.24E+30&65&32&VLA&1\\
060923A&52&$\cdots$&$\cdots$&$\cdots$&$\cdots$&110&55&VLA&1\\
060923C&76&$\cdots$&$\cdots$&$\cdots$&$\cdots$&100&50&VLA&1\\
060926&8&3.209&1.00E+52&$\cdots$&1.34E+31&94&56&VLA&1\\
061028&106&0.76&2.29E+51&$\cdots$&1.03E+30&80&40&VLA&1\\
061126&71&1.159&8.00E+52&1.00$\times10^{-8}$&2.81E+29&10&36&VLA&1,5\\
061210&85&0.41&9.00E+50&$\cdots$&2.62E+29&68&34&VLA&1\\
070220&129&$\cdots$&$\cdots$&$\cdots$&$\cdots$&15&50&VLA&1\\
070223&89&$\cdots$&$\cdots$&$\cdots$&$\cdots$&$\textless$11&47&VLA&1\\
070311&50&$\cdots$&$\cdots$&$\cdots$&$\cdots$&21&51&VLA&1\\
070429B&0.5&0.902&1.35E+50&$\cdots$&7.12E+28&4&100&VLA&1\\
070610&10&$\cdots$&$\cdots$&$\cdots$&$\cdots$&80&100&VLA&1\\
070714B&3&0.923&1.10E+52&0.056$_{-0.011}^{+0.024}$&1.67E+30&$\textless$48&45&VLA&1,2\\
070724B&50&$\cdots$&$\cdots$&$\cdots$&$\cdots$&$\textless$47&36&VLA&1\\
070729&0.9&$\cdots$&$\cdots$&$\cdots$&$\cdots$&$\textless$99&85&VLA&1\\
071010A&6&0.985&1.32E+51&3&1.47E+30&$\textless$66&35&VLA&1,5\\
071011&81&$\cdots$&$\cdots$&$\cdots$&$\cdots$&$\textless$106&60&VLA&1\\
071018&288&$\cdots$&$\cdots$&$\cdots$&$\cdots$&$\textless$3&39&VLA&1\\
071112C&15&0.823&1.95E+52&$\cdots$&1.14E+30&$\textless$57&38&VLA&1\\
080212&117&$\cdots$&$\cdots$&$\cdots$&$\cdots$&83&51&VLA&1\\
080413B&8&1.101&1.59E+52&$\cdots$&2.21E+30&86&36&VLA&1\\
080430&14&0.767&3.00E+51&$\cdots$&8.92E+29&68&46&VLA&1\\
080503&0.3&$\cdots$&$\cdots$&$\cdots$&$\cdots$&3&30&VLA&1\\
080506&152&$\cdots$&$\cdots$&$\cdots$&$\cdots$&$\textless$40&40&VLA&1\\
080507&30&$\cdots$&$\cdots$&$\cdots$&$\cdots$&44&49&VLA&1\\
080603B&59&2.689&7.70E+52&$\cdots$&1.22E+30&11&41&VLA&1\\
080604&69&1.417&7.08E+51&$\cdots$&3.12E+30&$\textless$70&39&VLA&1\\
080613A&30&$\cdots$&$\cdots$&$\cdots$&$\cdots$&7&46&VLA&1\\
080702A&0.5&$\cdots$&$\cdots$&$\cdots$&$\cdots$&$\textless$82&52&VLA&1\\
080721&176&2.591&1.23E+54&$\cdots$&9.75E+30&93&48&VLA&1\\
080913&8&6.733&6.46E+52&$\cdots$&4.07E+31&111&51&VLA&1\\
081024&2&$\cdots$&$\cdots$&8.1$_{-7.7}^{+150}$$\times10^{-5}$&$\cdots$&$\textless$68&52&VLA&1,2\\
081118&49&2.58&2.82E+52&$\cdots$&1.25E+31&$\textless$50&60&VLA&1\\
081126&58&$\cdots$&$\cdots$&$\cdots$&$\cdots$&24&64&VLA&1\\
081203A&223&2.05&3.47E+53&$\cdots$&5.57E+30&76&54&VLA&1\\
081222&33&2.77&3.54E+53&$\cdots$&6.24E+30&54&53&VLA&1\\
090102&29&1.547&1.99E+53&$\cdots$&4.22E+30&91&49&VLA&1\\
090205&9&4.65&2.95E+52&$\cdots$&4.91E+30&21&47&VLA&1\\
090429B&5.5&9.4&5.56E+52&$\cdots$&2.93E+31&55&37&VLA&1\\
090809&8&2.737&1.39E+52&$\cdots$&9.09E+29&8&39&VLA&1\\
090812&75&2.452&4.40E+53&$\cdots$&5.99E+30&104&43&VLA&1\\
100420&48&$\textless$20&$\cdots$&$\cdots$&$\cdots$&24&17&EVLA&1\\
100528A&25&$\cdots$&$\cdots$&$\cdots$&$\cdots$&$\textless$48&46&EVLA&1\\
101112A&35&$\cdots$&$\cdots$&$\cdots$&$\cdots$&149&54&EVLA&1\\
110106B&25&0.618&3.05E+52&$\cdots$&4.16E+29&$\textless$21&24&EVLA&1\\
110731A&38.8&2.83&1.18E+54&$\cdots$&3.82E+30&32&21&EVLA&1\\
120305A  &0.1&$\cdots$&$\cdots$&$\cdots$&$\cdots$&260$\pm$90&110&AMI&8,9\\
120311A  &3.5&$\cdots$&$\cdots$&$\cdots$&$\cdots$&210$\pm$80&80&AMI&8,9\\
120324A  &118&$\cdots$&$\cdots$&$\cdots$&$\cdots$&110$\pm$70&90&AMI&8,9\\
120308A  &60.6&$\cdots$&$\cdots$&$\cdots$&$\cdots$&80$\pm$50&60&AMI&8,9\\
120403A  &1.25&$\cdots$&$\cdots$&$\cdots$&$\cdots$&190$\pm$100&90&AMI&8,9\\
120404A  &38.7&2.876&$\cdots$&$\cdots$&6.71E+31&330$\pm$1090&100&AMI&8,9\\
120422A&5.35&0.28&4.40E+49&$\cdots$&1.79E+28&30$\pm$30&410&AMI&8,9\\
120521C  &26.7&$\cdots$&$\cdots$&$\cdots$&$\cdots$&150$\pm$250&130&AMI&8,9\\
120711B  &60&$\cdots$&$\cdots$&$\cdots$&$\cdots$&230$\pm$60&80&AMI&8,9\\
120722A  &42.4&$\cdots$&$\cdots$&$\cdots$&$\cdots$&670$\pm$1090&510&AMI&8,9\\
120724A  &72.8&1.48&6.02E+51&$\cdots$&4.30E+30&80$\pm$70&350&AMI&8,9\\
120729A  &71.5&0.8&2.30E+52&$\cdots$&2.56E+30&180$\pm$100&100&AMI&8,9\\
120802A  &50&3.796&$\cdots$&$\cdots$&1.40E+32&20$\pm$20&130&AMI&8,9\\
120803B  &37.5&$\cdots$&$\cdots$&$\cdots$&$\cdots$&210$\pm$110&120&AMI&8,9\\
120805A  &48&$\cdots$&$\cdots$&$\cdots$&$\cdots$&370$\pm$140&170&AMI&8,9\\
120811C  &26.8&2.671&8.80E+52&$\cdots$&1.86E+31&350$\pm$300&440&AMI&8,9\\
120816A  &7.6&$\cdots$&$\cdots$&$\cdots$&$\cdots$&440$\pm$730&190&AMI&8,9\\
120819A  &71&$\cdots$&$\cdots$&$\cdots$&$\cdots$&120$\pm$140&210&AMI&8,9\\
120907A  &16.9&0.97&$\cdots$&$\cdots$&3.46E+30&60$\pm$50&110&AMI&8,9\\
120911A  &17.8&$\cdots$&$\cdots$&$\cdots$&$\cdots$&100$\pm$50&90&AMI&8,9\\
120913A  &30.1&$\cdots$&$\cdots$&$\cdots$&$\cdots$&160$\pm$80&70&AMI&8,9\\
120923A  &27.2&$\cdots$&$\cdots$&$\cdots$&$\cdots$&50$\pm$60&100&AMI&8,9\\
120927A  &43&$\cdots$&$\cdots$&$\cdots$&$\cdots$&210$\pm$100&160&AMI&8,9\\
121001A  &147&$\cdots$&$\cdots$&$\cdots$&$\cdots$&190$\pm$130&230&AMI&8,9\\
121011A  &75.6&$\cdots$&$\cdots$&$\cdots$&$\cdots$&190$\pm$100&150&AMI&8,9\\
121017A  &4.2&$\cdots$&$\cdots$&$\cdots$&$\cdots$&310$\pm$870&150&AMI&8,9\\
121028A  &3.8&$\cdots$&$\cdots$&$\cdots$&$\cdots$&20$\pm$20&90&AMI&8,9\\
121108A  &89&$\cdots$&$\cdots$&$\cdots$&$\cdots$&50$\pm$70&90&AMI&8,9\\
121125A  &52.2&$\cdots$&$\cdots$&$\cdots$&$\cdots$&370$\pm$150&170&AMI&8,9\\
121202A  &17.7&$\cdots$&$\cdots$&$\cdots$&$\cdots$&550$\pm$240&290&AMI&8,9\\
121211A  &182&1.023&$\cdots$&$\cdots$&2.24E+30&20$\pm$20&70&AMI&8,9\\
121212A  &6.8&$\cdots$&$\cdots$&$\cdots$&$\cdots$&$\cdots$&70&AMI&8,9\\
130102A  &77.5&$\cdots$&$\cdots$&$\cdots$&$\cdots$&110$\pm$220&60&AMI&8,9\\
130122A  &64&$\cdots$&$\cdots$&$\cdots$&$\cdots$&10$\pm$10&70&AMI&8,9\\
130131A  &51.6&$\cdots$&$\cdots$&$\cdots$&$\cdots$&60$\pm$40&70&AMI&8,9\\
130131B  &4.3&2.539&$\cdots$&$\cdots$&1.02E+31&200$\pm$360&160&AMI&8,9\\
130327A  &9&$\cdots$&$\cdots$&$\cdots$&$\cdots$&50$\pm$40&90&AMI&8,9\\
130418A  &300&1.218&6.30E+52&$\cdots$&2.76E+30&180$\pm$100&130&AMI&8,9\\
130420A  &123.5&1.297&6.20E+52&$\cdots$&2.74E+30&160$\pm$100&120&AMI&8,9\\
130420B  &10.2&$\cdots$&$\cdots$&$\cdots$&$\cdots$&260$\pm$90&100&AMI&8,9\\
130502A  &3&$\cdots$&$\cdots$&$\cdots$&$\cdots$&300$\pm$180&180&AMI&8,9\\
130505A  &88&2.27&3.80E+54&$\cdots$&1.03E+31&40$\pm$30&110&AMI&8,9\\
130511A  &5.43&1.3&$\cdots$&$\cdots$&2.69E+31&780$\pm$1800&390&AMI&8,9\\
130514A  &204&3.6&4.95E+53&$\cdots$&$\cdots$&$\cdots$&1310&AMI&8,9\\
130521A  &11&$\cdots$&$\cdots$&$\cdots$&$\cdots$&220$\pm$520&110&AMI&8,9\\
130603B  &0.18&0.356&2.10E+51&$\cdots$&6.11E+29&90$\pm$50&60&AMI&8,9\\
130609A  &7&$\cdots$&$\cdots$&$\cdots$&$\cdots$&90$\pm$60&90&AMI&8,9\\
130610A  &46.4&2.092&5.78E+52&$\cdots$&1.06E+31&140$\pm$70&100&AMI&8,9\\
130701A  &4.38&1.155&2.10E+52&$\cdots$&4.75E+30&170$\pm$70&70&AMI&8,9\\
130806A&$\cdots$&$\cdots$&$\cdots$&$\cdots$&$\cdots$&640$\pm$2470&150&AMI&8,9\\
130829A  &42.56&$\cdots$&$\cdots$&$\cdots$&$\cdots$&20$\pm$20&110&AMI&8,9\\
130831A  &32.5&0.479&4.60E+51&$\cdots$&1.37E+30&120$\pm$70&70&AMI&8,9\\
130912A  &0.28&$\cdots$&$\cdots$&$\cdots$&$\cdots$&40$\pm$50&50&AMI&8,9\\
131002A  &55.59&$\cdots$&$\cdots$&$\cdots$&$\cdots$&170$\pm$80&90&AMI&8,9\\
131127A  &92.1&$\cdots$&$\cdots$&$\cdots$&$\cdots$&30$\pm$20&70&AMI&8,9\\
131128A  &3&$\cdots$&$\cdots$&$\cdots$&$\cdots$&70$\pm$50&70&AMI&8,9\\
140103A  &17.3&$\cdots$&$\cdots$&$\cdots$&$\cdots$&50$\pm$70&40&AMI&8,9\\
140114A  &139.7&$\cdots$&$\cdots$&$\cdots$&$\cdots$&90$\pm$60&60&AMI&8,9\\
140129B  &1.36&1.5&$\cdots$&$\cdots$&6.17E+30&50$\pm$30&40&AMI&8,9\\
140206A  &93.6&2.74&2.78E+54&$\cdots$&2.96E+31&180$\pm$80&90&AMI&8,9\\
140211A  &89.4&$\cdots$&$\cdots$&$\cdots$&$\cdots$&100$\pm$30&40&AMI&8,9\\
140311B  &70&$\cdots$&$\cdots$&$\cdots$&$\cdots$&20$\pm$30&60&AMI&8,9\\
140419A  &94.7&3.956&1.90E+52&$\cdots$&3.98E+31&100$\pm$50&60&AMI&8,9\\
140428A  &17.42&4.7&$\cdots$&$\cdots$&2.61E+31&100$\pm$40&40&AMI&8,9\\
140502A  &16.9&$\cdots$&$\cdots$&$\cdots$&$\cdots$&500$\pm$1750&60&AMI&8,9\\
140508A  &44.3&1.03&2.10E+53&$\cdots$&2.50E+30&80$\pm$40&50&AMI&8,9\\
140515A  &23.4&6.32&5.38E+52&$\cdots$&$\cdots$&460$\pm$960&120&AMI&8,9\\
140516A  &0.19&$\cdots$&$\cdots$&$\cdots$&$\cdots$&170$\pm$90&70&AMI&8,9\\
140518A  &60.5&4.707&5.98E+52&$\cdots$&4.51E+31&80$\pm$40&40&AMI&8,9\\
140521A  &9.88&$\cdots$&$\cdots$&$\cdots$&$\cdots$&120$\pm$90&40&AMI&8,9\\
140623A  &$\cdots$&$\cdots$&$\cdots$&$\cdots$&$\cdots$&140$\pm$40&50&AMI&8,9\\
140709B  &155&$\cdots$&$\cdots$&$\cdots$&$\cdots$&10$\pm$10&60&AMI&8,9\\
140710A  &3.52&0.558&$\cdots$&$\cdots$&5.68E+29&80$\pm$50&70&AMI&8,9\\
140801A  &7&1.32&4.90E+52&$\cdots$&4.51E+31&140$\pm$50&50&AMI&8,9\\
140817A  &244&$\cdots$&$\cdots$&$\cdots$&$\cdots$&20$\pm$20&50&AMI&8,9\\
140824A  &3.09&$\cdots$&$\cdots$&$\cdots$&$\cdots$&80$\pm$50&90&AMI&8,9\\
140907A  &79.2&1.21&2.71E+52&$\cdots$&2.12E+30&140$\pm$100&110&AMI&8,9\\
140930B  &$\cdots$&$\cdots$&$\cdots$&$\cdots$&$\cdots$&40$\pm$30&50&AMI&8,9\\
141005A  &4.34&$\cdots$&$\cdots$&$\cdots$&$\cdots$&270$\pm$900&70&AMI&8,9\\
141026A  &146&3.35&$\cdots$&$\cdots$&5.60E+31&70$\pm$40&40&AMI&8,9\\
141031B  &16&$\cdots$&$\cdots$&$\cdots$&$\cdots$&50$\pm$30&40&AMI&8,9\\
141130A  &62.9&$\cdots$&$\cdots$&$\cdots$&$\cdots$&$\cdots$&50&AMI&8,9\\
141220A  &7.21&1.3195&2.29E+52&$\cdots$&2.82E+30&20$\pm$30&40&AMI&8,9\\
141225A  &40.24&0.915&8.59E+51&$\cdots$&1.65E+30&30$\pm$30&150&AMI&8,9\\
150101A  &0.06&$\cdots$&$\cdots$&$\cdots$&$\cdots$&$\cdots$&80&AMI&8,9\\
150120A  &1.2&0.46&1.90E+50&$\cdots$&8.73E+29&50$\pm$30&40&AMI&8,9\\
150211A  &13.6&$\cdots$&$\cdots$&$\cdots$&$\cdots$&80$\pm$40&50&AMI&8,9\\
150212A  &11.4&$\cdots$&$\cdots$&$\cdots$&$\cdots$&370$\pm$1810&40&AMI&8,9\\
150302A  &23.74&$\cdots$&$\cdots$&$\cdots$&$\cdots$&$\cdots$&40&AMI&8,9\\
150309A  &242&$\cdots$&$\cdots$&$\cdots$&$\cdots$&70$\pm$30&40&AMI&8,9\\
150314A  &14.79&1.758&6.70E+53&$\cdots$&9.17E+30&110$\pm$60&70&AMI&8,9\\
150317A  &23.29&$\cdots$&$\cdots$&$\cdots$&$\cdots$&80$\pm$40&40&AMI&8,9\\
150323A  &149.6&0.593&1.00E+52&$\cdots$&5.59E+29&30$\pm$20&40&AMI&8,9\\
150323C  &159.4&$\cdots$&$\cdots$&$\cdots$&$\cdots$&10$\pm$10&50&AMI&8,9\\

\hline

\end{longtable}
	\footnotesize
{\textbf{Note.}\textbf{In Column 1, $\star$ presents the SN/GRB. The peak flux density in Column 7 is upper limit. References are given in order for duration time($T_{90}$), redshift($z$), isotropic equivalent energy($E_{\gamma,iso}$) and medium density($n$), respectively. [1]\citet{Chandra+0}; [2]\citet{Fong+0}; [3]\citet{Cusumano+0}; [4]\citet{Tashiro+0}; [5]\citet{Ghisellini+0};[6]\citet{de Ugarte Postigo+0}};[7]\cite{Bloom+03};[8]\textbf{\url{https://gcn.gsfc.nasa.gov/gcn3_archive.html}};[9]\citet{Anderson+0}.
}

\clearpage

\begin{longtable}{p{4cm}<{\centering}p{2cm}<{\centering}p{2cm}<{\centering}p{2cm}<{\centering}p{2cm}<{\centering}p{2cm}<{\centering}p{2cm}<{\centering}p{4cm}<{\centering}}
	
	\caption{Physical parameters of radio-none GRBs}
	\label{Table3:radio-none} \\
	\hline
	\hline
	\multicolumn{1}{c}{GRB}&\multicolumn{1}{c}{$T_{90}$}&\multicolumn{1}{c}{$z$}&\multicolumn{1}{c}{$E_{\gamma,iso}$}&\multicolumn{1}{c}{$n$}&\multicolumn{1}{c}{Radio Telescope}&\multicolumn{1}{c}{reference}\\
	\multicolumn{1}{c}{}&\multicolumn{1}{c}{$(s)$}&\multicolumn{1}{c}{}&\multicolumn{1}{c}{$(erg)$}&\multicolumn{1}{c}{($cm^{-3}$)}&\multicolumn{1}{c}{}&\multicolumn{1}{c}{}\\
	\hline
	\endhead
970111&31&$\cdots$&$\cdots$&$\cdots$&VLA&1 \\
970402&105&$\cdots$&$\cdots$&$\cdots$&ATCA&1 \\
970616&200&$\cdots$&$\cdots$&$\cdots$&VLA&1 \\
970815&130&$\cdots$&$\cdots$&$\cdots$&VLA&1\\
971227&7&$\cdots$&$\cdots$&$\cdots$&VLA&1\\
980109&20&$\cdots$&$\cdots$&$\cdots$&ATCA&1 \\
980326&9&0.9&5.60E+51&$\cdots$&VLA&1 \\
980515&15&$\cdots$&$\cdots$&$\cdots$&ATCA&1 \\
981220&15&$\cdots$&$\cdots$&$\cdots$&VLA&1 \\
990217&$\cdots$&$\cdots$&$\cdots$&$\cdots$&ATCA&-\\
990520&10&$\cdots$&$\cdots$&$\cdots$&VLA&1 \\
990627&59&$\cdots$&$\cdots$&$\cdots$&ATCA&1 \\
990704&23&$\cdots$&$\cdots$&$\cdots$&VLA&1 \\
991014&3&$\cdots$&$\cdots$&$\cdots$&VLA&1 \\
991106&$\cdots$&$\cdots$&$\cdots$&$\cdots$&VLA&-\\
000115&15&$\cdots$&$\cdots$&$\cdots$&VLA&1 \\
000126&70&$\cdots$&$\cdots$&$\cdots$&VLA&1 \\
000214&115&0.47&8.00E+51&$\cdots$&ATCA&1 \\
000226&131&$\cdots$&$\cdots$&$\cdots$&VLA&1 \\
000301A&6&$\cdots$&$\cdots$&$\cdots$&VLA&1\\
000315&60&$\cdots$&$\cdots$&$\cdots$&VLA&1 \\
000326&2&$\cdots$&$\cdots$&$\cdots$&VLA&1 \\
000424&$\cdots$&$\cdots$&$\cdots$&$\cdots$&VLA&-\\
000519&15&$\cdots$&$\cdots$&$\cdots$&VLA&1 \\
000528&80&$\cdots$&$\cdots$&$\cdots$&VLA&1 \\
000529&4&$\cdots$&$\cdots$&$\cdots$&ATCA&1\\
000604&15&$\cdots$&$\cdots$&$\cdots$&VLA&1 \\
000607&0.2&$\cdots$&$\cdots$&$\cdots$&VLA&1 \\
000615A&12&$\cdots$&$\cdots$&$\cdots$&VLA&1 \\
000620&15&$\cdots$&$\cdots$&$\cdots$&VLA&1 \\
000727&10&$\cdots$&$\cdots$&$\cdots$&VLA&1 \\
000801&30&$\cdots$&$\cdots$&$\cdots$&VLA&1 \\
000812&80&$\cdots$&$\cdots$&$\cdots$&VLA&1 \\
000830&9&$\cdots$&$\cdots$&$\cdots$&VLA&1 \\
001025B&0.3&$\cdots$&$\cdots$&$\cdots$&VLA&1 \\
001109&60&$\cdots$&$\cdots$&$\cdots$&VLA&1 \\
001204&0.5&$\cdots$&$\cdots$&$\cdots$&VLA&1 \\
010119&0.2&$\cdots$&$\cdots$&$\cdots$&VLA&1 \\
010213&34&$\cdots$&$\cdots$&$\cdots$&VLA&1 \\
010214&15&$\cdots$&$\cdots$&$\cdots$&VLA&1 \\
010220&40&$\cdots$&$\cdots$&$\cdots$&VLA&1 \\
010728&8&$\cdots$&$\cdots$&$\cdots$&VLA&1 \\
011130&$\textless$3&$\cdots$&$\cdots$&$\cdots$&VLA&1 \\
020127&26&1.9&3.57E+52&$\cdots$&VLA&1 \\
020321&70&$\cdots$&$\cdots$&$\cdots$&ATCA&1 \\
020322&75&$\cdots$&$\cdots$&$\cdots$&VLA&1 \\
020331&75&$\cdots$&$\cdots$&$\cdots$&VLA&1 \\
020409B&$\cdots$&$\cdots$&$\cdots$&$\cdots$&VLA&-\\
020427&66&$\cdots$&$\cdots$&$\cdots$&ATCA& 1\\
020525&25&$\cdots$&$\cdots$&$\cdots$&VLA&1 \\
020531&1&1&$\cdots$&$\cdots$&VLA& 1\\
021008&30&$\cdots$&$\cdots$&$\cdots$&VLA& 1\\
021020&20&$\cdots$&$\cdots$&$\cdots$&VLA&1 \\
021125&25&$\cdots$&$\cdots$&$\cdots$&VLA&1\\
021201&0.3&$\cdots$&$\cdots$&$\cdots$&VLA&1 \\
021219&6&$\cdots$&$\cdots$&$\cdots$&VLA&1\\
030306&20&$\cdots$&$\cdots$&$\cdots$&VLA&1\\
030324&16&&$\cdots$&$\cdots$&VLA&1\\
030528&84&0.782&3.04E+52&$\cdots$&VLA&1 \\
031111&10&$\cdots$&$\cdots$&$\cdots$&VLA&1 \\
040223&258&$\cdots$&$\cdots$&$\cdots$&VLA&1 \\
040701&60&0.21&8.02E+49&$\cdots$&VLA&1\\
040827&49&$\cdots$&$\cdots$&$\cdots$&VLA&1 \\
040912&127&1.563&1.65E+51&$\cdots$&VLA&1\\
040916&450&$\cdots$&$\cdots$&$\cdots$&VLA&1\\
040924&5&0.859&1.10E+52&$\cdots$&VLA&1 \\
041006&25&0.716&3.50E+52&$\cdots$&VLA&1 \\
041218&60&$\cdots$&$\cdots$&$\cdots$&VLA&1 \\
050117A&167&$\cdots$&$\cdots$&$\cdots$&VLA&1\\
050124&4&$\cdots$&$\cdots$&$\cdots$&VLA&1 \\
050126&25&1.29&8.00E+51&$\cdots$&VLA&1\\
050128&19&$\cdots$&$\cdots$&$\cdots$&VLA&1 \\
050202&0.3&$\cdots$&$\cdots$&$\cdots$&VLA&1\\
050319&153&3.24&4.60E+52&1.00E-08&VLA&1,3 \\
050410&43&$\cdots$&$\cdots$&$\cdots$&VLA&1 \\
050412&27&$\cdots$&$\cdots$&$\cdots$&VLA&1 \\
050421&15&$\cdots$&$\cdots$&$\cdots$&VLA&1 \\
050509B&0.07&0.225&2.40E+48&$\cdots$&VLA&1\\
050520&80&$\cdots$&$\cdots$&$\cdots$&VLA&1\\
050522&15&$\cdots$&$\cdots$&$\cdots$&VLA&1\\
050709&0.07&0.161&1.00E+51&1$_{-0.4}^{+0.5}$&VLA&1,2\\
050712&52&$\cdots$&$\cdots$&$\cdots$&VLA&1\\
050714B&54&$\cdots$&$\cdots$&$\cdots$&VLA&1\\
050803&88&$\cdots$&$\cdots$&$\cdots$&VLA&1 \\
050813&0.5&0.72&1.50E+50&$\cdots$&VLA&1\\
050906&0.3&$\cdots$&$\cdots$&$\cdots$&VLA&1\\
050911&16.2&0.165&2.69E+49&$\cdots$&VLA&1\\
050922B&151&$\cdots$&$\cdots$&$\cdots$&VLA&1\\
051006&35&$\cdots$&$\cdots$&$\cdots$&VLA&1\\
051008&280&$\cdots$&$\cdots$&$\cdots$&VLA&1\\
051016A&23&$\cdots$&$\cdots$&$\cdots$&VLA&1\\
051103&0.2&$\cdots$&$\cdots$&$\cdots$&VLA& 1\\
051105A&0.09&$\cdots$&$\cdots$&$\cdots$&VLA&1\\
051114&$\cdots$&$\cdots$&$\cdots$&$\cdots$&VLA&-\\
051117A&136&$\cdots$&$\cdots$&$\cdots$&VLA&1\\
051117B&9&$\cdots$&$\cdots$&$\cdots$&VLA&1\\
060110&26&$\cdots$&$\cdots$&$\cdots$&VLA&1\\
060123&900&1.099&$\cdots$&$\cdots$&VLA&1\\
060206&8&4.05&4.07E+52&2&VLA&1,3\\
060210&255&3.91&4.20E+53&1.00E-08&VLA&1,3 \\
060213&60&$\cdots$&$\cdots$&$\cdots$&VLA&1\\
060313&0.7&$\cdots$&$\cdots$&0.0033$_{-0.5}^{+1}$&VLA&1,2\\
060421&12&$\cdots$&$\cdots$&$\cdots$&VLA&1\\
060428B&58&$\cdots$&$\cdots$&$\cdots$&VLA&1\\
060502B&0.13&0.287&3.00E+49&$\cdots$&VLA&1\\
060505&4&0.089&4.37E+49&$\cdots$&VLA&1\\
070306&210&1.497&6.00E+52&$\cdots$&VLA&1\\
070518&5.5&$\cdots$&$\cdots$&$\cdots$&VLA&1 \\
070721B&32&3.63&3.13E+53&$\cdots$&VLA&1\\
070724B&0.4&0.457&2.45E+49&1.90$_{-1.6}^{+12}$$\times10^{-05}$&VLA&1,2\\
070923&0.2&$\cdots$&$\cdots$&$\cdots$&VLA&1\\
071018&288&$\cdots$&$\cdots$&$\cdots$&VLA&1\\
071112B&0.3&$\cdots$&$\cdots$&$\cdots$&VLA&1\\
080120&15&$\cdots$&$\cdots$&$\cdots$&VLA&1\\
080723B&95&$\cdots$&$\cdots$&$\cdots$&ATCA&1\\
081024B&0.8&$\cdots$&$\cdots$&$\cdots$&VLA&1\\
090417A&0.07&$\cdots$&$\cdots$&$\cdots$&VLA&1\\
090417B&283&0.345&1.10E+51&$\cdots$&VLA&1\\
100424A&104&2.465&$\cdots$&$\cdots$&EVLA&1\\
110721A&$\cdots$&$\cdots$&$\cdots$&$\cdots$&EVLA&-\\

	\hline	
	
\end{longtable}
\footnotesize
{\textbf{Note.} References are given in order for duration time($T_{90}$), redshift($z$), isotropic equivalent energy($E_{\gamma,iso}$) and medium density($n$), respectively. [1]\citet{Chandra+0}; [2]\citet{Fong+0}; [3]\citet{Ghisellini+0}}

\clearpage

\begin{table}
	\centering
	\caption{Statistical parameters of the distributions of $z$, $T_{int}$ and $E_{\gamma,iso}$ }
	\begin{tabular}{cccccccccc}
		\hline
		sample&$\langle logz\rangle$&$\sigma_{logz}$&$\chi^{2}/dof$&$\langle logE_{\gamma,iso}\rangle$&$\sigma_{logE_{\gamma,iso}}$&$\chi^{2}/dof$&$\langle logT_{int}\rangle$ &$\sigma_{logT_{int}}$&$\chi^{2}/dof$\\
		\hline
		radio-loud(N=100)&0.18$\pm$0.014&0.69$\pm$0.031&1.84&53.0$\pm$0.04&1.80$\pm$0.09&1.94&1.48$\pm$0.02&0.86$\pm$0.04&1.72\\
		radio-quiet(N=82)&0.16$\pm$0.01&0.70$\pm$0.025&1.054&52.33$\pm$0.029&1.76$\pm$0.063&0.85&1.08$\pm$0.02&1.06$\pm$0.048&1.82\\
		radio-none(N=25)&-0.12&0.48&-- &51.36&1.46&-- &0.8&1.13&-- \\
		\hline
	\end{tabular}
\label{Table4:statistical para}
\end{table}

\begin{longtable}{p{2cm}<{\centering}p{1cm}<{\centering}p{2cm}<{\centering}p{2cm}<{\centering}p{2cm}<{\centering}p{2cm}<{\centering}p{2cm}<{\centering}}
	
	\caption{The relevant parameters of K-S tests}
	\label{Table5:k-s test} \\
	\hline
	\hline
	\multicolumn{1}{c}{Fig.$^{\dag}$}&\multicolumn{1}{c}{$N_{1}$}&\multicolumn{1}{c}{$N_{2}$}&\multicolumn{1}{c}{$D$}&\multicolumn{1}{c}{$P$}&\multicolumn{1}{c}{$D_{\alpha}(N_1,N_2)$}&\multicolumn{1}{c}{$Note^{\ddag}$}\\
	\hline
	\endhead
1L&84$^a$&63$^b$&0.88&3.84 $\times$ $10^{-19}$&0.23&rejected\\
1R&45$^a$&74$^b$&0.71&3.87$ \times 10^{-14}$&0.26&rejected\\
2L&31$^c$&46$^d$&0.26&0.17&0.32&accepted\\
2L&46$^d$&48$^e$&0.33&8.5$\times$10$^{-3}$&0.28&\textbf{rejected}\\
2L&31$^c$&48$^e$&0.41&2.5$\times$10$^{-4}$&0.31&rejected\\
2R&31$^c$&43$^d$&0.23&0.26&0.32&accepted\\
2R&43$^d$&45$^e$&0.40&1$\times$10$^{-4}$&0.29&rejected\\
2R&31$^c$&45$^e$&0.35&2.7$\times$10$^{-2}$&0.32&\textbf{rejected}\\
3L&77$^f$&48$^e$&0.34&1$\times$10$^{-3}$&0.25&rejected\\
3L&77$^f$&\textbf{25}$^g$&0.30&5$\times10^{-2}$&0.31&\textbf{rejected}\\
3L&48$^e$&25$^g$&0.23&0.24&0.33&accepted\\
3L&21$^h$&34$^i$&0.30&0.15&0.37&accepted\\
3R&74$^f$&45$^e$&0.32&3.9$\times10^{-4}$&0.26&rejected\\
3R&74$^f$&25$^g$&0.50&7.7$\times10^{-5}$&0.31&rejected\\
3R&45$^e$&25$^g$&0.35&3.2$\times10^{-2}$&0.34&\textbf{rejected}\\
3R&21$^h$&34$^i$&0.25&0.31&0.38&accepted\\
4&25$^f$&21$^e$&0.31&0.19& 0.40&accepted\\
6&34$^f$&15$^e$&0.49&1.4$\times10^{-2}$& 0.45&rejected\\
7L&79$^f$&48$^e$&0.40&6.6$\times10^{-5}$&0.25&rejected\\
7R&21$^h$&31$^i$&0.27&0.27&0.38&accepted\\
\hline		
\end{longtable}
\footnotesize
{Note: $N_{1}$ and $N_{2}$ are two sample sizes. $D$ is the K-S test statistic with a $P$ value showing whether the two samples are taken from the same parent distribution. $D_{\alpha}(N_1,N_2)$ is the critical value in contrast with $D$ for a significant level (SL) of $\alpha=0.05$. The diverse samples characterized by whether the radio afterglows are detected or not are denoted by $^a$ for detection, $^b$ for upper limit, $^c$ for radio-loud I, $^d$ for radio-loud II, $^e$ for radio-quiet, $^f$ for radio-loud all, $^g$ for radio-none, $^h$ for AMI radio-loud and $^i$ for AMI radio-quiet.\\ $^{\dag}$ The capital letters represent the right (R) and left (L) panels in the corresponding figures. \\$^{\ddag}$ The bold face indicates those sample pairs with poor K-S test in a lower confidence level.}
\appendix

\clearpage

\section{Comparisons of timescales and energies between different radio-selected GRBs}

\begin{figure*}
	\centering
	\subfigure{
		\includegraphics[width=0.3\textwidth]{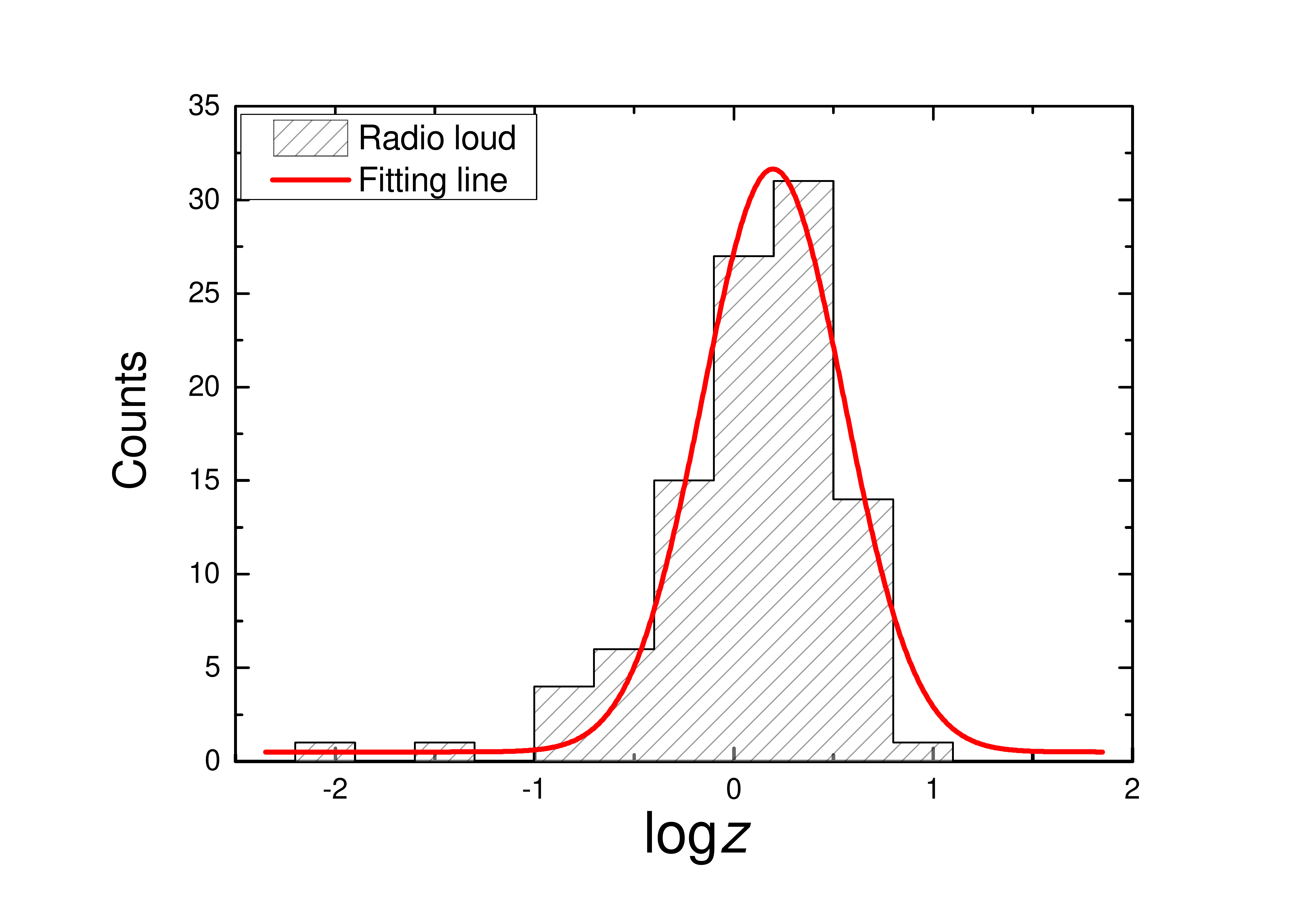}
	}
	\subfigure{
		\includegraphics[width=0.3\textwidth]{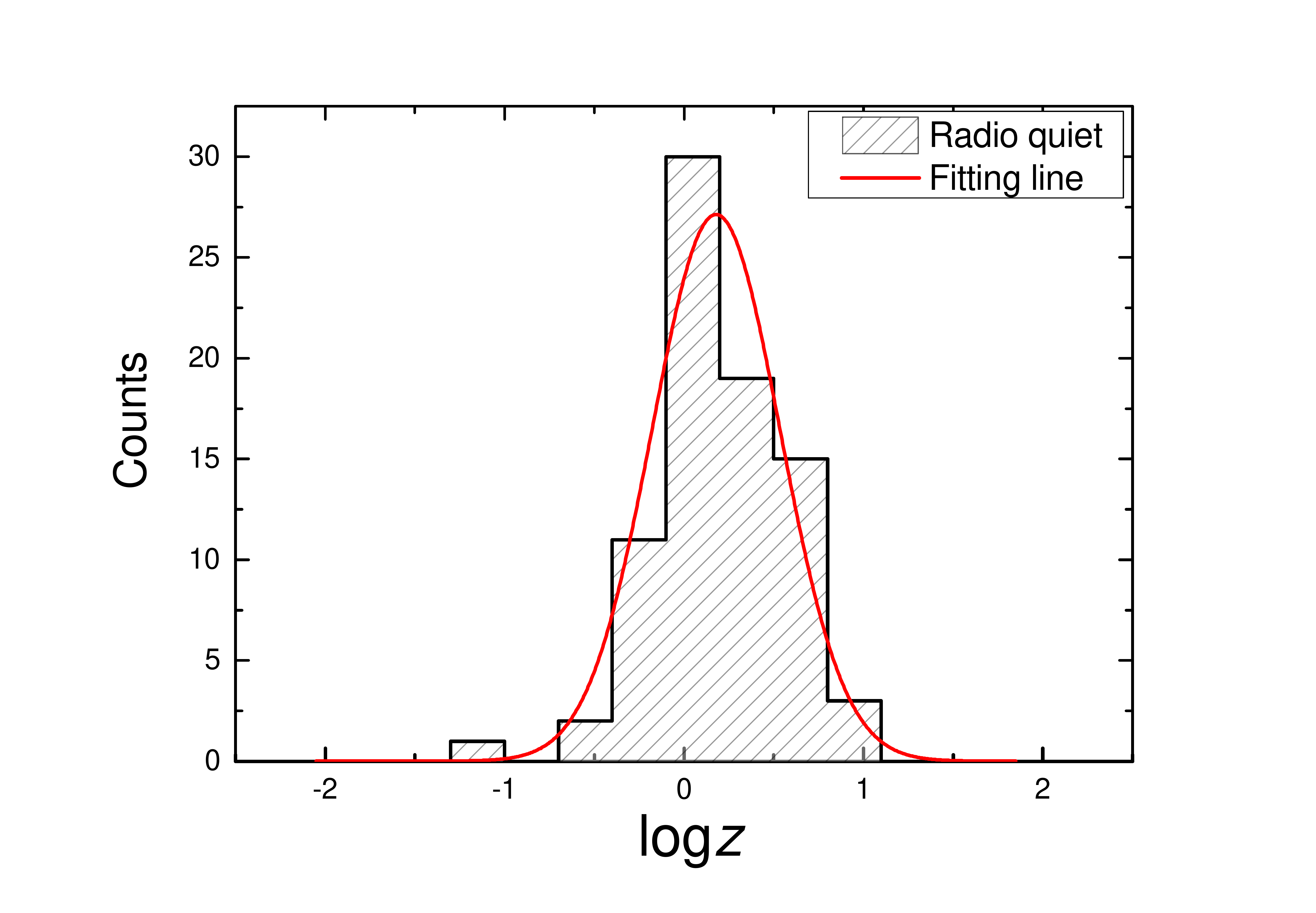}
	}
	\subfigure{
		\includegraphics[width=0.3\textwidth]{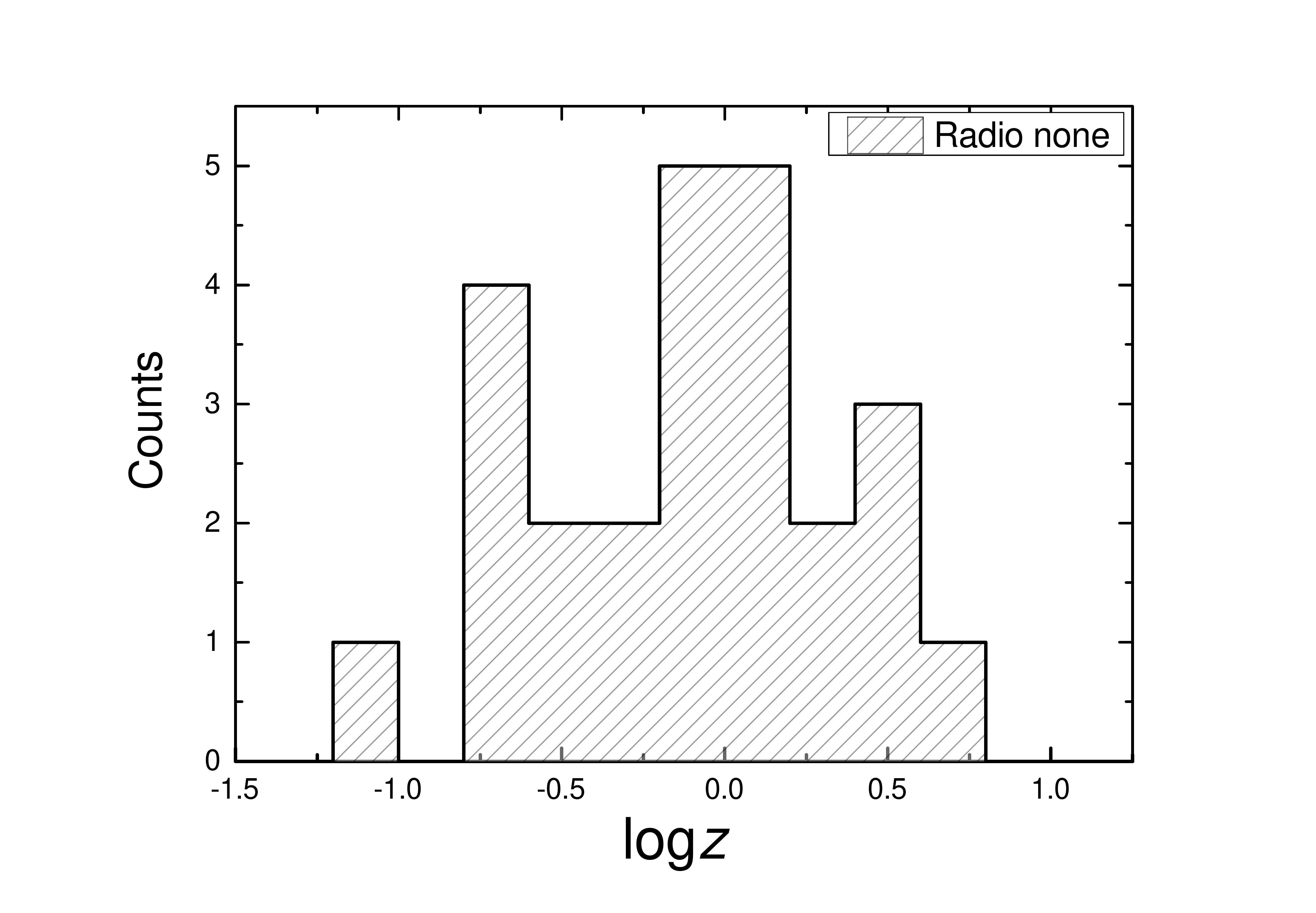}}
	\subfigure{
		\includegraphics[width=0.3\textwidth]{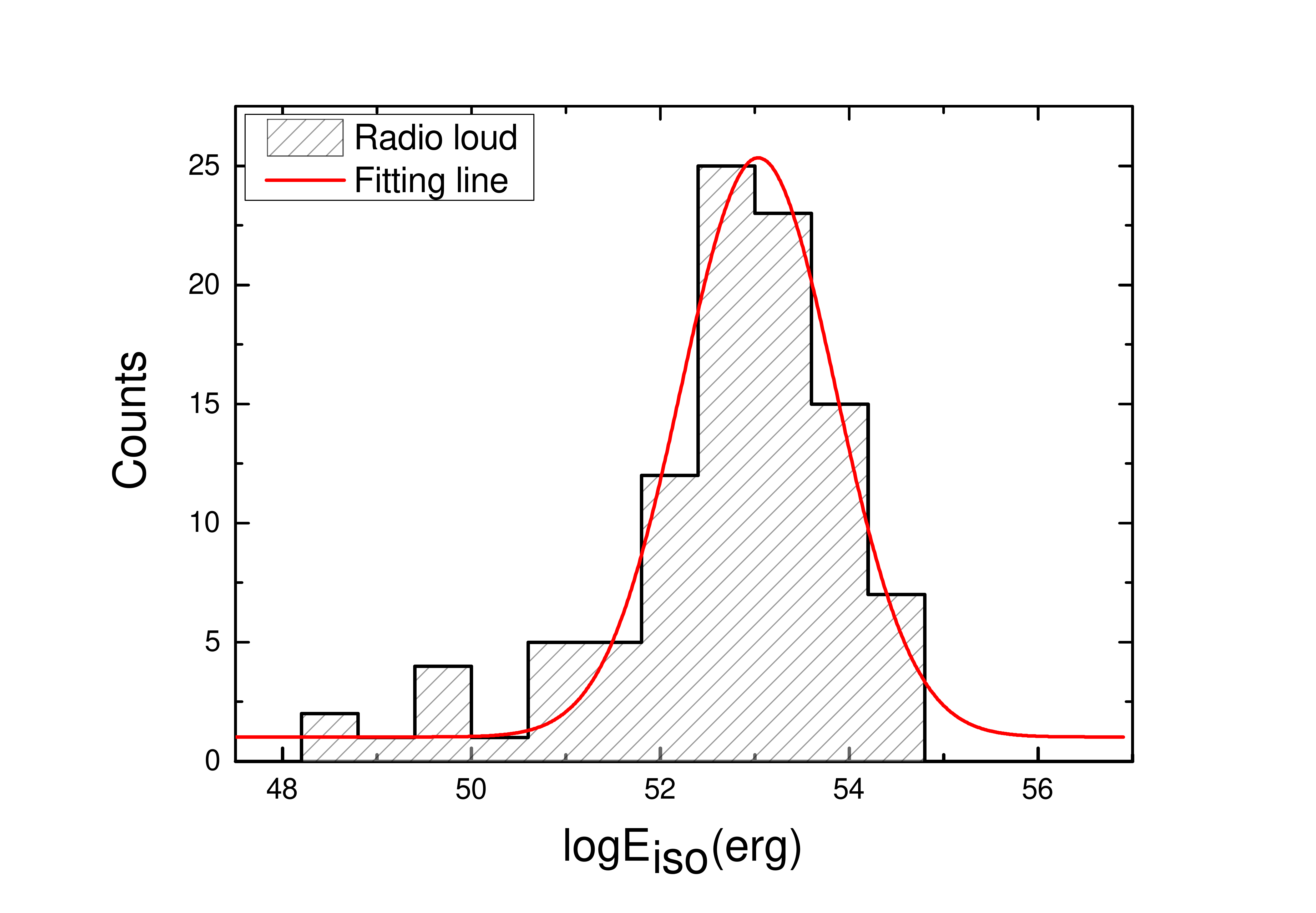}
	}
	\subfigure{
		\includegraphics[width=0.3\textwidth]{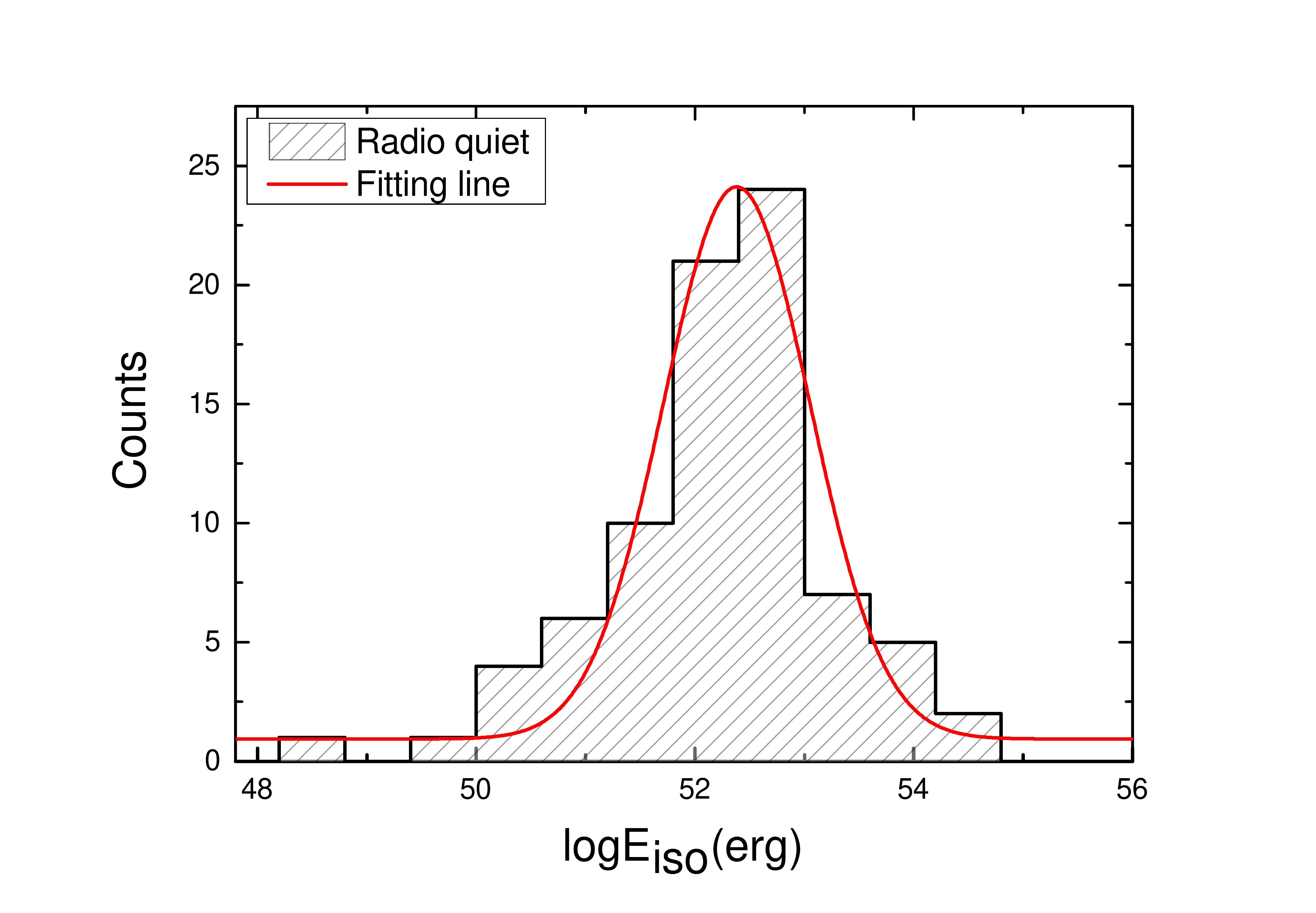}
	}
	\subfigure{
		\includegraphics[width=0.3\textwidth]{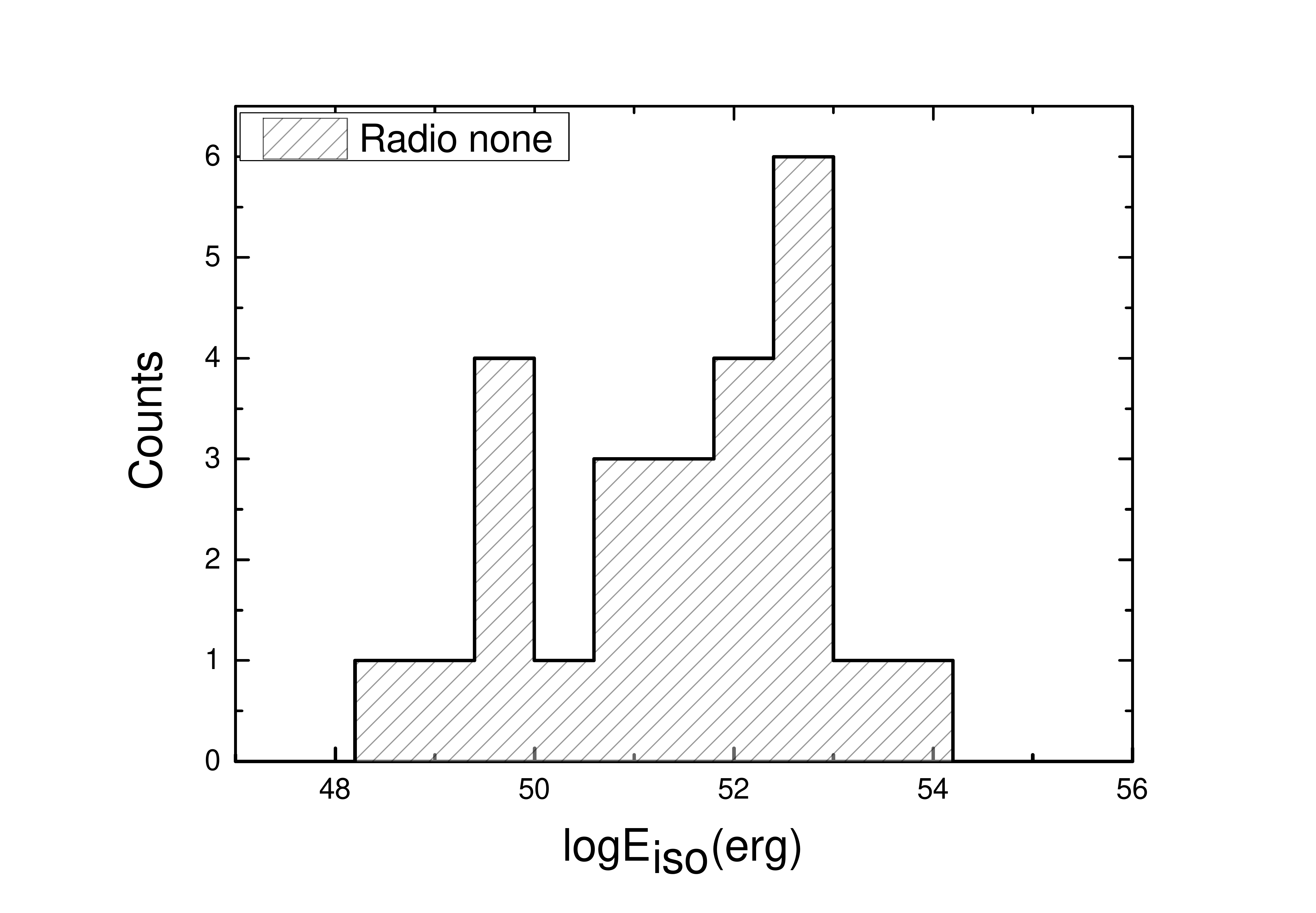}}
	\subfigure{
		\includegraphics[width=0.3\textwidth]{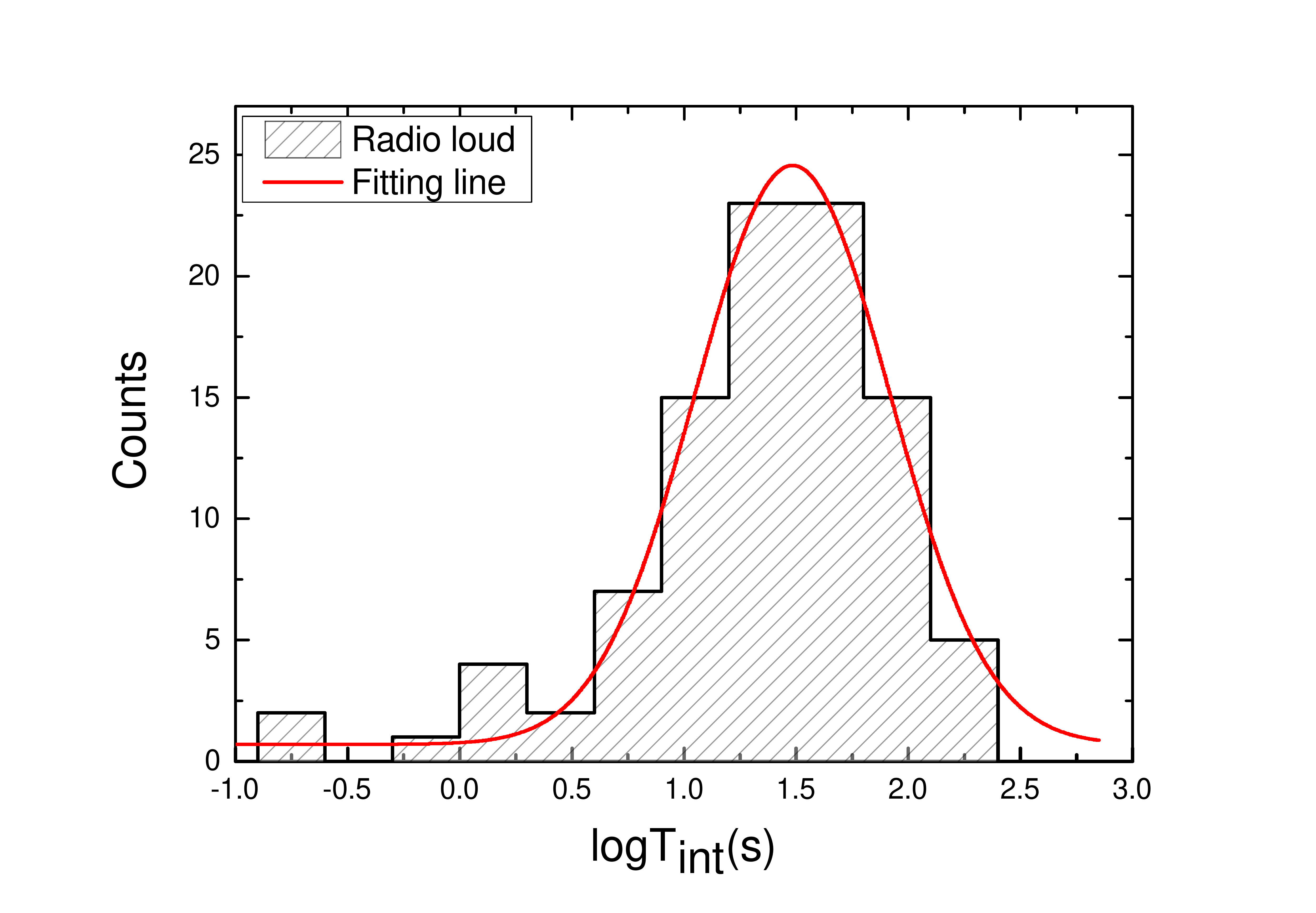}
	}
	\subfigure{
		\includegraphics[width=0.3\textwidth]{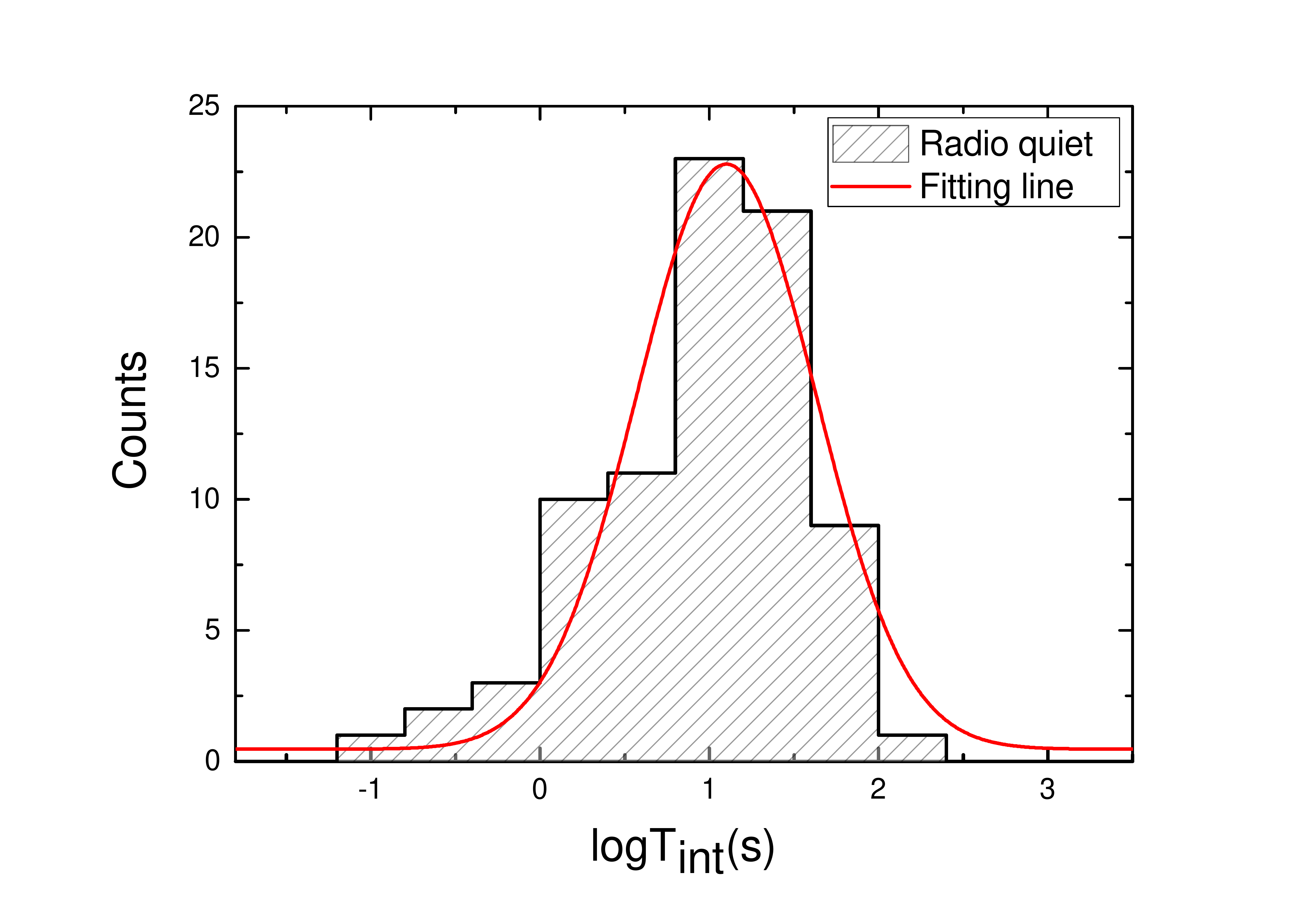}
	}
	\subfigure{
		\includegraphics[width=0.3\textwidth]{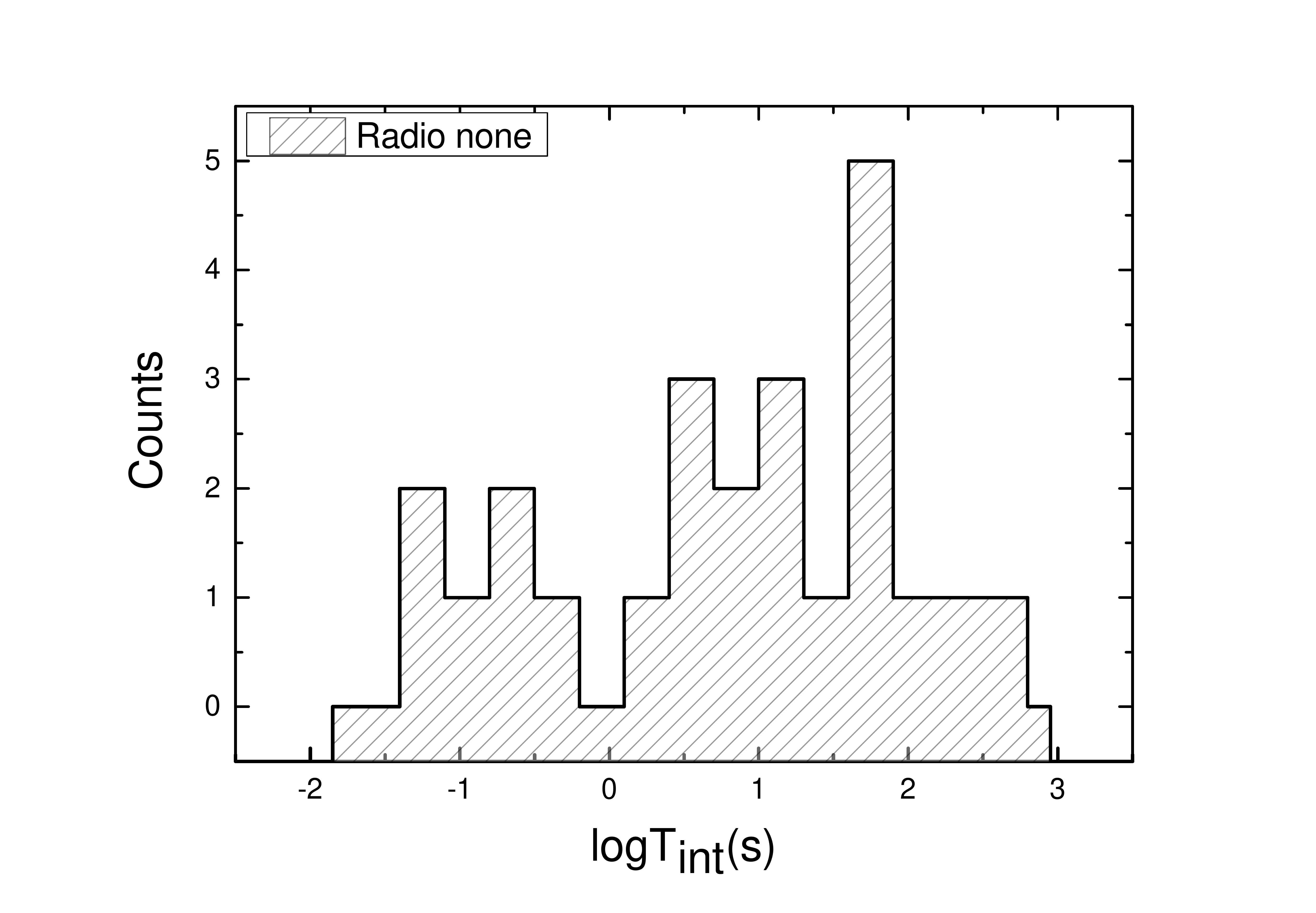}}
	\caption{The histograms of $z$, $E_{\gamma,iso}$ and $T_{int}$ for the radio-loud, radio-quiet and radio-none samples, respectively. The solid lines represent the best fit with a Gaussian function. }
	\label{Figure12-histogram}
\end{figure*}

Here, we combine the VLA-based and the AMI GRBs to expand our sample and explore how the basic parameters of three radio-selected GRBs with known redshift are distributed. In total, 100 radio-loud, 81 radio-quiet and 25 radio-none bursts have been included and compared in Figure \ref{Figure12-histogram}. Interestingly, the mean values of $z$, $T_{int}$ and $E_{\gamma,iso}$ become smaller and smaller and are ranked in order for radio-loud, radio-quiet and radio-none GRBs. More importantly, this implies that radio-none GRBs with the lower $\gamma$-ray energy output and the shorter intrinsic duration time often occur in the nearby universe in contrast with other two kinds of GRBs with radio afterglows. However, only 24\% of radio-none sources belong to short GRBs, which hints that a significant fraction of long GRBs without any radio detections have lower values of $T_{int}$ and $E_{\gamma,iso}$. The new type of long GRBs is obviously different from most SN-associated GRBs with bright radio afterglow but lower $E_{\gamma,iso}$ as shown in Figure \ref{Fig9-LpvsEiso}.

\bsp	
\label{lastpage}
\end{document}